\documentclass[preprintnumbers,superscriptaddress,endnote,nofootinbib,aps,prd,floatfix]{revtex4}
\usepackage{amssymb,amsmath,multirow,graphicx,tabularx,epsfig}
\usepackage{color}
\usepackage{float}
\usepackage{multirow}
\usepackage{verbatim}
\usepackage[normalem]{ulem}
\usepackage{subfigure}
\usepackage{hyperref}
\usepackage{pgf}
\usepackage{tikz}
\usetikzlibrary{arrows,automata}
\usetikzlibrary{positioning}

\tikzset{
    state/.style={
           rectangle,
           draw=black, thin,
           minimum height=2em,
           inner sep=2pt,
           text centered,
           },
}

\newcommand\one{\leavevmode\hbox{\small1\normalsize\kern-.33em1}}

\newcommand{\qqquad}{\qquad \qquad}

\newcommand{\gev}{{\ensuremath\rm GeV}}

\def\slashchar#1{\setbox0=\hbox{$#1$}           
   \dimen0=\wd0                                 
   \setbox1=\hbox{/} \dimen1=\wd1               
   \ifdim\dimen0>\dimen1                        
      \rlap{\hbox to \dimen0{\hfil/\hfil}}      
      #1                                        
   \else                                        
      \rlap{\hbox to \dimen1{\hfil$#1$\hfil}}   
      /                                         
   \fi}

\def\eg{{\sl e.g.} \,}
\def\ie{{\sl i.e.} \,}

\begin{document}

\title{Fox--Wolfram Moments in Higgs Physics}

\author{Catherine Bernaciak}
\affiliation{Institut f\"ur Theoretische Physik, Universit\"at Heidelberg, Germany}

\author{Malte Se\'{a}n Andreas Buschmann}
\affiliation{Institut f\"ur Theoretische Physik, Universit\"at Heidelberg, Germany}

\author{Anja Butter}
\affiliation{Institut f\"ur Theoretische Physik, Universit\"at Heidelberg, Germany}

\author{Tilman Plehn}
\affiliation{Institut f\"ur Theoretische Physik, Universit\"at Heidelberg, Germany}

\begin{abstract}
  Geometric correlations between jets as part of hard processes or in
  addition to hard processes are key ingredients to many LHC analyses.
  Fox--Wolfram moments systematically describe these correlations in
  terms of spherical harmonics. These moments, either computed from
  the tagging jets or from all jets in each event, can significantly
  improve Higgs searches in weak boson fusion. Applications of
  Fox--Wolfram moments in LHC analyses obviously surpass jets as
  analysis objects as well as Higgs searches in terms of analyses.
\end{abstract}

\maketitle

\tableofcontents
\newpage

\newpage

\section{Introduction}
\label{sec:intro}

Fox--Wolfram moments are an established tool to analyze geometric
patterns in QCD~\cite{fwm_orig}, but have never been employed at the
LHC~\cite{field}.  The structure of jet activity in association with a
hard process is a crucial feature for many LHC analyses, in the Higgs
sector~\cite{higgs,review} as well as in new physics
searches~\cite{jamie}. For example, the Higgs discovery by
ATLAS~\cite{atlas} and CMS~\cite{cms} relied on information about jets
produced in association with the Higgs boson in several different
channels. The most established LHC search based on information from
additional jets is weak-boson-fusion (WBF) Higgs
production~\cite{wbf_tau,wbf_w,wbf_gamma}, \ie Higgs production
together with two relatively hard forward tagging
jets~\cite{tagging}. In addition to the two forward tagging jets WBF
Higgs production predicts a lack of central jet activity between
them~\cite{manchester,jetveto1,jetveto2}.  We propose to analyze the
jet activity for example in WBF Higgs production based on the general
framework of Fox--Wolfram moments~\cite{fwm_orig,field}. For our first
study of the jet geometry this process has the advantage that it
includes jets with very different origin: the WBF Higgs signal uses
two tagging jets from the hard process, its $Z+$jets background comes
with two radiated QCD jets, and in the $t\bar{t}$ background at least
one decay jet acts as a tagging jet. \bigskip

Fox--Wolfram moments originate in an expansion of jet--jet
correlations in terms of spherical harmonics. They are constructed by
summing all jet--jet correlations over all $2\ell+1$ directions with a
momentum dependent weight. This way they are sensitive to the number
of jets in the final state, their angular correlation, and their
energy distribution.  Historically, Fox--Wolfram moments have been an
alternative to the usual event shapes~\cite{event_shapes}.
They were tested in the {\sc Aleph} Higgs search in the four-jet channel,
but did not get used in the final analysis. On the other hand, with
some definition of the weight they are available in {\sc
  Pythia}~\cite{pythia} and have been used in $B$ physics. The
question is what we can learn from them for Higgs physics at the
LHC.\bigskip

The objects which enter the computation of Fox--Wolfram moments do not
have to be jets. Because they are closely related to event shapes it
is even likely that some kind of calorimeter entry, particle flow
object, or topocluster will eventually turn out to be more efficient. We
make our first case based on jets because jets allow us to relatively
easily reduce effects from underlying event and pile-up on the
moments. Keeping all jets hard and well separated also ensures that
perturbative QCD is applicable, so we can trust the predictions from
standard QCD Monte Carlo generators with a parton shower.\bigskip

The structure of this paper is simple. First, we introduce the
Fox--Wolfram moments and determine what kind of weight factors are most
appropriate for our purpose. Then we compute Fox--Wolfram moments from
the two tagging jets alone and test how they can improve the WBF Higgs
analysis. Finally, we repeat this analysis with moments including
correlations between all observed jets.

\section{Fox--Wolfram moments}
\label{sec:moments}

For many decades we have known that QCD events can be very efficiently
described by the geometry of the partons. This geometry can be
analyzed at the level of energy deposition, using event shapes, or
based on reconstructed jets. At the LHC the latter have the advantage
that effects from underlying event and pile-up should be easier to
remove. Therefore, we will in this paper focus on the geometric
correlations between jets.\bigskip

For example, cosmological analyses parameterize angular correlations in
terms of spherical harmonics.  Fox--Wolfram moments are based on a
superposition of spherical harmonics, $Y_\ell^m(\theta,\phi)$, where
$\theta$ and $\phi$ are the usual spherical coordinates. They were
originally defined as~\cite{fwm_orig}
\begin{alignat}{1}
H_\ell = \frac{4\pi}{2\ell+1}
       \sum_{m=-\ell}^\ell \;
       \left| \sum_{i=1}^N Y_\ell^m(\Omega_i) \; \frac{|\vec{p}_i|}{\sqrt{s}}
       \right|^2 \; ,
\label{eq:fwm_def1}
\end{alignat}
where the index $i$ sums over all final state objects which can be
defined anywhere at the detector or jet level.  In
Sec.~\ref{sec:tagging} we will start by limiting ourselves to
Fox--Wolfram moments computed from the two tagging jets ($N=2$) while
in Sec.~\ref{sec:qcd} we will also include additional QCD jets with an
event-by-event choice of $N$.  The angular distance $\Omega_i$ assumes
a reference axis which as we will see drops out. The denominator
$\sqrt{s}$ is the energy of all states $i$, ensuring the normalization
$0< H_\ell <1$.  The weight factor $|\vec{p_i}|/\sqrt{s}$ is
only one possible choice so we will adapt it to hadron collider
physics later.

We can rewrite Eq.\eqref{eq:fwm_def1} to express the dependence on the
total angle between each final state object using the addition theorem
for spherical harmonics
\begin{alignat}{1}
H_\ell 
=& \sum_{i,j=1}^N \frac{|\vec{p}_i|}{\sqrt{s}} \frac{|\vec{p}_j|}{\sqrt{s}} \;
  \frac{4\pi}{2\ell+1} \; \sum_{m=-\ell}^\ell
  Y_\ell^m(\Omega_i) Y_\ell^{m*}(\Omega_j)
\notag \\
=& \sum_{i,j=1}^N \; \frac{|\vec{p}_i||\vec{p}_j|}{s} P_\ell(\cos\Omega_{ij}) \; ,
\label{eq:fwm_def2}
\end{alignat}
with the distance measure $\cos\Omega_{ij} = \cos\theta_i\cos\theta_j
+ \sin\theta_i\sin\theta_j\cos(\phi_i-\phi_j)$.\bigskip

At hadron colliders it may be better to use a weight based on the
products of transverse momenta and normalized to their squared
sum~\cite{field}. We study different weight factors $W_{ij}$, namely
\begin{alignat}{1}
H^x_\ell 
=& \sum_{i,j=1}^N \; W_{ij}^x \; P_\ell(\cos\Omega_{ij}) \; ,
\label{eq:fwm_def3}
\end{alignat}
with the specific choices 
\begin{alignat}{3}
W_{ij}^s 
&= \frac{|\vec{p}_i||\vec{p}_j|}{s}  
= \frac{|\vec{p}_i||\vec{p}_j|}{\left(\sum p_i \right)^2}  
\qqquad 
&W_{ij}^p 
&= \frac{|\vec{p}_i||\vec{p}_j|}{|\vec{p}|_\text{tot}^2}
= \frac{|\vec{p}_i||\vec{p}_j|}{\left(\sum |\vec{p}_i|\right)^2}
\notag \\
W_{ij}^T   
&= \frac{p_{T i}\,p_{T j}}{p_{T,\text{tot}}^2}           
= \frac{p_{T i}\,p_{T j}}{\left(\sum p_{T i}\right)^2}  
\qqquad
&W_{ij}^z   
&= \frac{p_{z i}\,p_{z j}}{p_{z,\text{tot}}^2}           
= \frac{p_{z i}\,p_{z j}}{\left(\sum p_{z i}\right)^2}  
\notag \\
W_{ij}^y 
&= \frac{|y_i-\bar{y}|^{-1} |y_j-\bar{y}|^{-1}}
        {\left(\sum |y_i-\bar{y}|^{-1} \right)^2} 
&W_{ij}^1 
&= 1
\label{eq:weights}
\end{alignat}
For the rapidity-based weight we use $\bar{y}$ for the average
rapidity of the two tagging jets.\bigskip

To compare the performance of the different weights listed in
Eq.\eqref{eq:weights} we need a reference process at the LHC.  Higgs
searches use the QCD structure of signal events in the search for
weak-boson-fusion Higgs production~\cite{wbf_tau,wbf_w,wbf_ex}. This
is why we use it, with a decay $H \to \tau^+ \tau^-$ as the reference
channel for our feasibility study. The Higgs decay products do not
enter our analysis; to define the backgrounds we assume that both of
the taus decay leptonically. Two major background processes with
distinctly different jet geometries are $Z+n$~jets production at order
$\alpha \alpha_s^n$ and top pair production~\cite{wbf_tau}. Again, we
do not include the $Z \to \tau^+\ \tau^-$ decays or the $W \to \tau
\nu$ decays, but we do include the corresponding branching ratios for
$H,Z \to \tau^+ \tau^-$ and $t\bar{t} \to b \bar{b} \ell^+ \ell^-
\bar{\nu} \nu$ in all total cross section results shown. Our 
focus is on the jets from the hard process of the Higgs signal, the
QCD jet radiation in the $Z$+jet process and the decay jets in
addition to the QCD jet radiation in top pair production. From jet
scaling studies we know that the number of jets and their transverse
momentum spectra are very different~\cite{jetveto1,jetveto2}. Using
the Fox--Wolfram moments we focus on their angular correlations.

We use {\sc Sherpa}~\cite{sherpa} with {\sc Ckkw} merging~\cite{ckkw}
to generate merged samples of WBF $H$ plus up to three hard jets, $Z$
plus up to two hard jets and $t\bar{t}$ plus up to one hard jet.
Subsequent parton showering and hadronization is modeled also with
{\sc Sherpa}. For jet clustering, we use an anti-k$_T$ algorithm as
implemented in {\sc Fastjet}~\cite{fastjet} with $R=0.4$, \ie the size
of the jets will be small compared to their separation.\bigskip

Our signature consists of two (central) Higgs decay products plus two
tagging jets. The acceptance cuts for the two tagging jets are
\begin{alignat}{5}
p_{T j} > 20~\gev 
 \qqquad 
|y_j| < 5.0
 \qqquad 
\Delta R_{j_1j_2} > 0.7 \; .
\label{eq:jj1}
\end{alignat}
In addition to these minimal cuts we account for the tagging jet
geometry by placing an additional cut on their invariant mass,
\begin{alignat}{2}
m_{j_1 j_2} > 600~\gev \; .  
\label{eq:jj2}
\end{alignat}
Usually, WBF analyses apply two additional conditions on the tagging
jets, namely
\begin{alignat}{2}
y_{j_1} \cdot y_{j_2} &< 0  \qqquad &&\text{(jets in opposite hemispheres)} 
\notag \\
|y_{j_1} - y_{j_2}| &> 4.4  \qqquad &&\text{(jets widely separated)} 
\label{eq:jj3}
\end{alignat}
These latter two cuts have little impact after a hard invariant mass
criterion of Eq.\eqref{eq:jj2} and harm the extraction of the
underling coupling structure~\cite{spin_wbf}.\bigskip

\begin{figure}[t]
\includegraphics[width=0.30\textwidth]{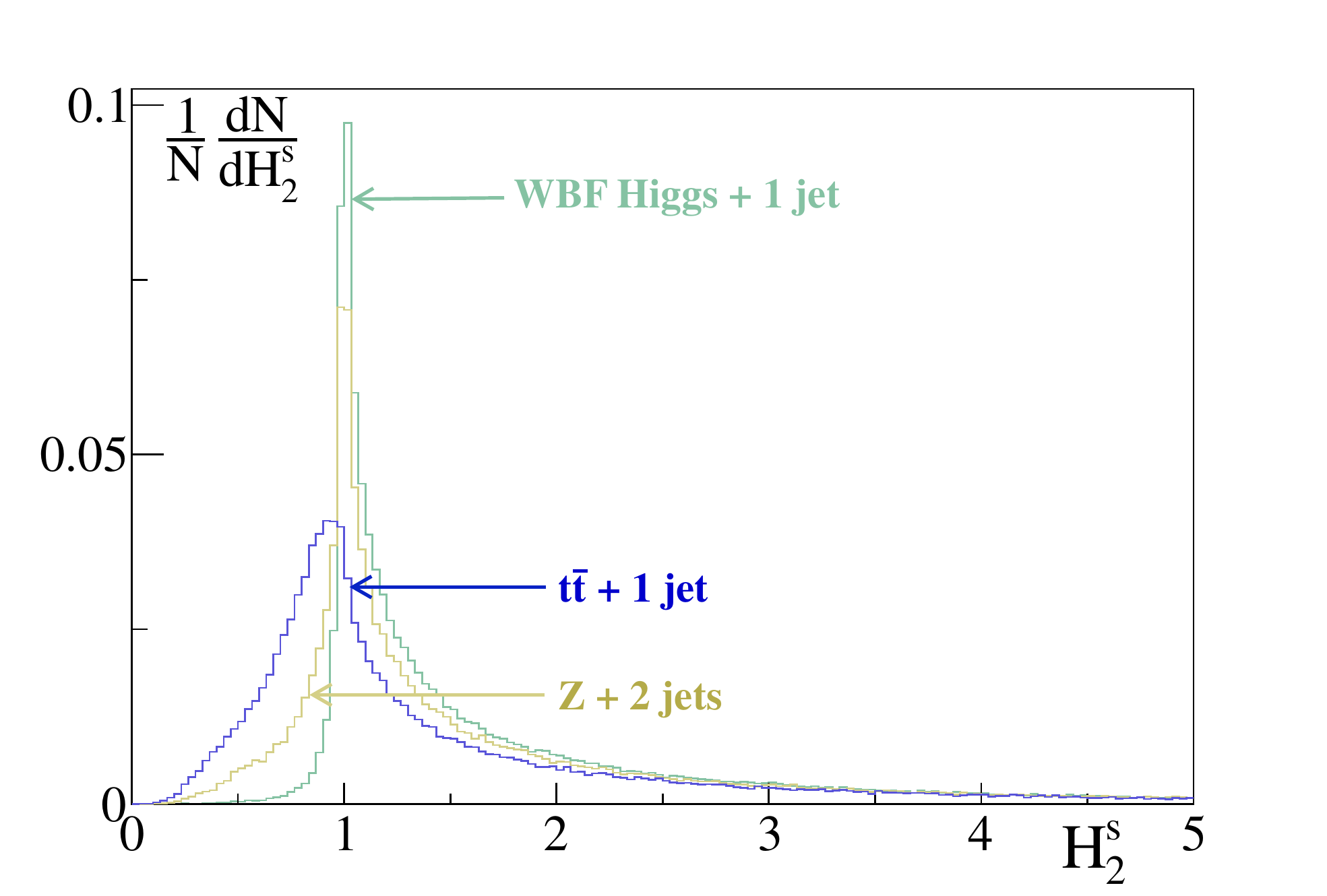}
\includegraphics[width=0.30\textwidth]{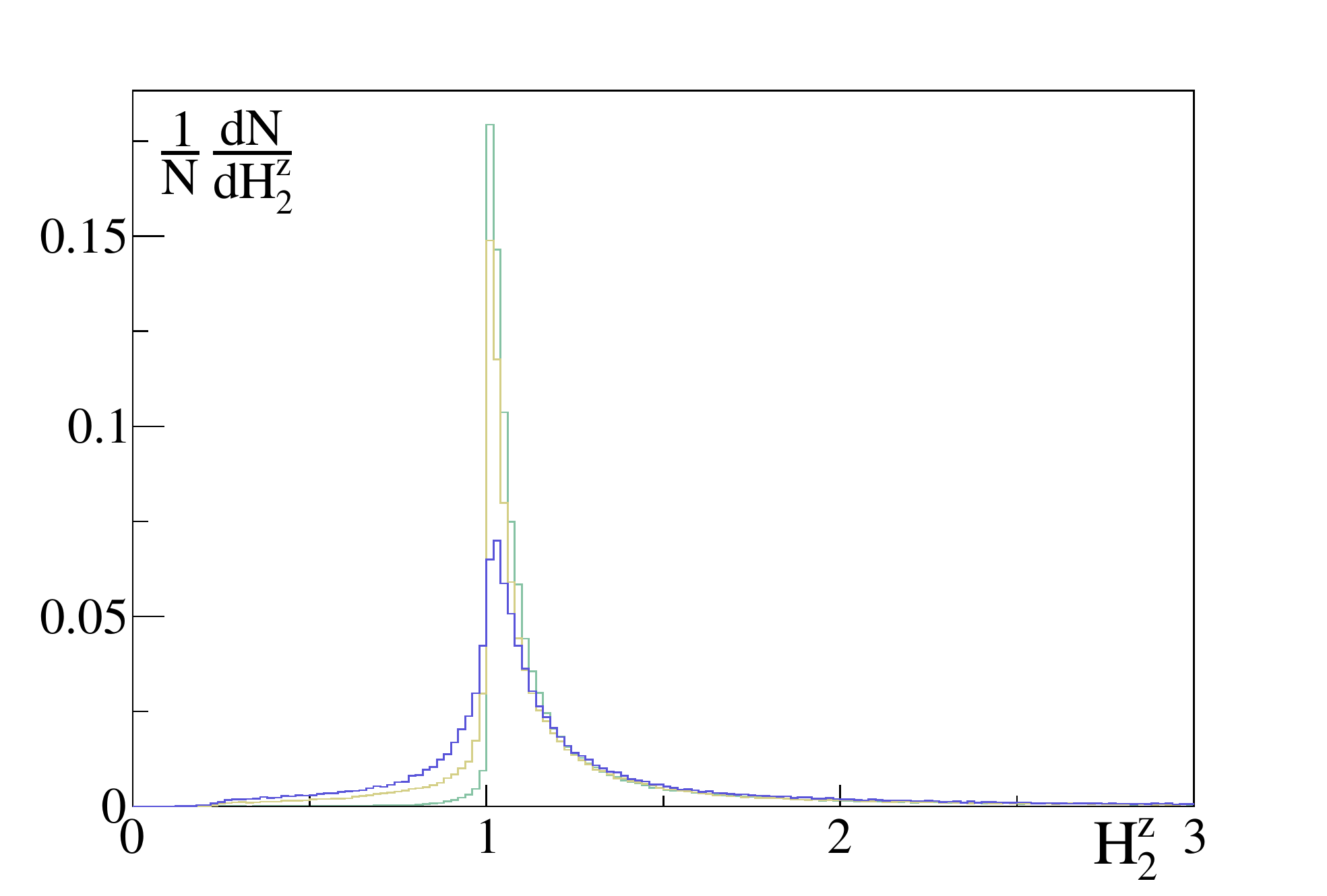}
\includegraphics[width=0.30\textwidth]{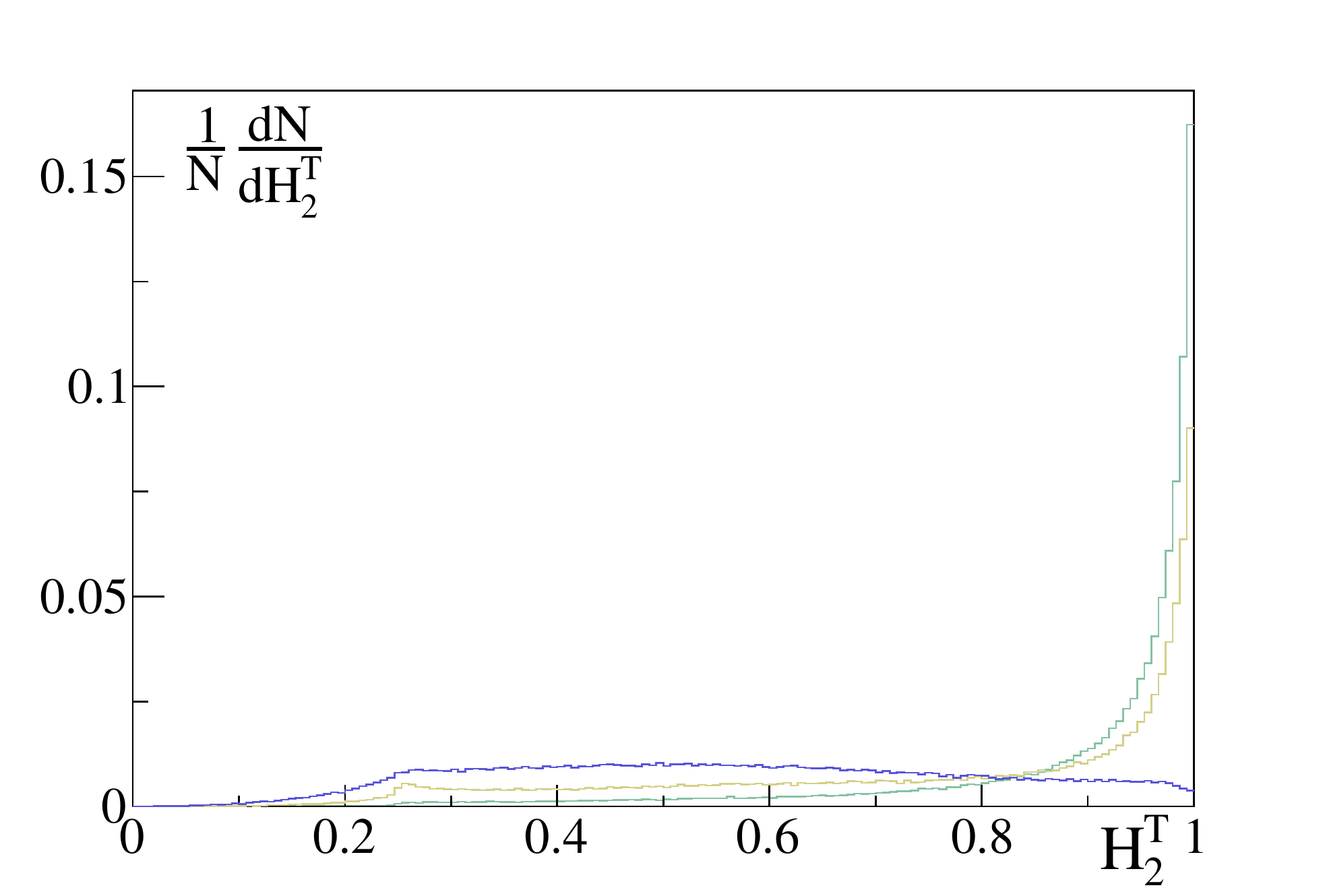} \\
\includegraphics[width=0.30\textwidth]{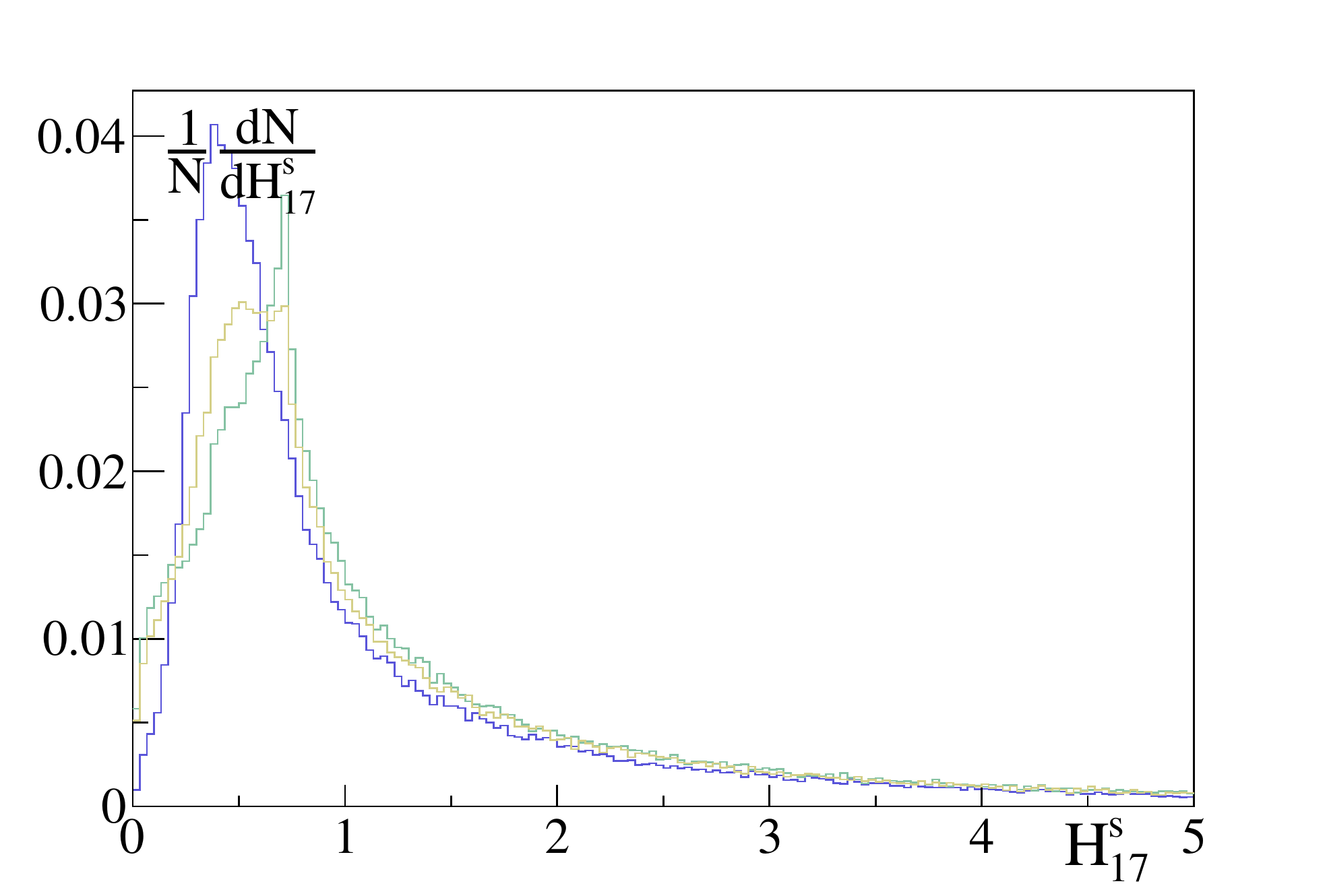}
\includegraphics[width=0.30\textwidth]{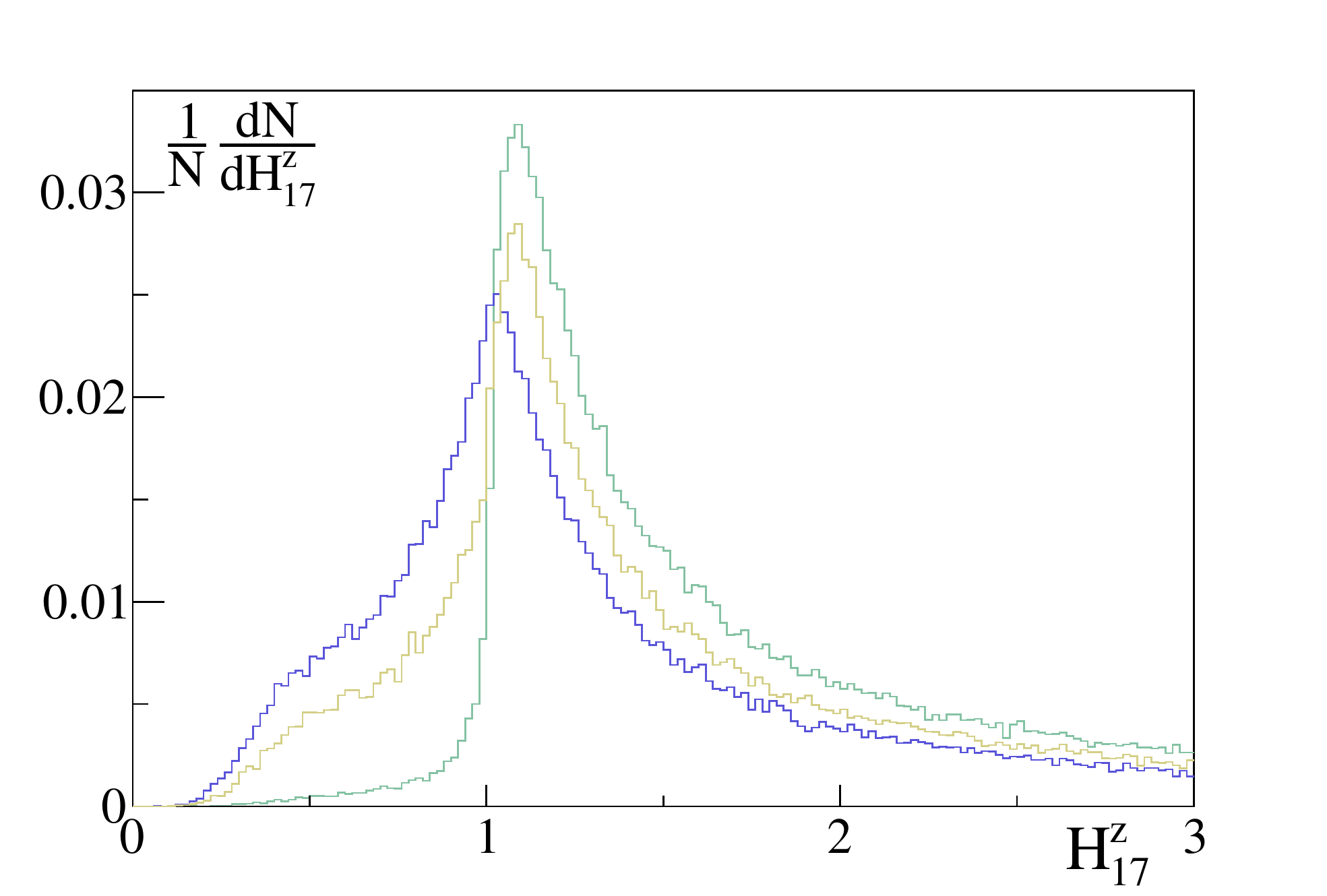}
\includegraphics[width=0.30\textwidth]{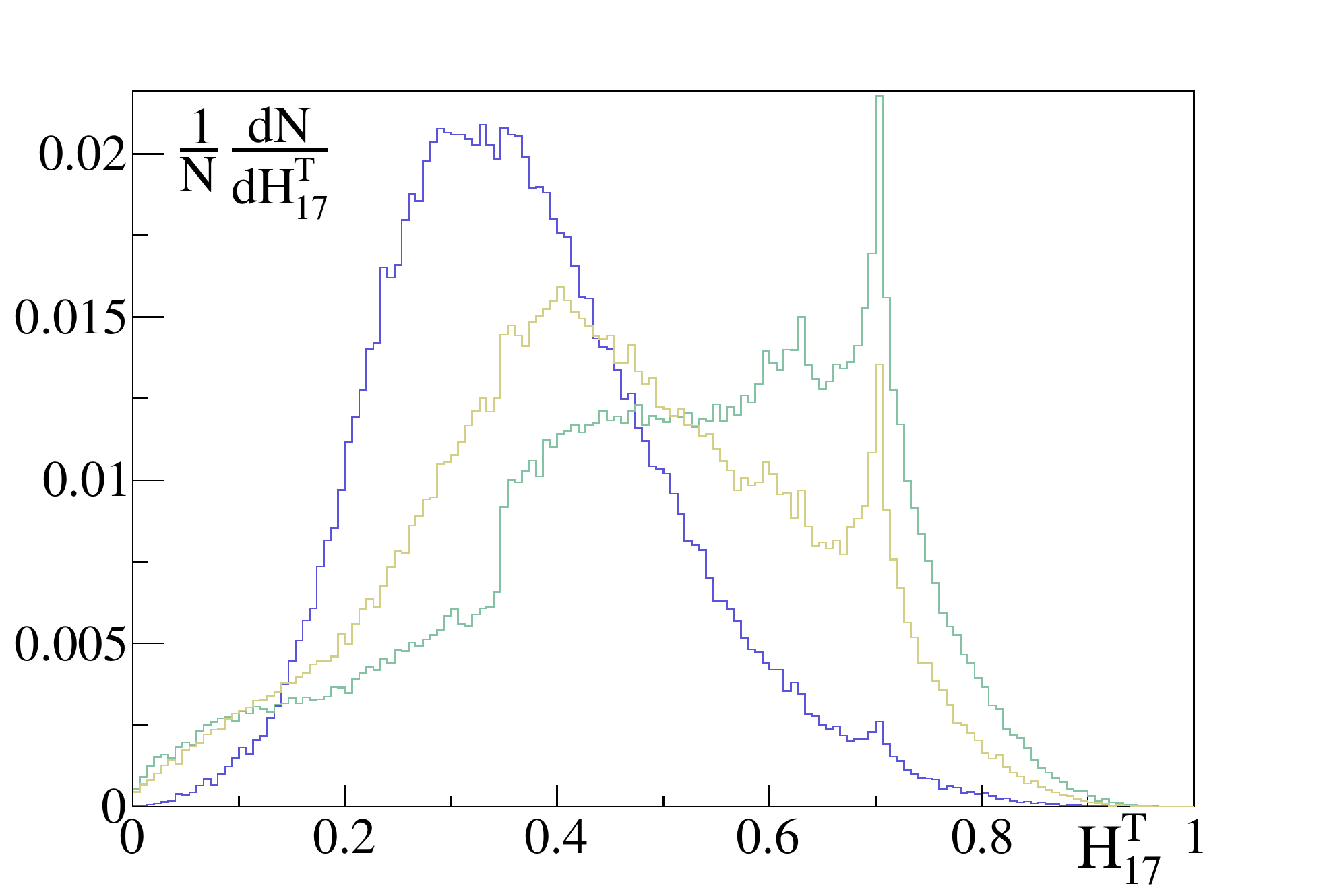} 
\caption{Fox--Wolfram moments for $\ell = 2,17$ with the weight
  factors $W_{ij}^s$ (left), $W_{ij}^z$ (center), and $W_{ij}^T$
  (right).  All jets entering the moments pass the basic acceptance
  cuts of Eq.\eqref{eq:jj1} and Eq.\eqref{eq:jj2}.}
\label{fig:test_roots}
\end{figure}

In Fig.~\ref{fig:test_roots} we show selected Fox--Wolfram moments for
the Higgs signal and the two background processes for three of the
weights shown in Eq.\eqref{eq:weights}. All events pass the acceptance
cuts Eq.\eqref{eq:jj1} as well as the $m_{jj}^\text{min}$ condition
which significantly improves the signal-to-background ratio. All jets
surviving Eq.\eqref{eq:jj1} and Eq.\eqref{eq:jj2} are included in the
moments.  That is, in Fig.~\ref{fig:test_roots} and in
Sec.~\ref{sec:qcd} we do not limit the moments to just two tagging
jets.  From the curves it is clear that $W_{ij}^s$ is less efficient
at discerning between the signal and backgrounds. This trend persists
for other moments not shown.  All weights except for $W_{ij}^s$ and
$W_{ij}^z$ ensure that the range for the moments is $0 \leq H_\ell
\leq 1$ and preserve the different shapes of $H_\ell$ for even or odd
moments. Without showing detailed results we conclude that $W_{ij}^y$
is not very efficient in extracting the WBF signal.  $W_{ij}^T$ and
$W_{ij}^p$, and to some degree $W_{ij}^1$, have more power to separate
the different background from the signal, so we focus on them for the
rest of this paper..\bigskip

For different values of $\ell$ the Fox--Wolfram moments $H_\ell$
reflect the strongly oscillatory behavior of the Legendre
polynomials. This can be viewed as a change in the resolution with
which the Fox--Wolfram moments probe the structure of the QCD jet
geometry. We illustrate these patterns based on a toy model with only
two final state objects, $N=2$. This corresponds to the distinctive
tagging jets in WBF Higgs production.

\begin{figure}[t]
\centering
\includegraphics[width=0.7\textwidth]{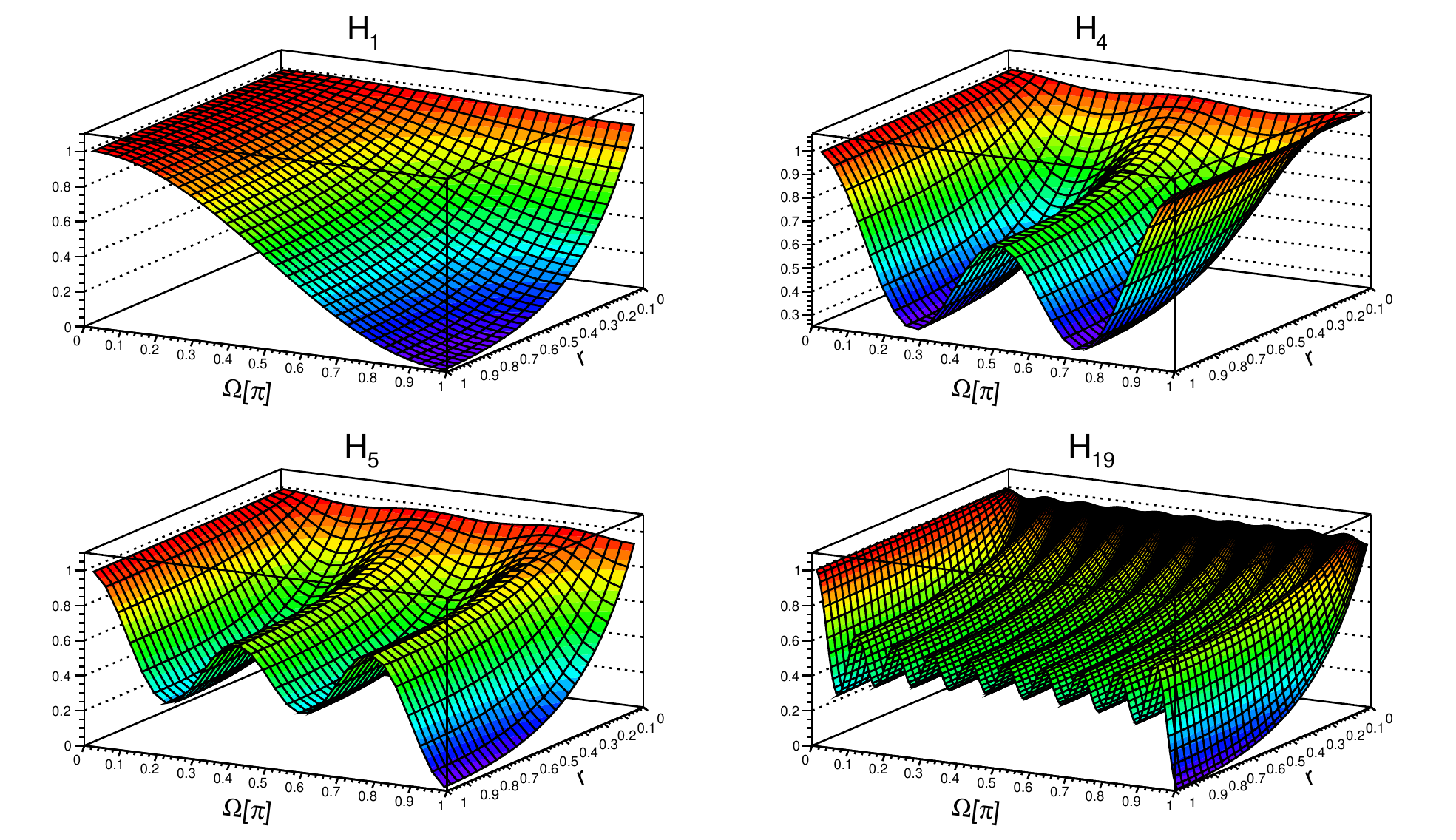}  
\caption{Fox--Wolfram moments $H^{p,T}_\ell$ for two jets as a
  function of $\Omega_{12}$ and $r = r_{p,T}$. We show the analytic
  result in our toy model. Eq.\eqref{eq:fwm_toy}}
\label{fig:3D}
\end{figure}

To simplify the functional form of the weights $W_{ij}$ we denote
$|\vec{p}_2| = r_p \; |\vec{p}_1|$ and $p_{T 2} = r_T \; p_{T 1}$
with $r_{p,T} = 0...1$. Expanding the sum in the definition
Eq.\eqref{eq:fwm_def3} yields
\begin{alignat}{5} 
H_\ell(\Omega_{12},r) = \frac{1+2rP_\ell(\cos\Omega_{12})+r^2}{1+2r+r^2}
\qqquad 
(r = r_{p,T}) \; .
\label{eq:fwm_toy}
\end{alignat}
When the two jets are back-to-back the even and odd moments each
display general properties.  For all even moments $\Omega_{12} \to
\pi$ implies $H_\ell \to 1$, independent of $r$.  For all odd
moments, we find $H_\ell \rightarrow 0$ in the limit
$r\rightarrow 0$.  These general trends are apparent in
Fig.~\ref{fig:3D} which shows selected Fox--Wolfram moments given by
Eq.\eqref{eq:fwm_toy} as a function of $\Omega_{12}$ and $r$.

Also in the limit $r\rightarrow 0$, the Fox--Wolfram moments depend on
the angle only weakly and essentially become independent of
$\ell$. That is to say, for strongly hierarchical jets the moments
defined including a momentum dependent weight are not a good
descriptor of jet geometry, as all values of $\ell$ for all values of
$\Omega_{12}$ will tend towards unity.\bigskip

The power of the moments based only on two back-to-back tagging jets
lies in the $r \gtrsim 0.4$ regime. While for $\ell = 1$ the shape of
the Legendre Polynomial dominates, higher values of $\ell$ make the
Fox--Wolfram moments more sensitive to larger and larger angles
between the two (tagging) jets.  For moderate even $\ell$ a pair of
WBF tagging jet will typically give $H_4 \sim 1$, independent of $r$.
Less widely separated jet are limited to $H_4 \lesssim 0.5$.  Odd
values, for instance $H_5$, have a distinct dependence on $r$, where
WBF tagging jets will give $H_5 \sim 1$ for hierarchical jets with $r
\to 0$ and $H_5 \to 0$ for balanced jets $r \to 1$. We will see that
for typical tagging jets the maximum ranges around $H_5 \sim 0.7$.

Finally, large $\ell$ values increase the sensitivity to the details
of forward jet emission, for example distinguishing collinearly
enhanced QCD jet radiation in $Z$+jets events from finite-$p_T$
tagging jets in Higgs production.  As we can see in Fig.~\ref{fig:3D},
$H_{19}$ is highly sensitive to signal events peaking around
$\Omega_{12} = \pi$. For background processes with a broad range of
central jets this region has a discernibly smaller impact.  Because
$H_{19}$ is an odd moment its $r$ dependence will still lead to a
strong peak around values of 0.7 for WBF events.  To take away from
Fig.~\ref{fig:3D}, different regions in the $(\Omega_{12},r)$ plane
translate into distinct regimes in $H_\ell$:
\begin{itemize}
\item for even $\ell$ small values of $H_\ell$ are not allowed;
  intermediate values $0.3 \lesssim H_\ell \lesssim 0.7$ appear for
  democratic jet radiation $r \gtrsim 0.4$; large values $H_\ell
  \gtrsim 0.7$ come from three regimes: strongly ordered jets $r
  \lesssim 0.4$, collinear jets $\Omega \lesssim 0.1$ or back-to-back
  jets $\Omega \gtrsim 0.9$.
\item for odd $\ell$ small values $H_\ell \lesssim 0.3$ indicate
  symmetric back-to-back jets; intermediate values $0.3 \lesssim
  H_\ell \lesssim 0.7$ correspond to relatively large $r$ values with
  a sizeable angular separation; large $H_\ell \gtrsim 0.7$ values are
  only possible for collinear or very hierarchical jets.
\end{itemize}

The distinction between collinear and back-to-back jets through the
low-$H_\ell$ regime suggests that odd Fox--Wolfram moments are more
sensitive to WBF production processes.  We will test the quantitative
impact of both patterns in Sec.~\ref{sec:tagging} based on an
appropriate simulation.

Once we include more than two jets, even moments can take small
values.  Consider for example an event with three democratic planar
jets, \ie $p_{T2}/p_{T1} = 0.7$ and $p_{T3}/p_{T1} =0.3$.  If the
hardest two jets are mostly forward-backward and the third jet is
central, we obtain $H^T_4 \sim 0.2$.\bigskip

In the definition, Eq.\eqref{eq:fwm_def3}, we see that the angular
dependence of the Fox--Wolfram moments alone does not allow us to
separate azimuthal and polar angular separation of the jets. This
means that $H^p$ will be insensitive to the details of the opening
angle. However, for $H^T$ the weight introduces a sensitivity to the
polar vs azimuthal separation.  This may make it possible to study the
CP nature of the resonance based on the well-known differences in
azimuthal angle and in rapidity between the tagging
jets~\cite{spin_wbf}.

We can simply apply a few useful trigonometric identities and see how
the Legendre polynomials $P_\ell$ are a function of
\begin{alignat}{5}
\cos \Omega_{ij} =  \frac{1}{2}
\left[ \cos \theta_{ij}^{(+)} \left( 1-\cos \phi_{ij}^{(-)} \right)
     + \cos \theta_{ij}^{(-)} \left( 1+\cos \phi_{ij}^{(-)} \right )
\right] \; .
\label{eq:cosine}
\end{alignat}
We introduce the compact notation $\theta_{ij}^{(+)} \equiv \theta_i +
\theta_j$ and $\theta_{ij}^{(-)} \equiv \theta_i - \theta_j$ and
likewise for $\phi_{ij}^{(\pm)}$.  Weighting $H_\ell^T$ with $p_T$
introduces a dependence on the polar angle through 
\begin{alignat}{5}
p_{Ti}\,p_{Tj} 
= |\vec{p}_i||\vec{p}_j| \sin\theta_i \sin\theta_j 
= \frac{|\vec{p}_i||\vec{p}_j|}{2}
  \left( \cos \theta_{ij}^{(-)} - \cos \theta_{ij}^{(+)}
  \right)  \; ,
\end{alignat}
so the full dependence of $H_\ell$ on the angular correlations becomes
\begin{alignat}{5}
H^T_\ell 
= \frac{1}{2p^2_{T,tot}} \sum_{i,j=1}^N |\vec{p}_i||\vec{p}_j|
  \left( \cos \theta_{ij}^{(-)} - \cos \theta_{ij}^{(+)}
  \right) \; 
  P_\ell(\theta_{ij}^{(\pm)},\phi_{ij}^{(-)}) \; .
\label{eq:fwm_angles}
\end{alignat}
%

\section{Two tagging jets}
\label{sec:tagging}

\begin{table}[!b]
\centering
\begin{tabular}{l|rr|rr|rr|r}
\hline
& \multicolumn{2}{c|}{WBF $H+2$~jets}  
& \multicolumn{2}{c|}{QCD $Z+2$~jets}
& \multicolumn{2}{c|}{$t\bar{t}+1$~jet}
&  S/B \\
acceptance &\% fail & $\sigma$(fb)  
           &\% fail & $\sigma$(fb) 
           &\% fail & $\sigma$(fb) &  \\
\hline
                         &      & 18.7 &      & 115000 &       & 17200  & 1/7070 \\
$p_{Tj}>20$~GeV          & 29.4 & 13.2 & 93.2 & 7820   & 9.63  & 15500  & 1/1767 \\
$|y_j|< 5.0$             & 1.49 & 13.0 & 0.97 & 7740   & 0.182 & 15500  & 1/1788 \\
$\Delta R_{jj}>0.7$      & 2.73 & 12.6 & 3.84 & 7440   & 2.32  & 15100  & 1/1789 \\
$m_{jj} > 600$~GeV       & 68.9 & 3.92 & 96.6 & 253    & 95.8  & 634    & 1/226 \\
$b$-veto                 &  0   & 3.92 &  0   & 253    & 54.0  & 292    & 1/139 \\
\hline
$y_1\cdot y_2<0$         & 1.41 & 3.86 & 9.17 & 230    & 13.8  & 252    & 1/125 \\
$\Delta y_{jj}\geq 4.4$  & 13.9 & 3.32 & 31.8 & 157    & 66.1  & 85.4   & 1/73 \\ 
\hline
\end{tabular}
\caption{Cut flow of the signal and background processes after the
  cuts Eq.\eqref{eq:jj1}-\eqref{eq:jj3} and the $b$ veto defined in
  Eq.\eqref{eq:bVETO}. The rates include the branching ratios $H,Z \to
  \tau^+ \tau^-$ and two leptonic top quarks, but no requirements on
  the leptons. The usual cuts requiring central Higgs decay products
  of the central jet veto are not applied in this table.}
\label{tab:CFclassic}
\end{table}

Tagging jets are the key to identifying Higgs bosons with a decay to
tau pairs. In addition, they allow us to extract weak-boson-fusion
events from the large gluon-fusion background, one of the key
ingredients to a Higgs coupling
measurement~\cite{higgs_couplings}. Usually, we require (at least) two
additional jets fulfilling the conditions in
Eqs.\eqref{eq:jj1}-\eqref{eq:jj3}. In addition, the Higgs decay
products have to lie centrally between the two tagging jets, a
requirement we do not make explicit because we omit any information
about the Higgs decay product for the sake of a most general
study. Eventually, it can be added to further improve the
signal-to-background ratios quoted in our analysis.  From the study of
Higgs coupling structures we know that Eq.\eqref{eq:jj3} with an
explicit cut on $\Delta y_{jj}$ removes one of the most promising
observables from the analysis~\cite{spin_wbf}.\bigskip

\begin{figure}[t]
\includegraphics[width=0.24\textwidth]{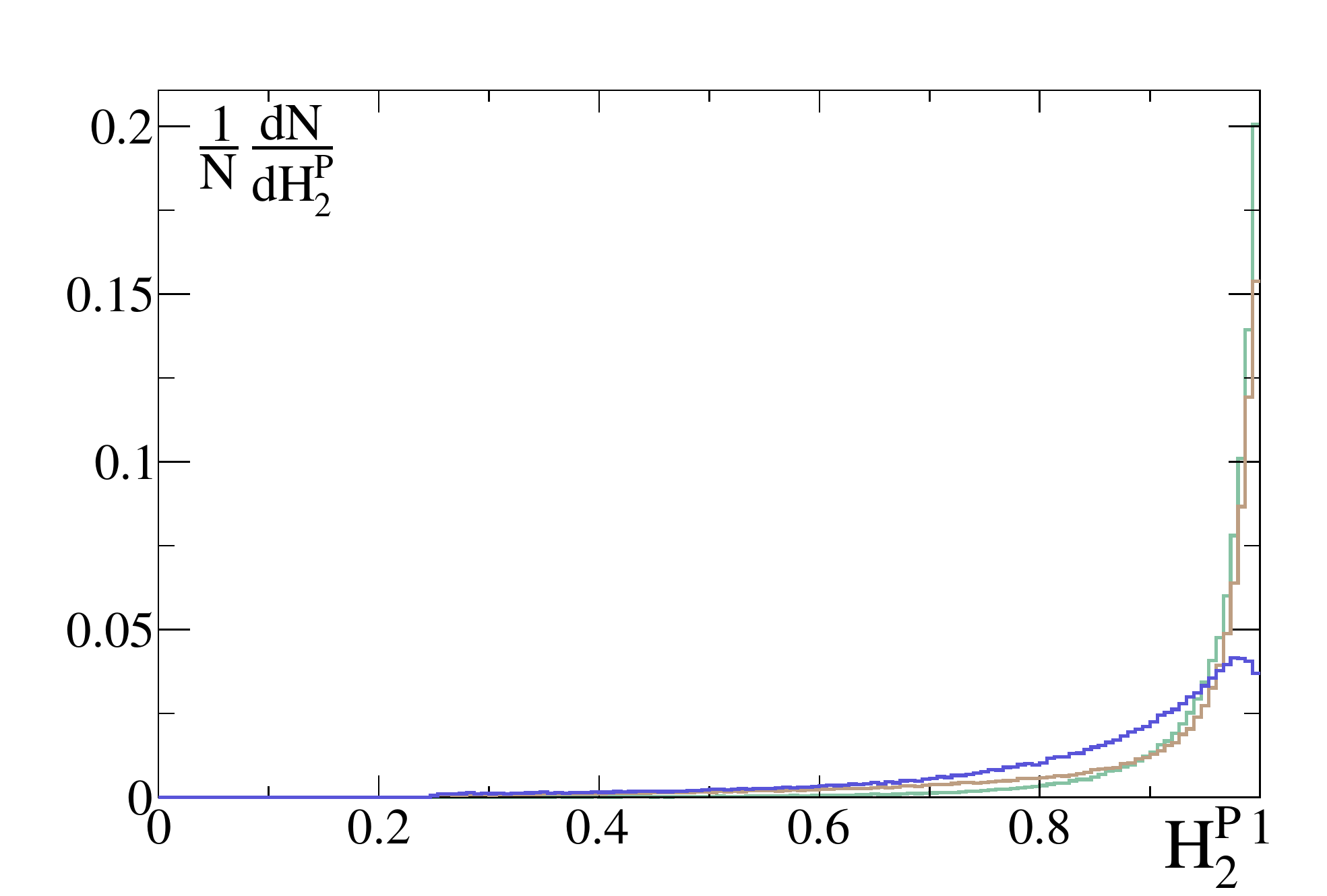}  
\includegraphics[width=0.24\textwidth]{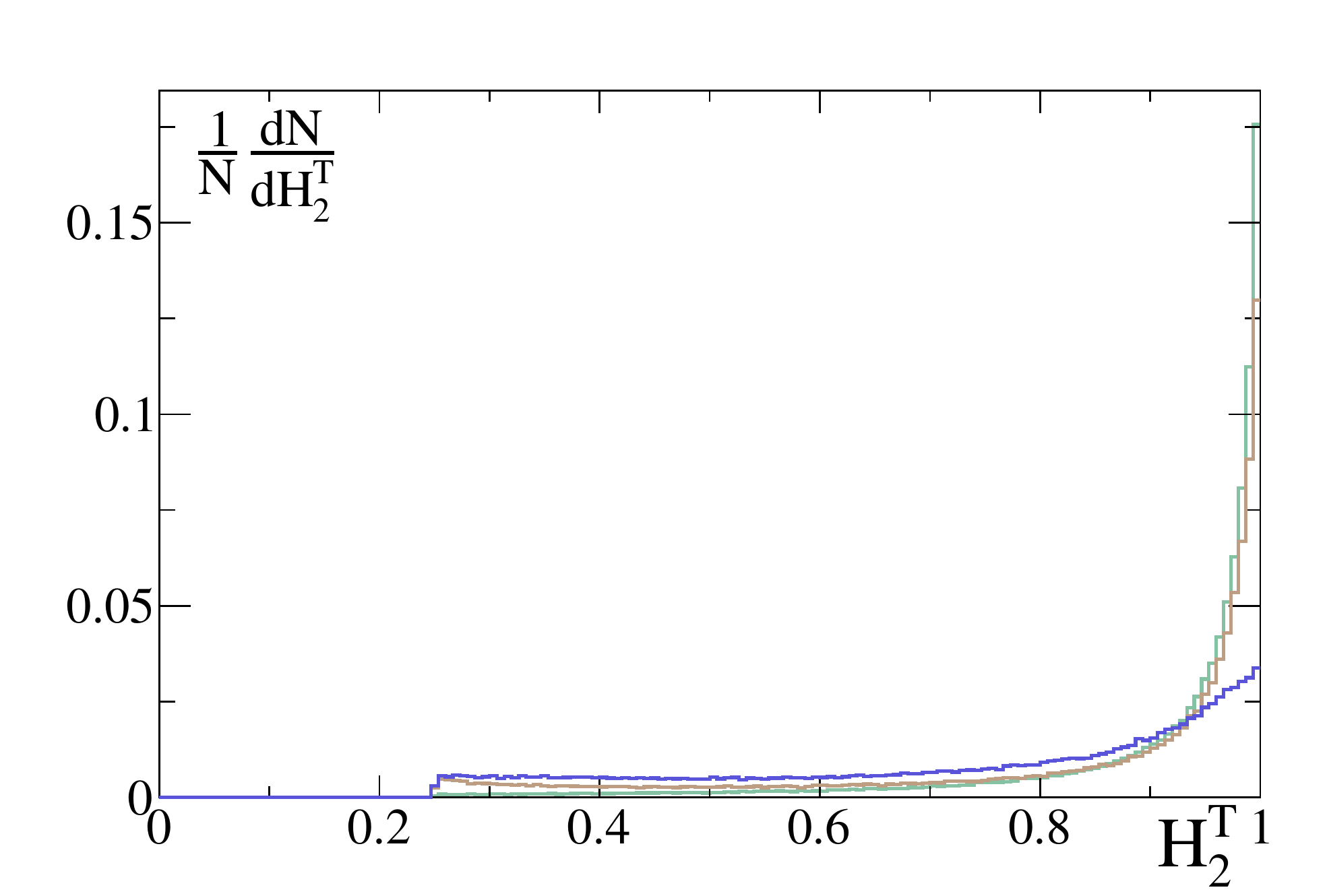} 
\includegraphics[width=0.24\textwidth]{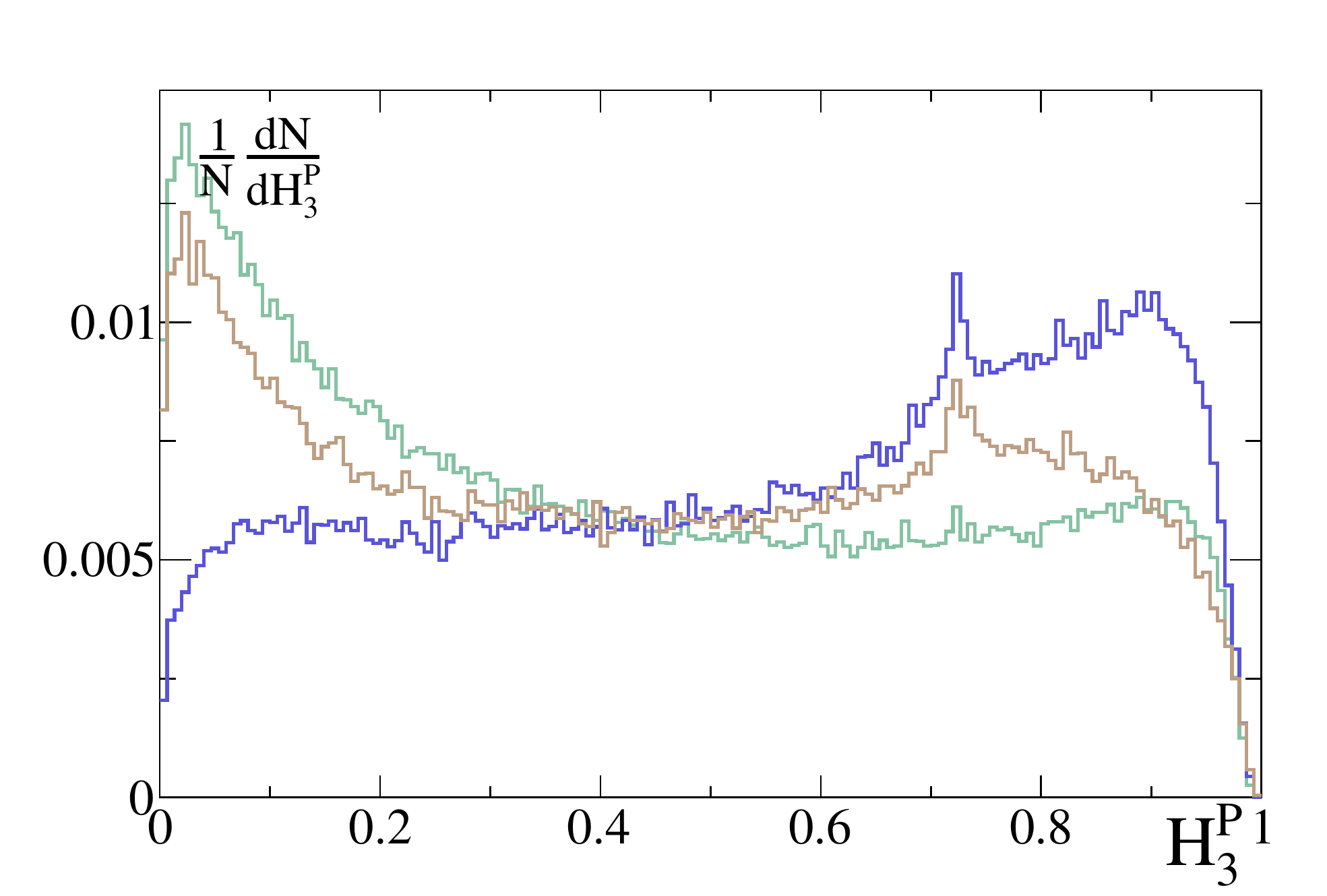}  
\includegraphics[width=0.24\textwidth]{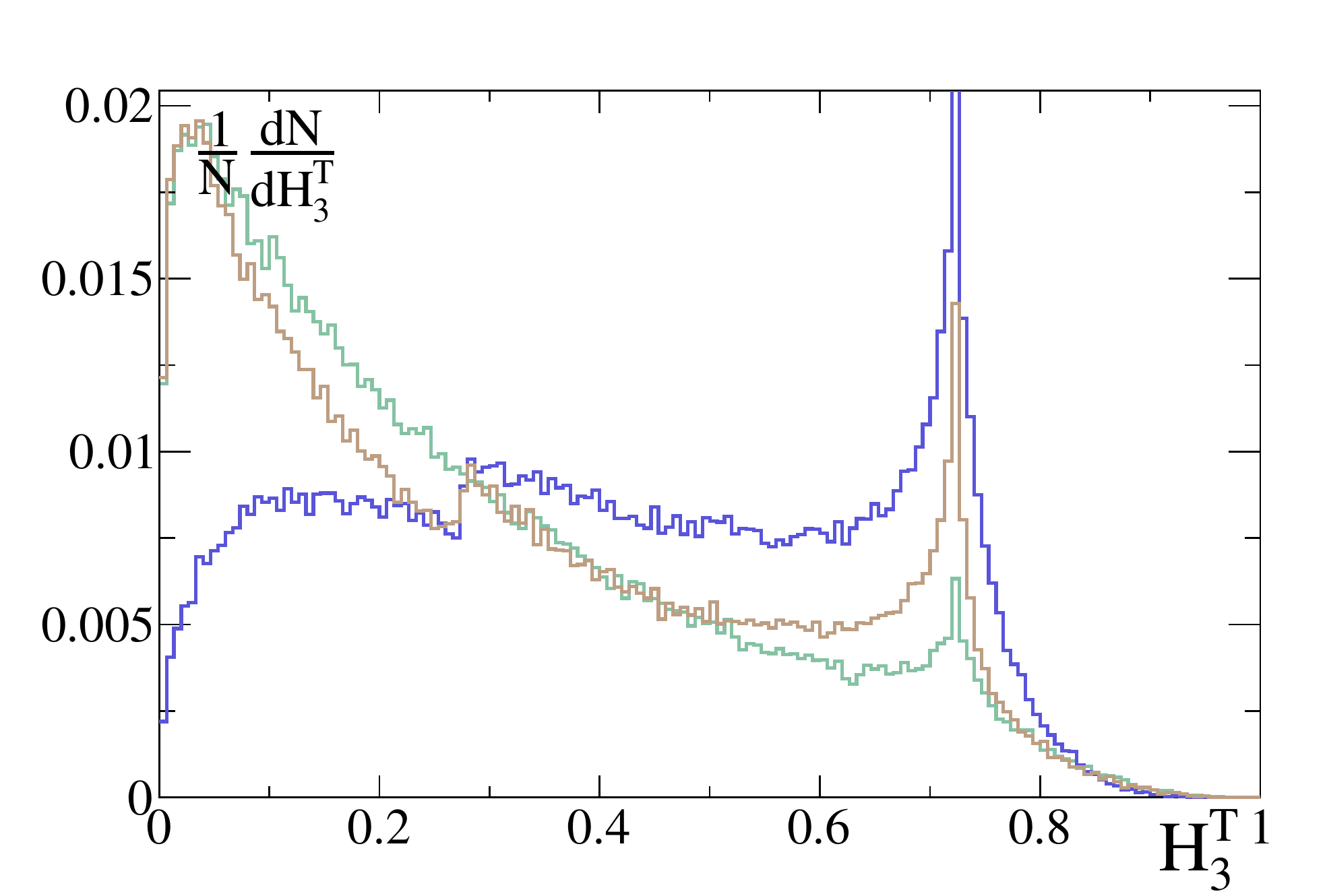}\\ 
\includegraphics[width=0.24\textwidth]{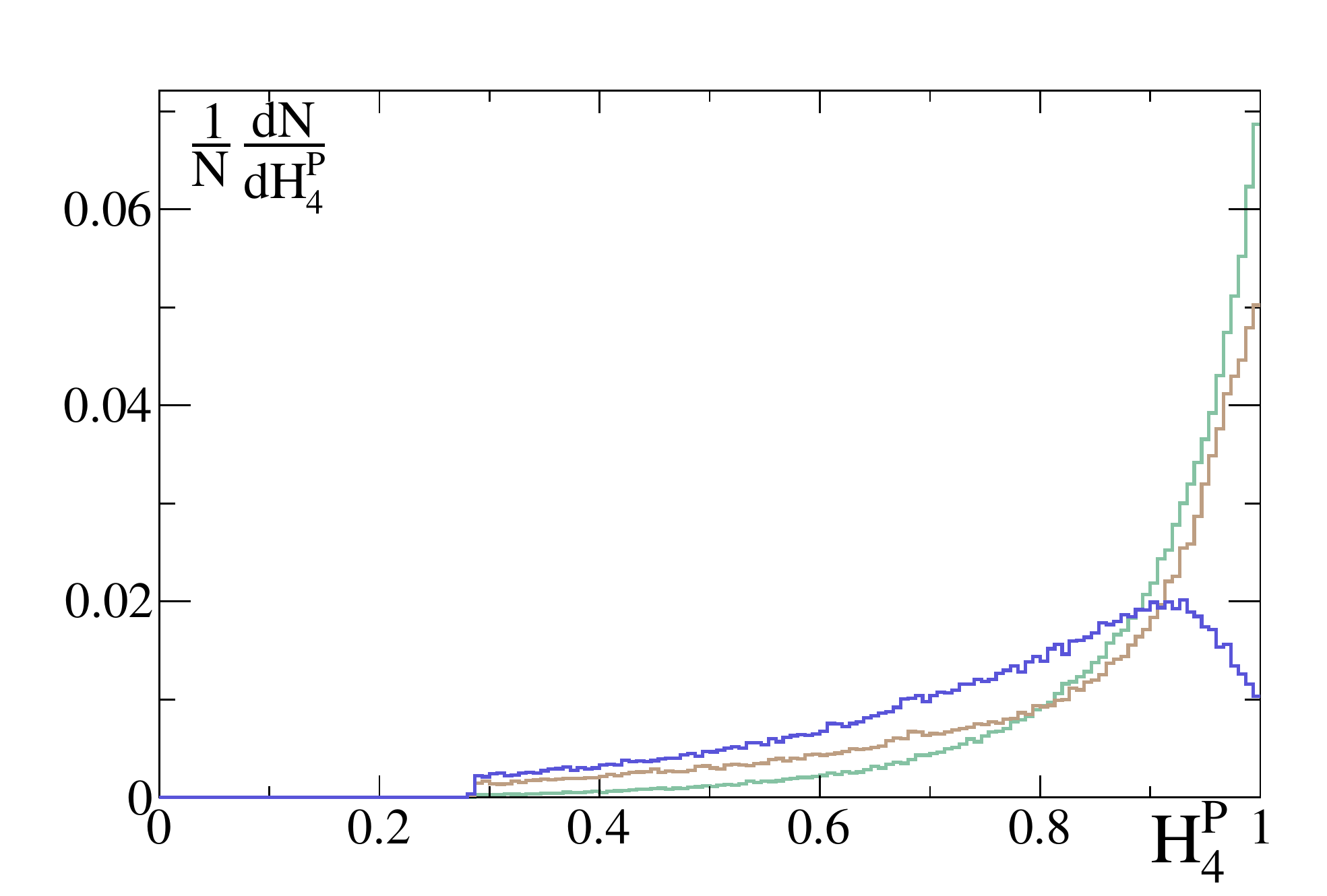}  
\includegraphics[width=0.24\textwidth]{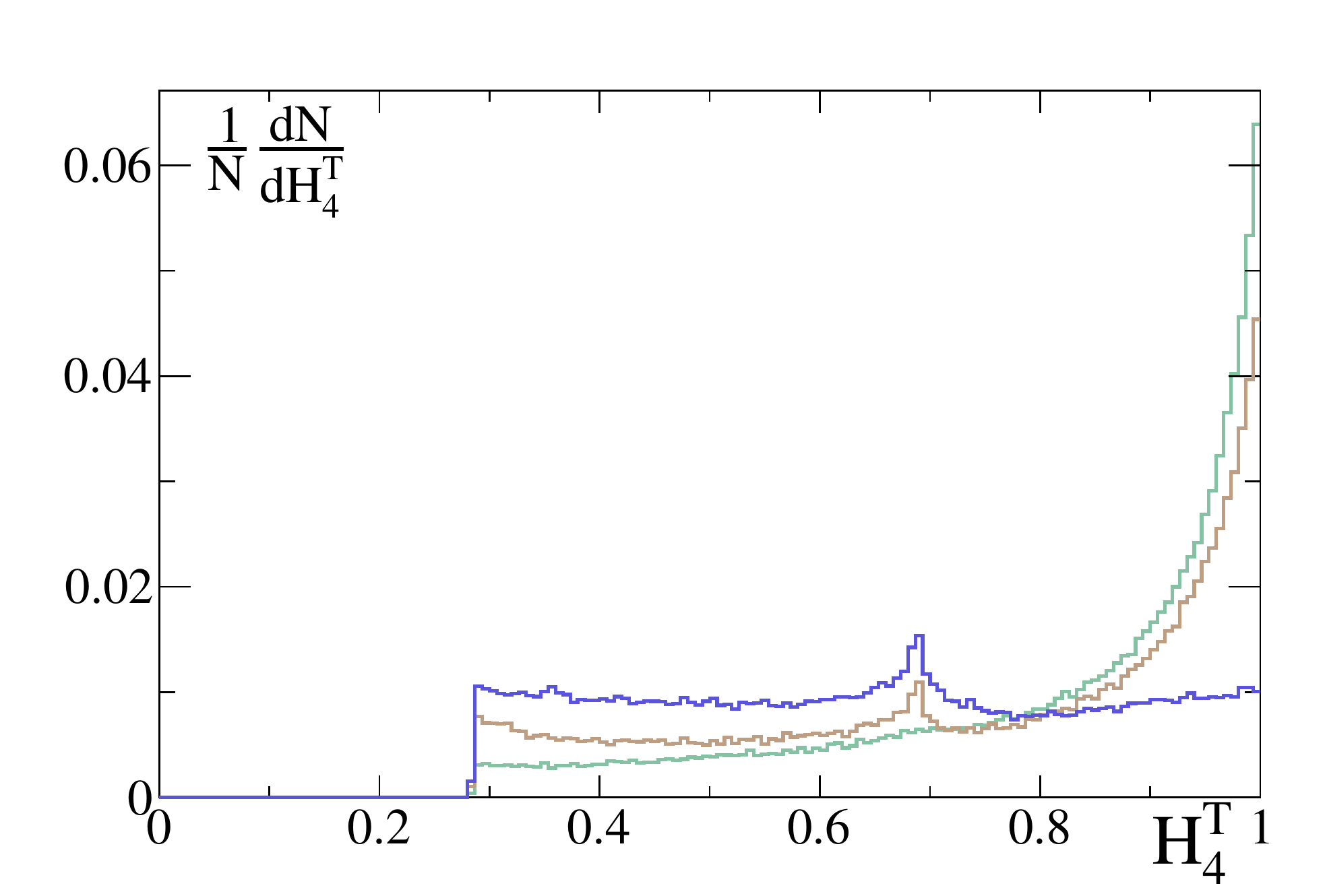} 
\includegraphics[width=0.24\textwidth]{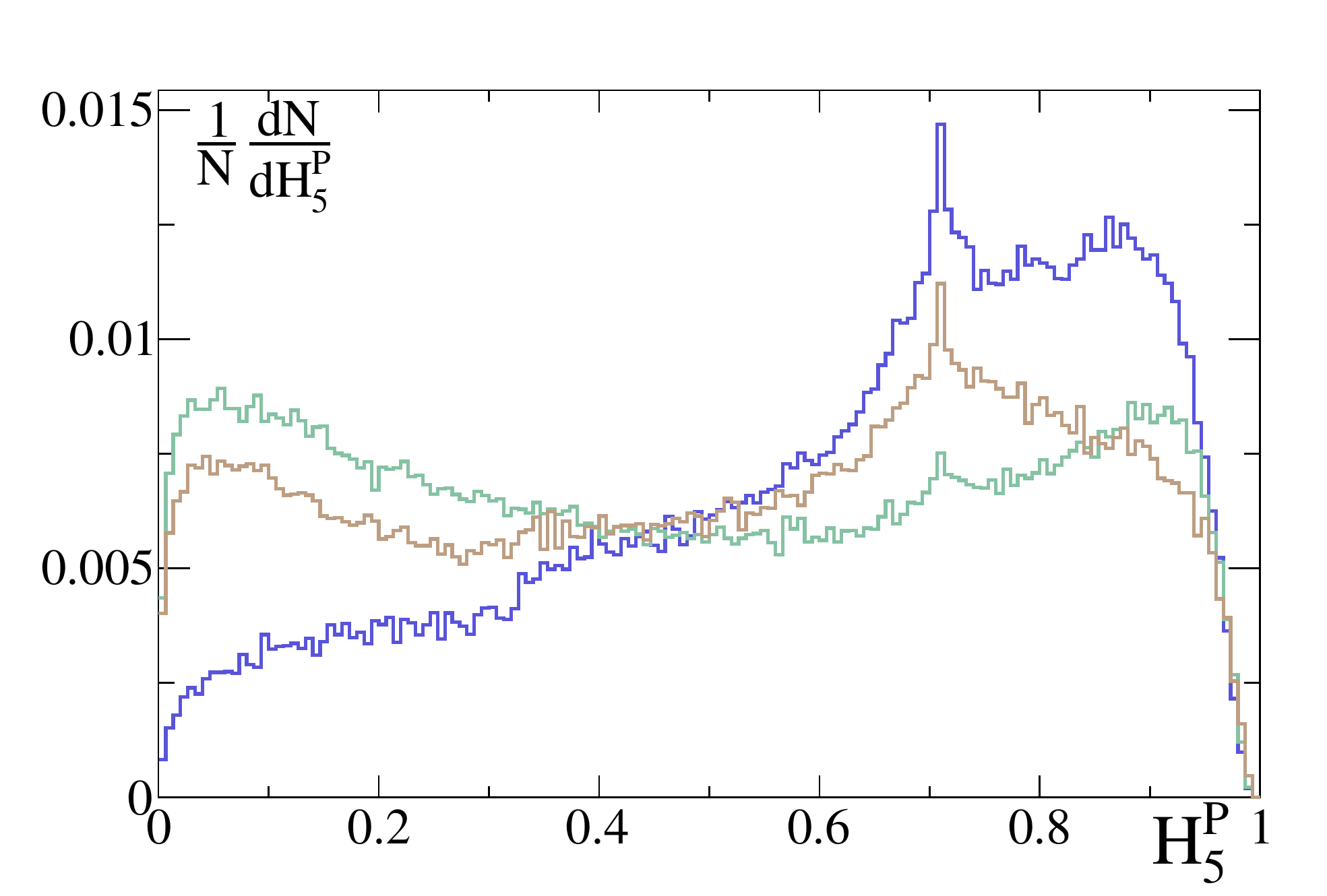}  
\includegraphics[width=0.24\textwidth]{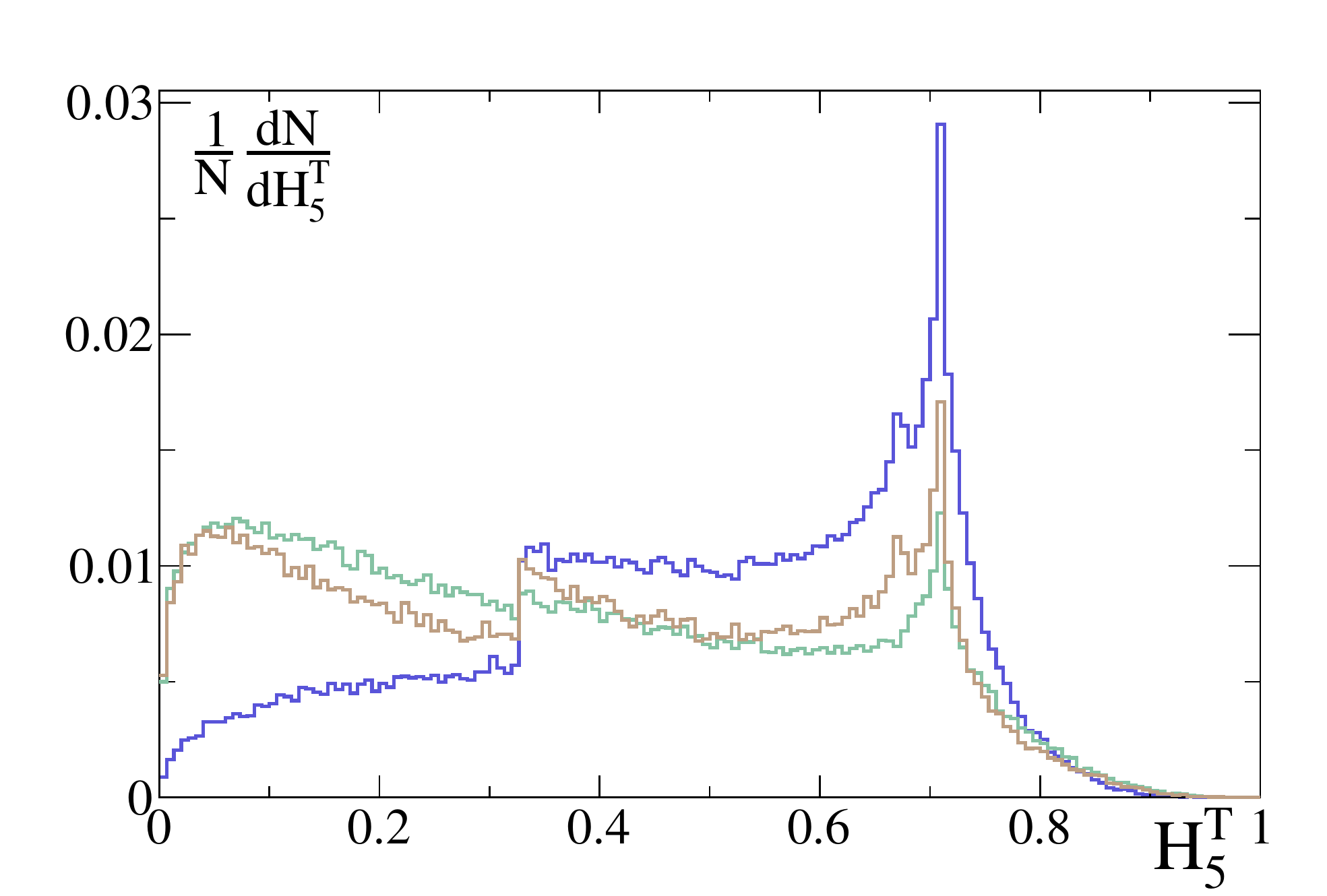}\\ 
\includegraphics[width=0.24\textwidth]{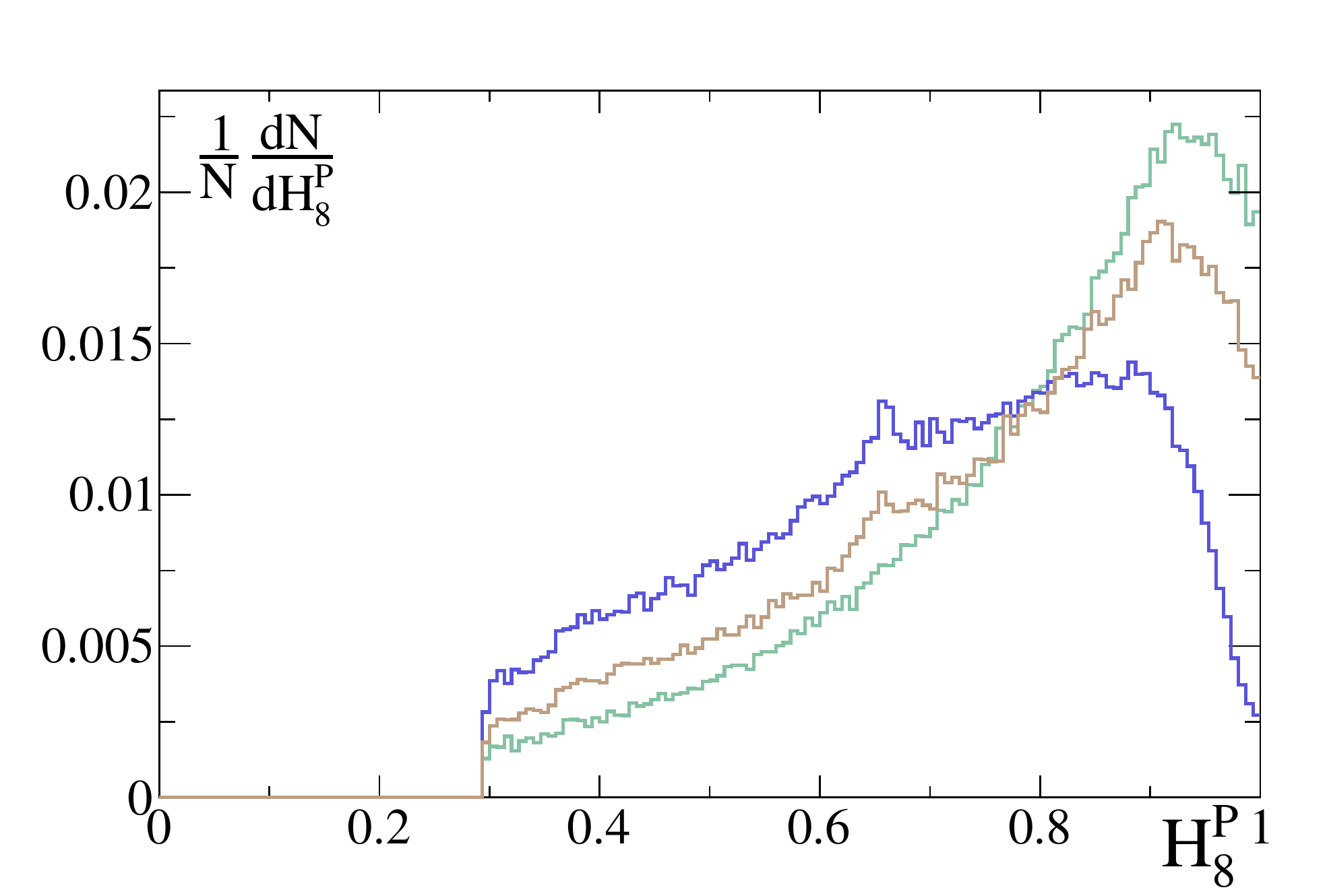}  
\includegraphics[width=0.24\textwidth]{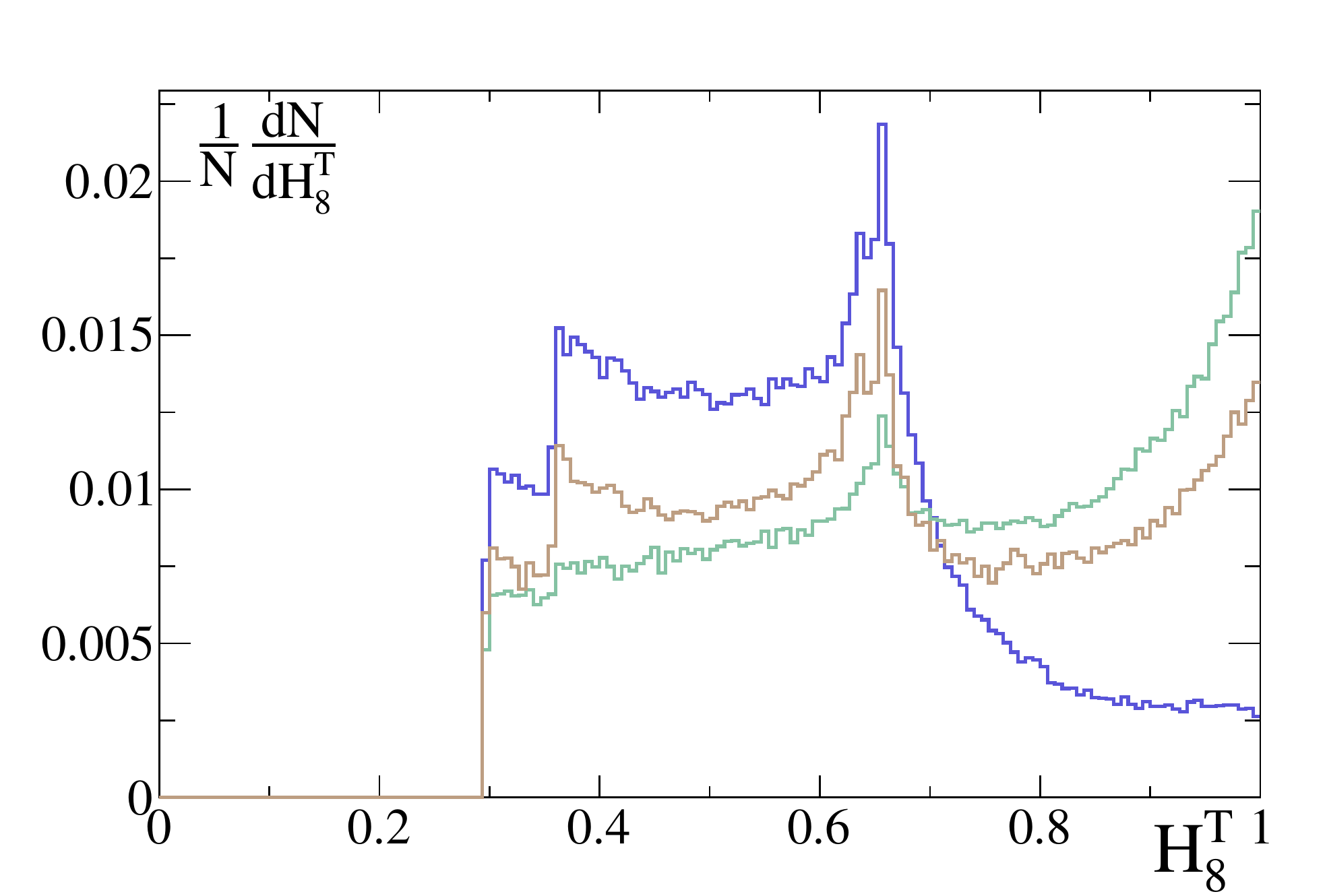} 
\includegraphics[width=0.24\textwidth]{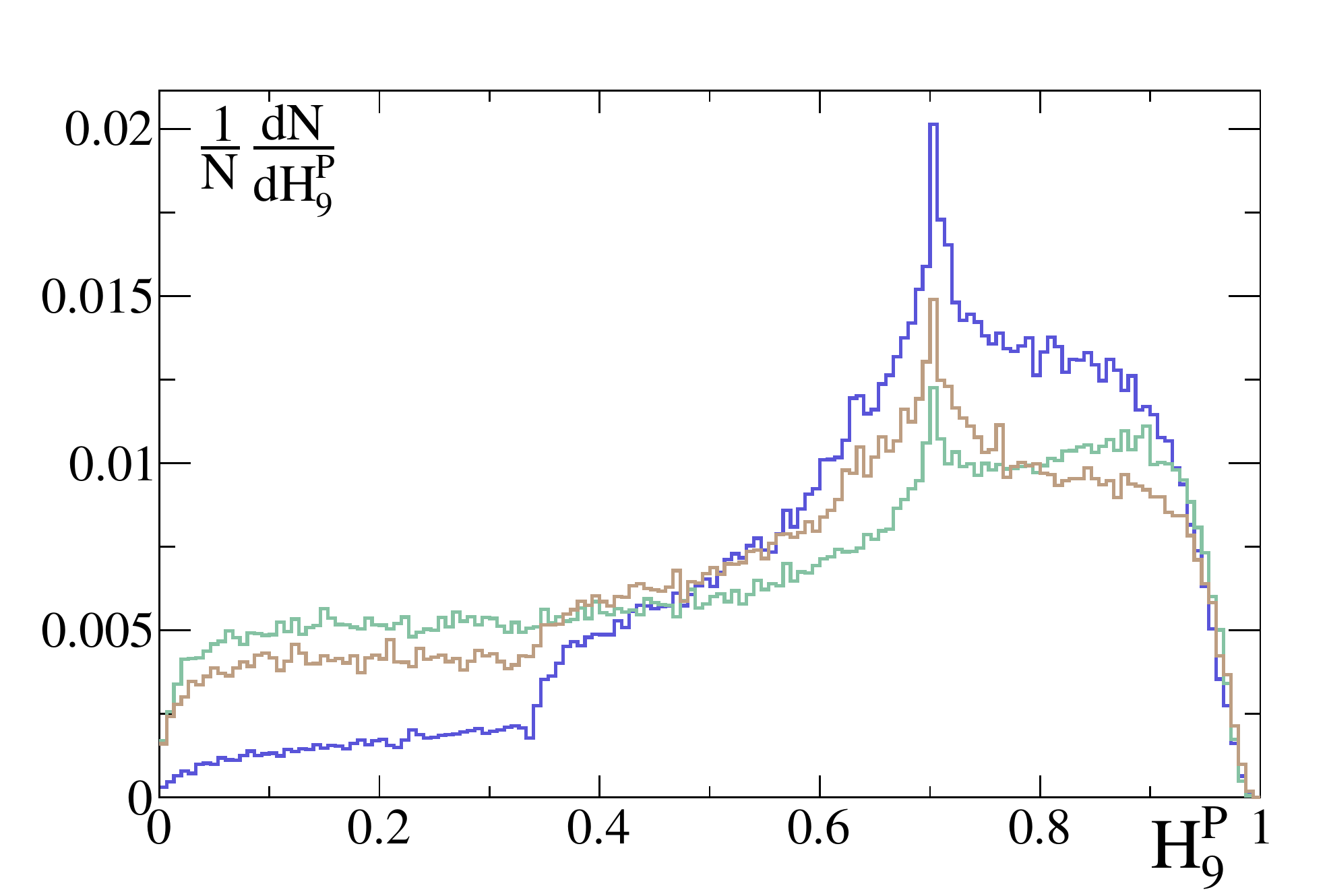}  
\includegraphics[width=0.24\textwidth]{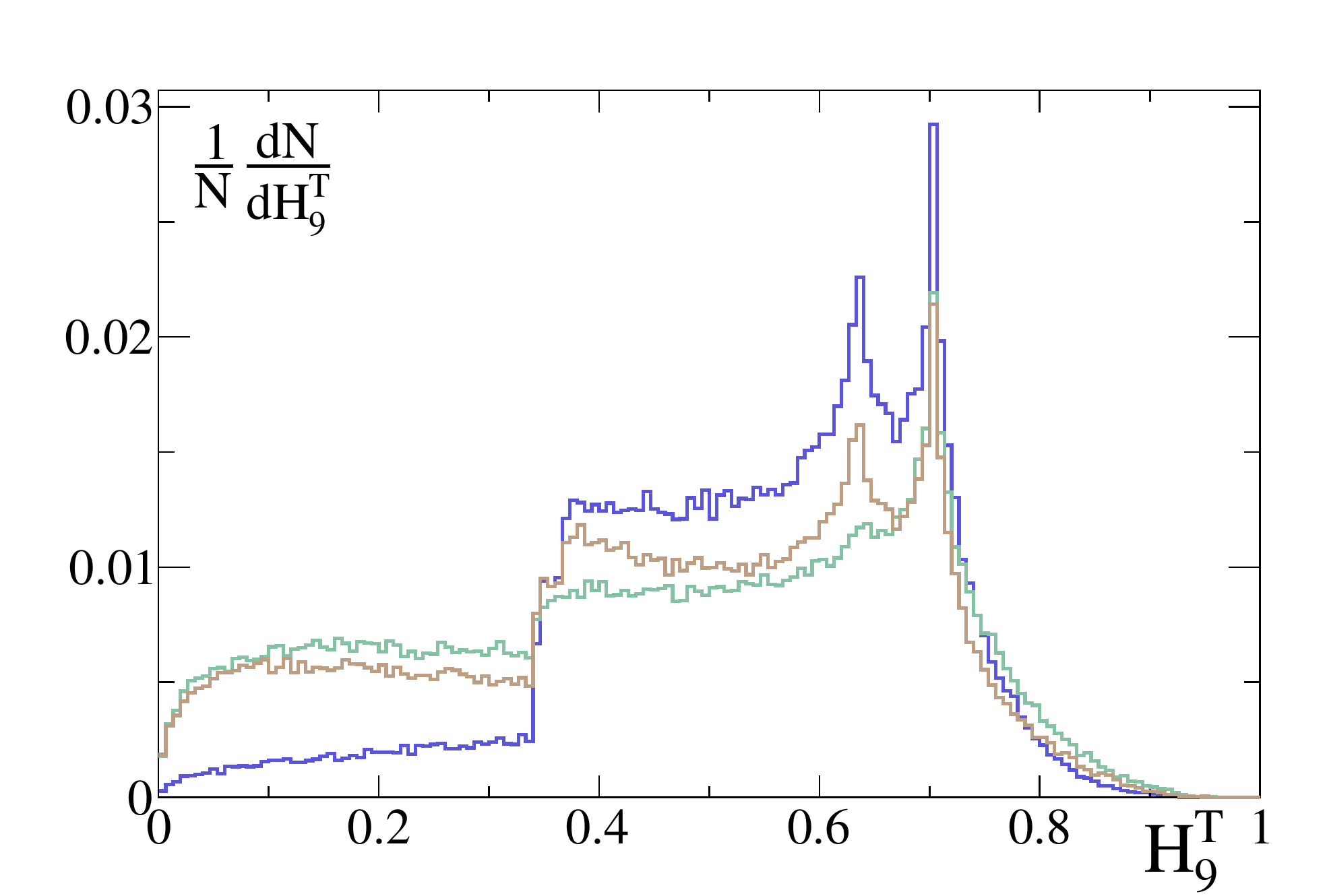}\\ 
\includegraphics[width=0.24\textwidth]{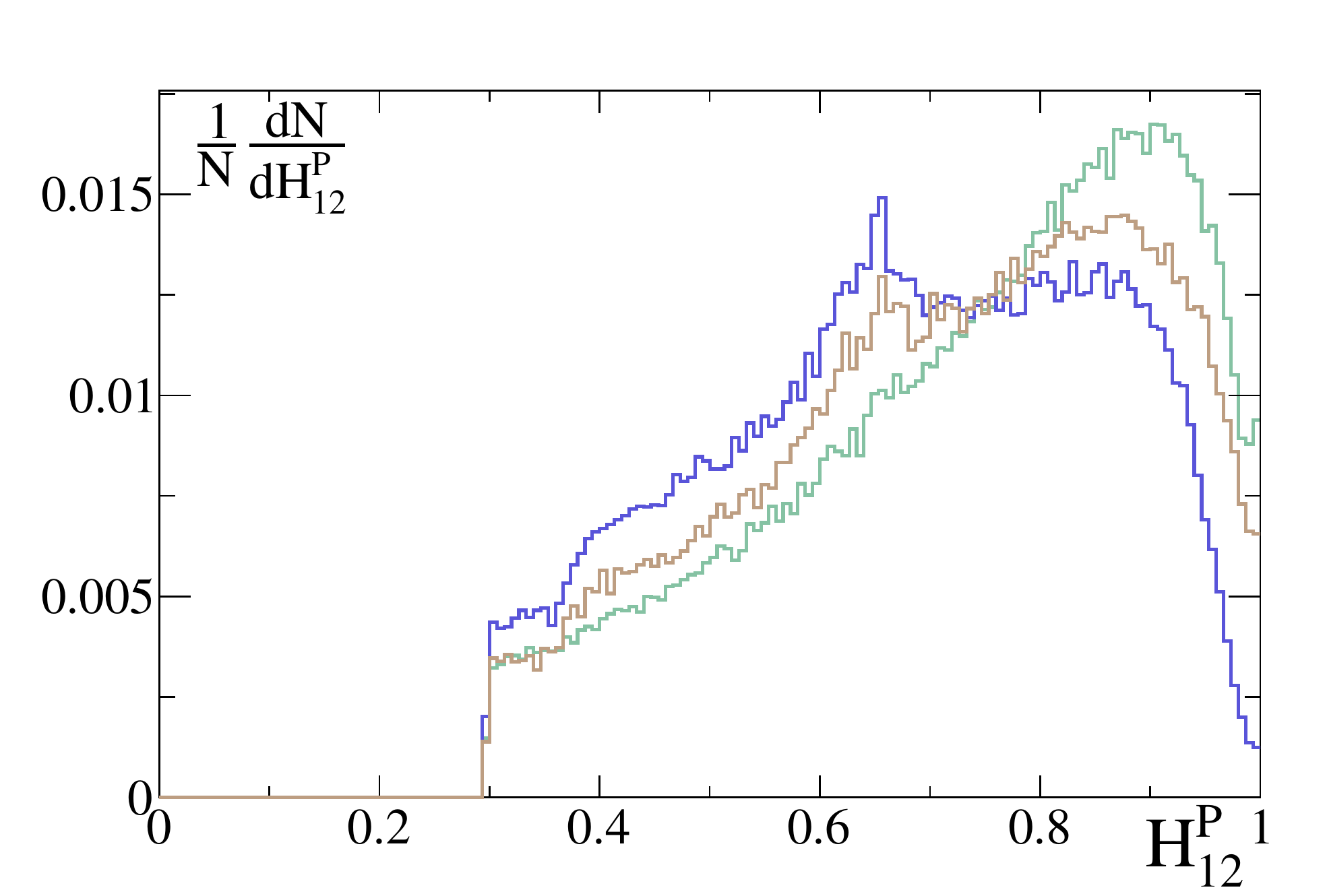}  
\includegraphics[width=0.24\textwidth]{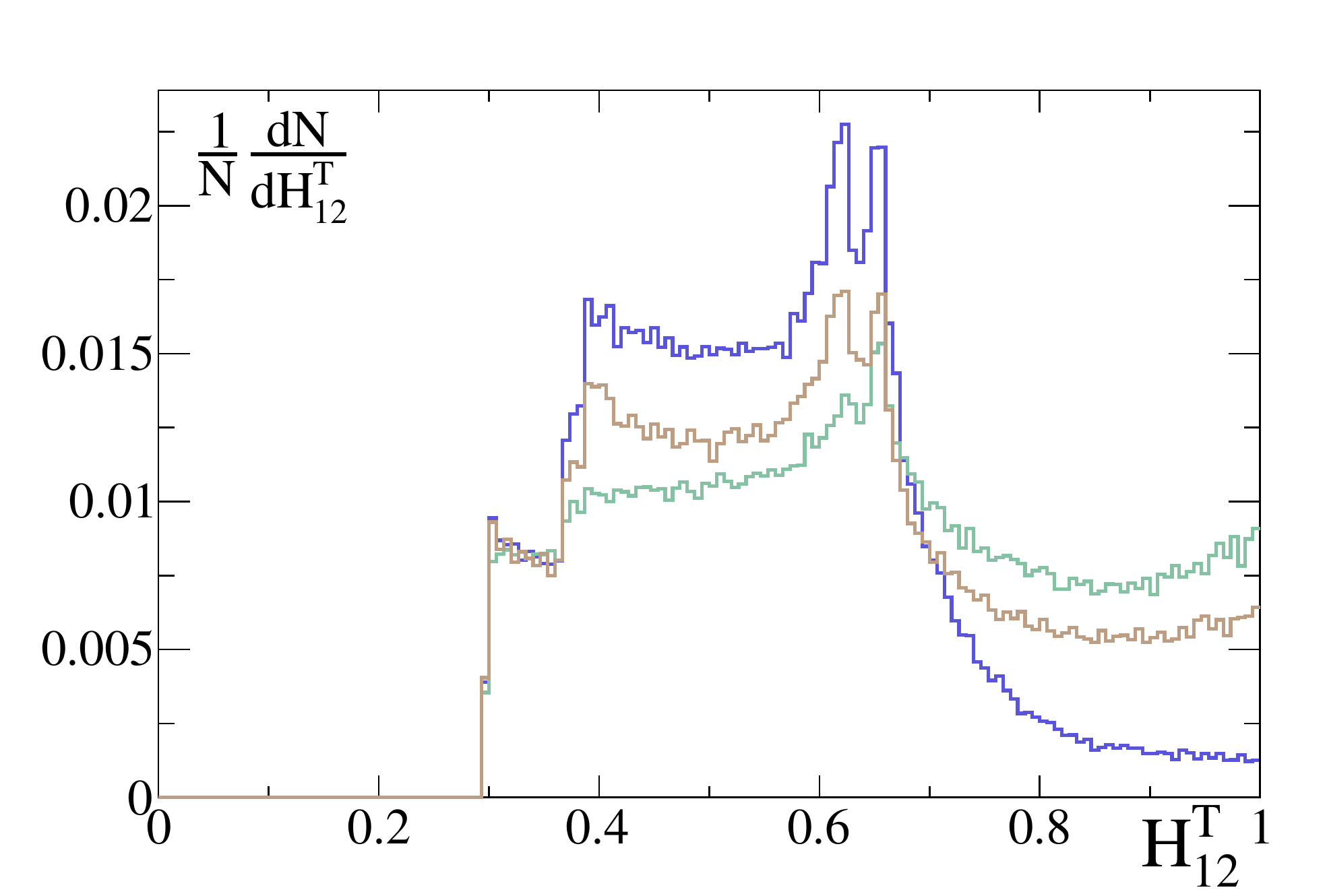} 
\includegraphics[width=0.24\textwidth]{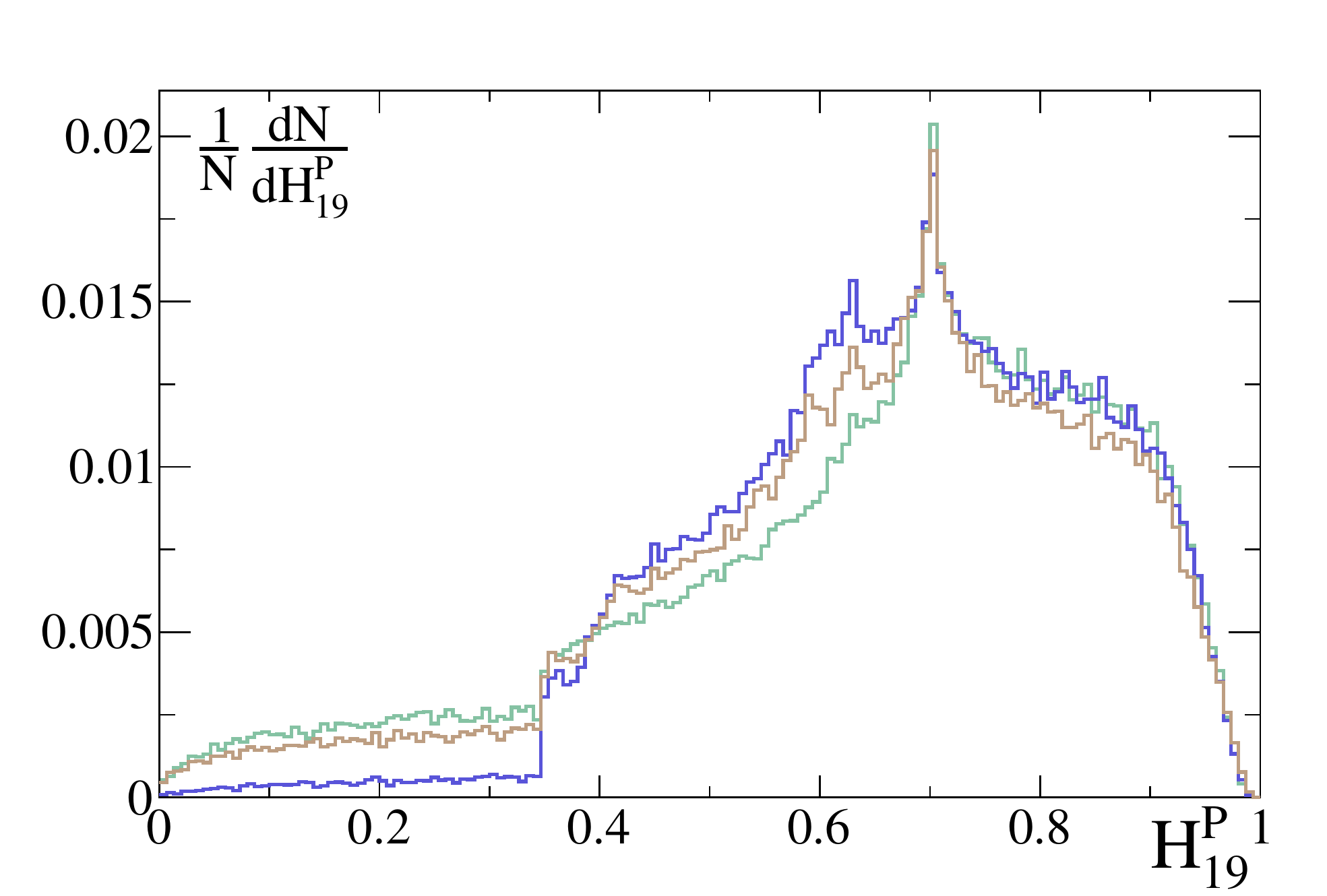}  
\includegraphics[width=0.24\textwidth]{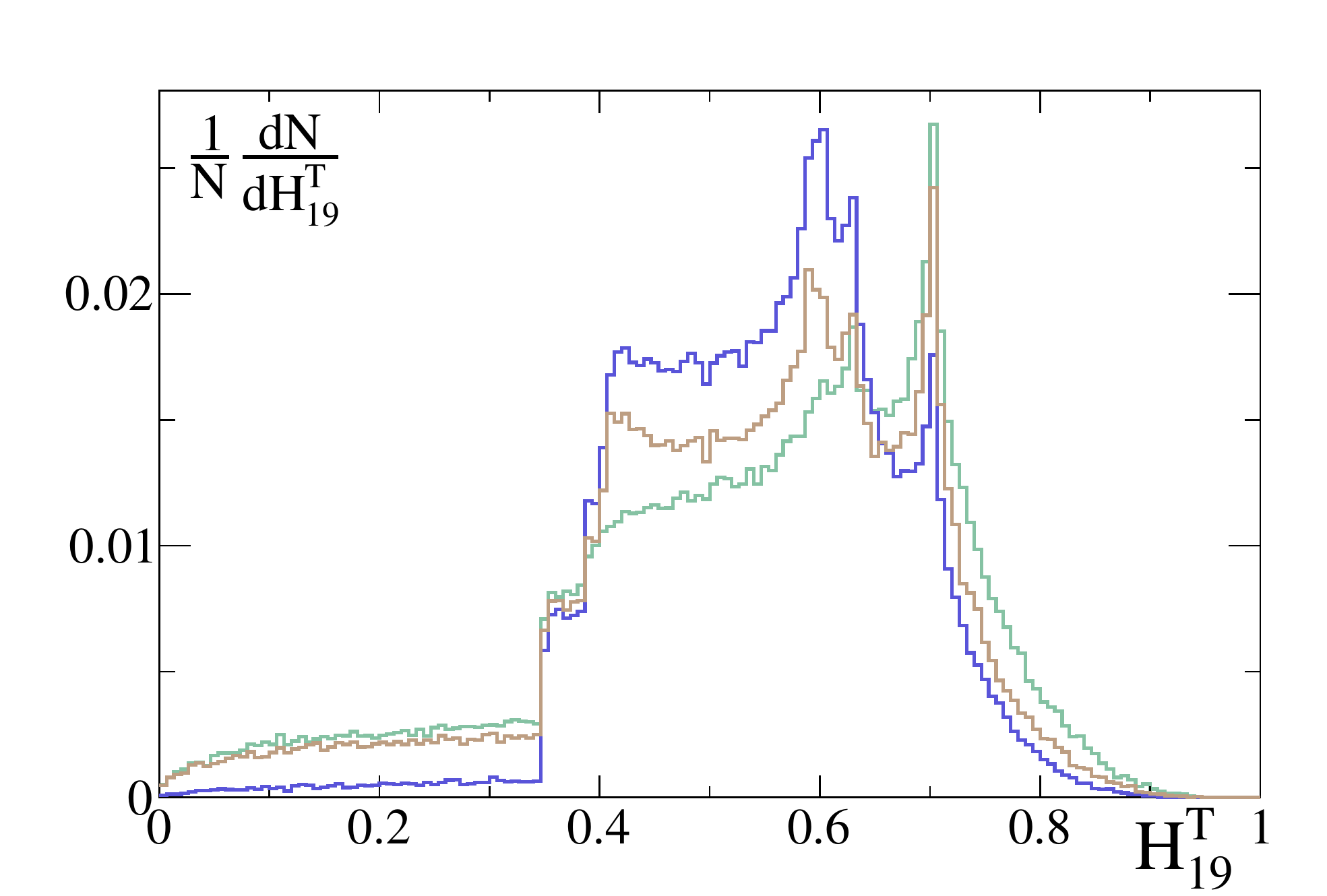} 
\caption{Normalized distributions of Fox--Wolfram moments computed
  only from the two tagging jets for $\ell = 2-5,8,9,12,19$ with
  weight factors $W_{ij}^p$ (left) and $W_{ij}^T$ (right) for 
 WBF H signal (green), Z+2 jets (brown) and $t\bar{t}$+1 jet (blue).  All events
  pass the cuts Eq.\eqref{eq:jj1} and Eq.\eqref{eq:jj2}.}
\label{fig:all_moments}
\end{figure}

In this section we use Fox--Wolfram moments {\sl only for the leading
  two tagging jets} in the Higgs signal and the $Z$+jets and
$t\bar{t}$+jets backgrounds, \ie we replace Eq.\eqref{eq:fwm_def3} by
\begin{alignat}{1}
H^{p,T}_\ell 
= \sum_{i,j=1}^2 \; W_{ij}^{p,T} \; P_\ell(\cos\Omega_{ij}) \; .
\label{eq:Hcuts}
\end{alignat}
Additional contributions from QCD jets
will enter in Sec.~\ref{sec:qcd}.  Without touching a general jet
veto, we nevertheless apply a $b$ veto to the top pair background, \ie
we veto $b$ jets from the top decays fulfilling
\begin{alignat}{2}
p_{Tb}>20~\gev \qqquad  |y_b|<2.5
\label{eq:bVETO}
\end{alignat}
with an efficiency of 60\%. We show the cut flow for this very basic
jet-only analysis in Tab.~\ref{tab:CFclassic}. It will serve as a
baseline to evaluate the performance of cutting on Fox--Wolfram
moments in addition to the standard WBF cuts.  All numbers in
Tab.~\ref{tab:CFclassic} should be taken with a grain of salt. They
only include part of the information from the actual WBF $H \to \tau
\tau$ analysis~\cite{wbf_tau}, where we achieve a signal-to-background
ratio $S/B = \mathcal{O}(1)$. What is important for our considerations
is only the efficiency of the shown cuts in rejecting
backgrounds.\bigskip

Using all events passing the $m_{jj}^\text{min}$ cut in
Tab.~\ref{tab:CFclassic} we evaluate a set of Fox--Wolfram moments in
order to estimate their power in improving the weak-boson-fusion Higgs
search.  The first question is if we can replace the cuts on $y_1
\cdot y_2$ and on $\Delta y_{jj}$ with Fox--Wolfram moments computed
from the two tagging jets. In Fig.~\ref{fig:all_moments} we show
$H^{p,T}_\ell$ for a set of even (left two columns) and odd (right two
columns) moments.  In general, both of them distinguish equally well
between signal and backgrounds, with a slight quantitative advantage
for even $H^T_\ell$.  Only looking at the two tagging jets the
$Z$+jets background with its forward jet radiation tends to be similar
to the signal.  The $m_{jj}^\text{min}$ cut has removed most of the
$Z$+jets events which look significantly different from the Higgs
signal while the hard and central decay jets in top pair production
tend to have a more unique shape. As we will see in Sec.~\ref{sec:qcd}
the key to distinguishing WBF $H+$jets production from QCD $Z+$jets
production is the jet activity in addition to the two tagging
jets~\cite{manchester,jetveto1,jetveto2}.

$H_0$ and $H_1$ do not have sufficient resolution to discriminate
signal and backgrounds and are not shown. The first non-trivial even
moment $H_2$ shows a narrow peak for the Higgs signal, but even the
top pair background does not look significantly different. As
discussed in Sec.~\ref{sec:moments} even moments do not take values
$H^{p,T}_\ell \lesssim 0.3$.  The lowest odd moment, $H_3$, shows the
events with back-to-back jets in the low-$H^T_3$ regime. The Higgs and
$Z$ topologies clearly prefer this region, with 62\% and 55\% of their
events giving $H^T_3 < 0.3$. Note that this percentage is nowhere
close to 100\%, which means that the back-to-back criterion imposed by
$H^T_3$ is harder than even the typical signal events can pass.  In
contrast, top pairs reside in the intermediate regime with a peak at
$H_3 \sim 0.7$, only 35\% of them lie below $H^T_3 < 0.3$. Large
values of $H_3$ only appear for $H_3^p$, independent of the production
mechanism. The reason is that the entire momentum instead of just its
transverse direction gives a smaller ratio $r_p$ for not quite balanced
jets in the beam direction.

\begin{table}[!b]
\centering
\begin{tabular}{l|rr|rr|rr|r}
\hline
& \multicolumn{2}{c|}{WBF $H+2$~jets}  
& \multicolumn{2}{c|}{QCD $Z+2$~jets}
& \multicolumn{2}{c|}{$t\bar{t}+1$~jet}
&  S/B \\
acceptance &\% fail & $\sigma$(fb)  
           &\% fail & $\sigma$(fb) 
           &\% fail & $\sigma$(fb) &  \\
\hline
$b$-veto \rule{0pt}{3ex}        &      &  3.92 &      & 253  &      & 292  & 1/139 \\
\hline
$H_3^T < 0.3$ \rule{0pt}{3ex}   & 38.4 &  2.41 & 44.4 & 141  & 64.6 & 103  & 1/101 \\   
\hline
$H_4^T > 0.8$ \rule{0pt}{3ex}   & 35.8 &  2.52 & 48.1 & 131  & 73.3 & 78.0 & 1/83  \\   
\hline
$H_8^T > 0.7$ \rule{0pt}{3ex}   & 50.1 &  1.96 & 60.5 & 100  & 81.6 & 53.7 & 1/78  \\   
\hline
$H_{12}^T > 0.7$ \rule{0pt}{3ex}& 64.5 &  1.39 & 73.0 & 68.3 & 88.0 & 35.0 & 1/74  \\   
\hline
\end{tabular}
\caption{Cut flow of the signal and background processes applying cuts
  on the Fox--Wolfram moments of Eq.\eqref{eq:Hcuts} weighted with
  $p_T$, calculated with
  only the two leading jets. The first line starts from the event
  numbers after the $m_{jj}^\text{min}$ cut and the $b$-veto in
  Tab.\ref{tab:CFclassic}.}
\label{tab:PvsTcuts}
\end{table}

Moving towards larger values of $\ell$ the odd moments become less
sensitive because $H_\ell$ resolves forward jets better. Eventually,
most of the WBF Higgs are not sufficiently back-to-back to contribute
to the low-$H_\ell$ regime. In that situation the even moments $H^T_4$
to $H^T_8$ turn out more useful. Their regimes $r \lesssim 0.4$ and
$\Omega > 0.9$ are merged and allow us to reject top pairs not only
based on the geometry of the tagging jets but also on the $p_T$
hierarchy between them. In the range $H^T_4 > 0.7$ which includes
back-to-back configurations as well as strongly ordered jets, we
find 76\% of the Higgs events, 62\% of the $Z+$jets background, and
40\% of the top pair events. Because the resolution of $H_3$ and $H_4$
is similar, cutting on $H^T_4$ does not only extract the typical WBF
back-to-back tagging jet geometry. This correlation is further watered
down when we replace $H^T$ by $H^p$.\bigskip

To answer the first question about replacing the $\Delta y_{jj}$ cut
we show results for cuts on the different Fox--Wolfram moments in
Tab.~\ref{tab:PvsTcuts}. As suggested above we start by removing the
large-$H^T_3$ events. Alternatively, we can require large values for
medium-$\ell$ even moments.  Indeed, it is possible to separate signal
and background events using Fox--Wolfram moments. $Z+$jets events are
more similar to the Higgs signal than top pair production, just as
expected.  Compared to the geometric cut benchmarks shown in
Tab.~\ref{tab:CFclassic} the Fox--Wolfram moments based on two tagging
jets achieve a similar improvement of $S/B$ but with significantly
smaller efficiencies. While they are indeed sensitive to the geometry
of the jets from the hard production process, QCD jet radiation, or
decays, they cannot entirely replace the well-known geometric cuts in
WBF analyses~\cite{wbf_tau,wbf_w,wbf_gamma}.

\begin{figure}[t]
\includegraphics[width=0.24\textwidth]{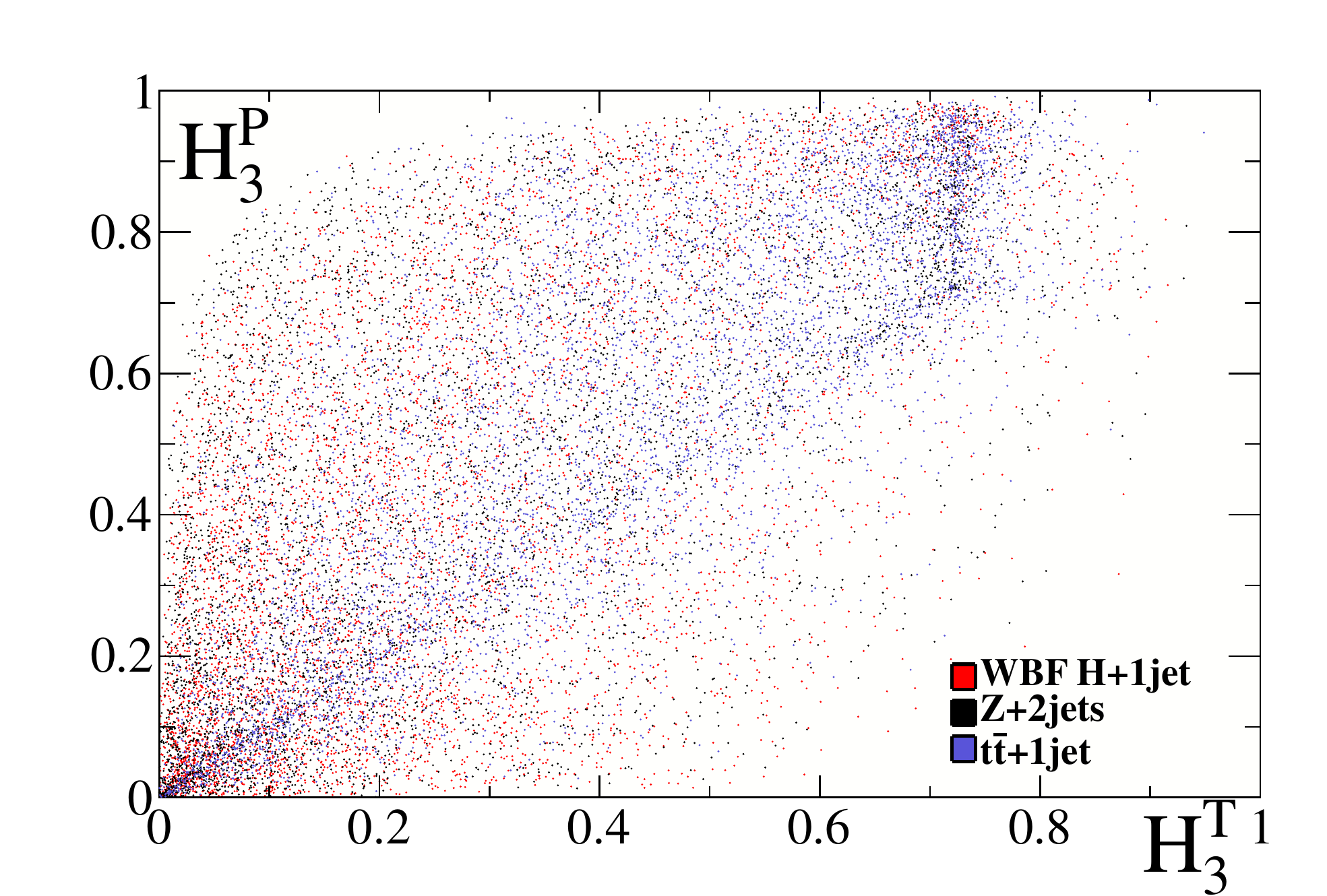} 
\hspace*{0.0\textwidth}
\includegraphics[width=0.24\textwidth]{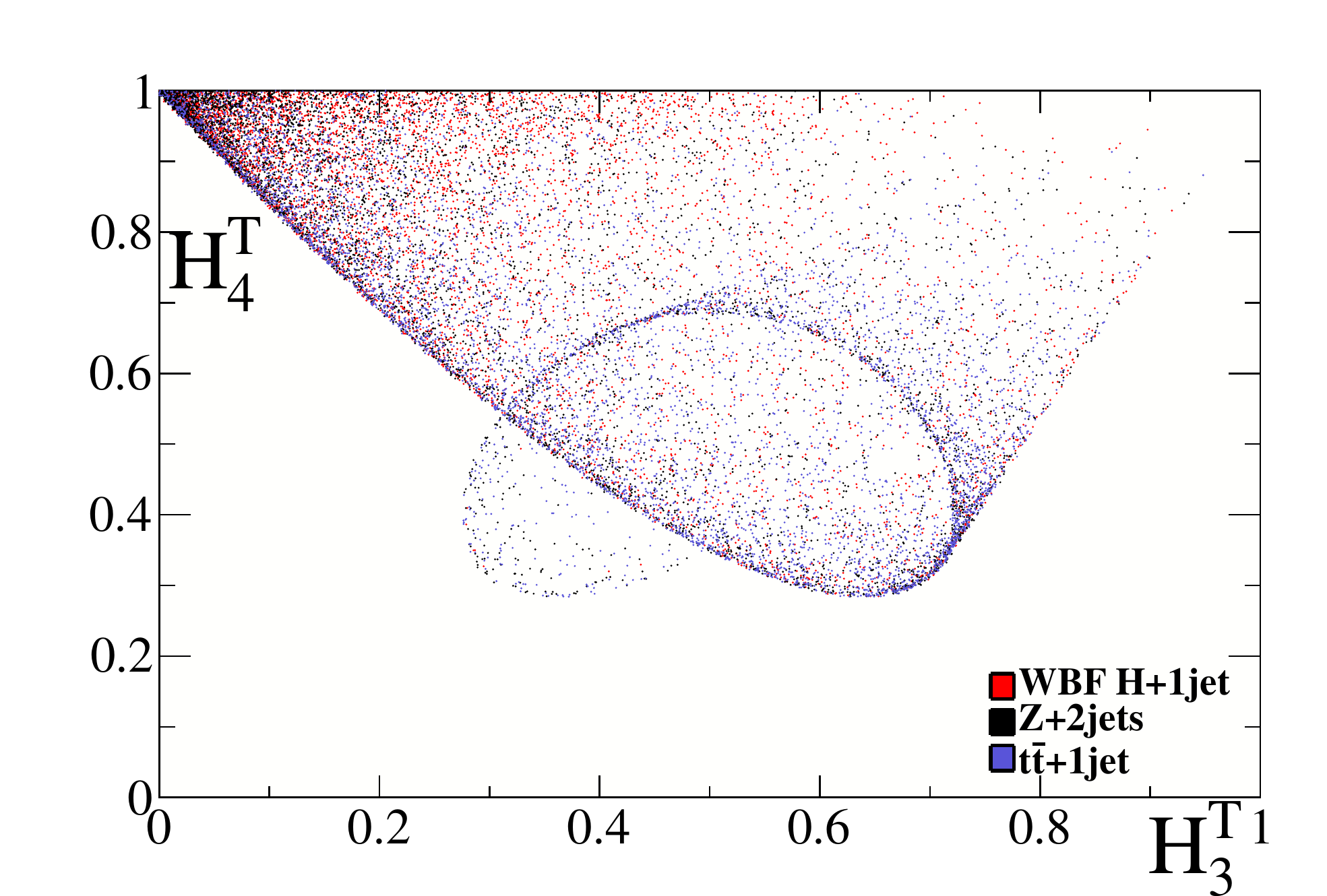} 
\includegraphics[width=0.24\textwidth]{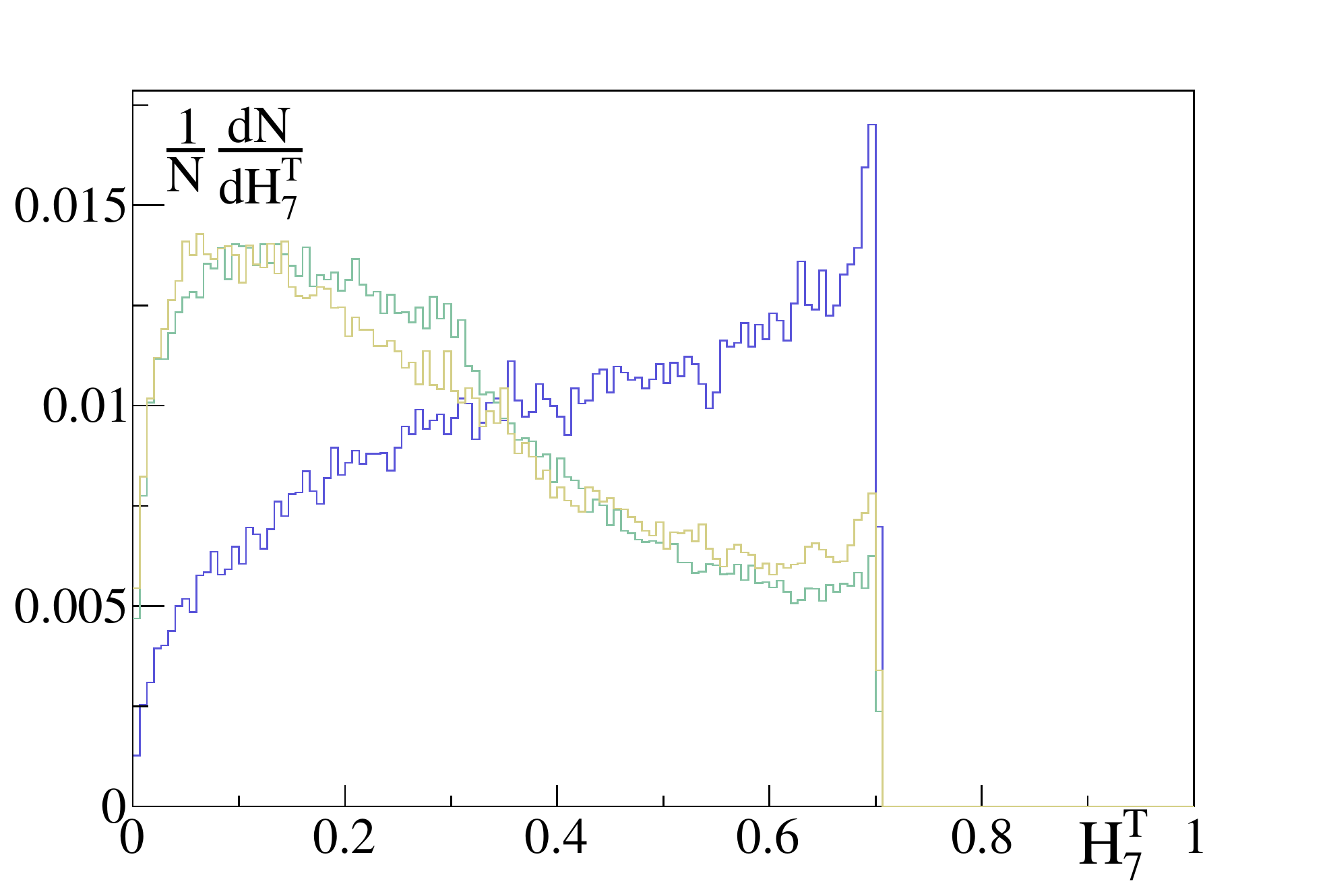} 
\includegraphics[width=0.24\textwidth]{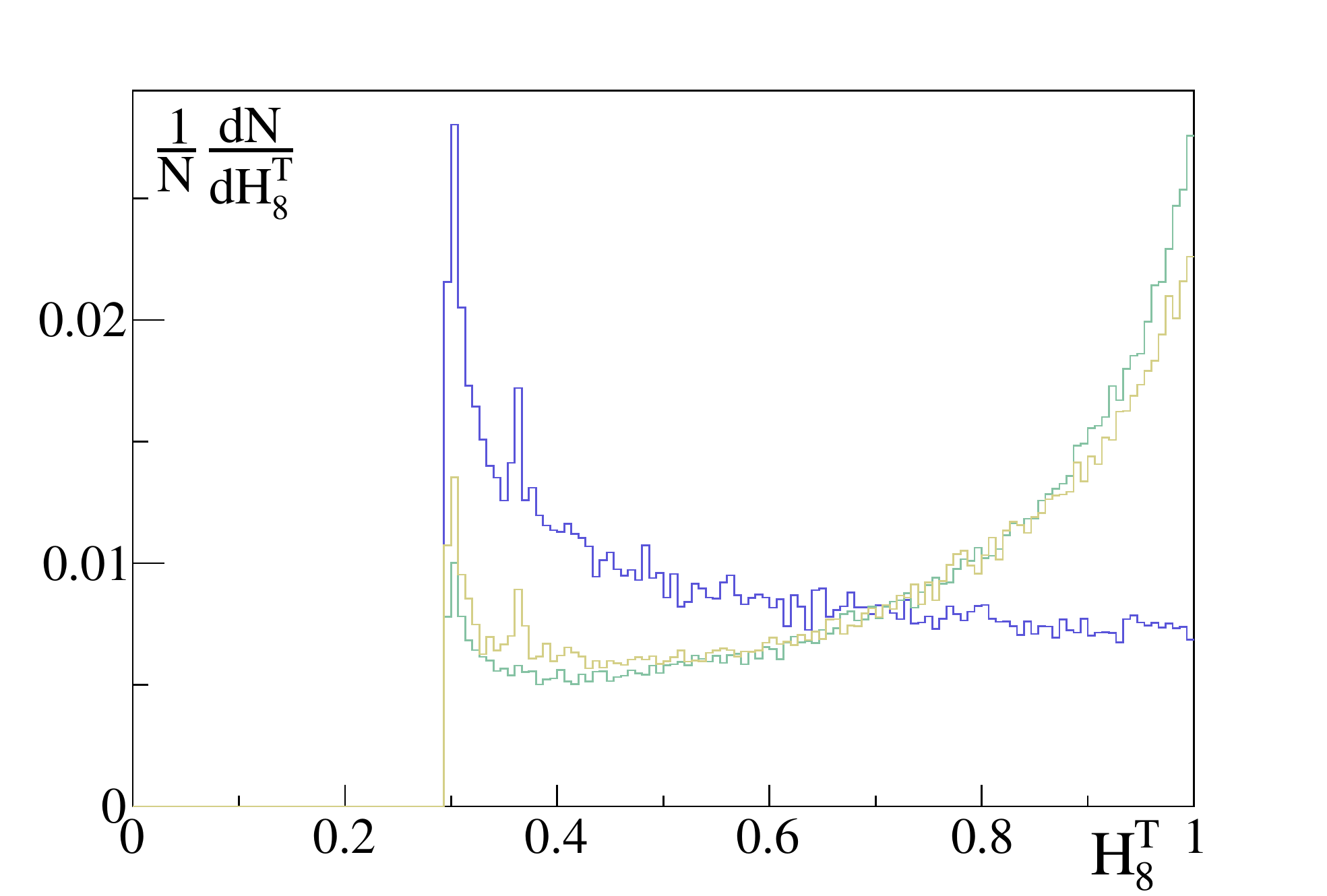} 
\caption{Left: correlations between $H^T_3$ with $H^p_3$ and $H^T_4$
  computed from the two tagging jets alone.  Right: Normalized
  distributions for $H^T_7$ and $H^T_8$ computed from the two tagging
  jets alone and after requiring $H^T_3 < 0.3$. All events in this
  figure pass the basic cuts Eq.\eqref{eq:jj1} and Eq.\eqref{eq:jj2},
  but not the final $\Delta y_{jj}$ requirement.}
\label{fig:moments_corr}
\end{figure}

Before moving on, we need to study correlations between the different
moments, because we know from Sec.~\ref{sec:moments} that different
kinds of events populate well-defined regions for different
Fox--Wolfram moments. It is fairly obvious that even or odd moments
are correlated among themselves. Following Sec.~\ref{sec:moments} an
increase in $\ell$ for even or odd moments dominantly increases the
resolution for example in the back-to-back regime. The answer is less
obvious when we consider correlations between even and odd moments or
between different weights $W^{p,T}$. In Fig.~\ref{fig:moments_corr} we
show the correlations between the useful $H^T_3$ and its full-momentum
counter part $H^p_3$ as well as the closest even moment $H^T_4$. Both
show a clear correlation, but with significant deviations from the
dominant pattern.  To quantify the effect of correlations on the analysis
we first require $H^T_3 < 0.4$ and then show higher Fox--Wolfram
moments only based on the remaining events. Surprisingly, the just
slightly higher odd moment $H^T_5$ still shows significant potential
in separating signal and backgrounds. The even moment $H^T_8$ retains
essentially all its distinguishing features, no matter if we cut on
$H^T_3$ or not. This suggests that we should treat the different
Fox--Wolfram moments as correlated, but by no means reducible to one
even and one odd pattern.\bigskip

\begin{figure}[!h]
\includegraphics[width=0.24\textwidth]{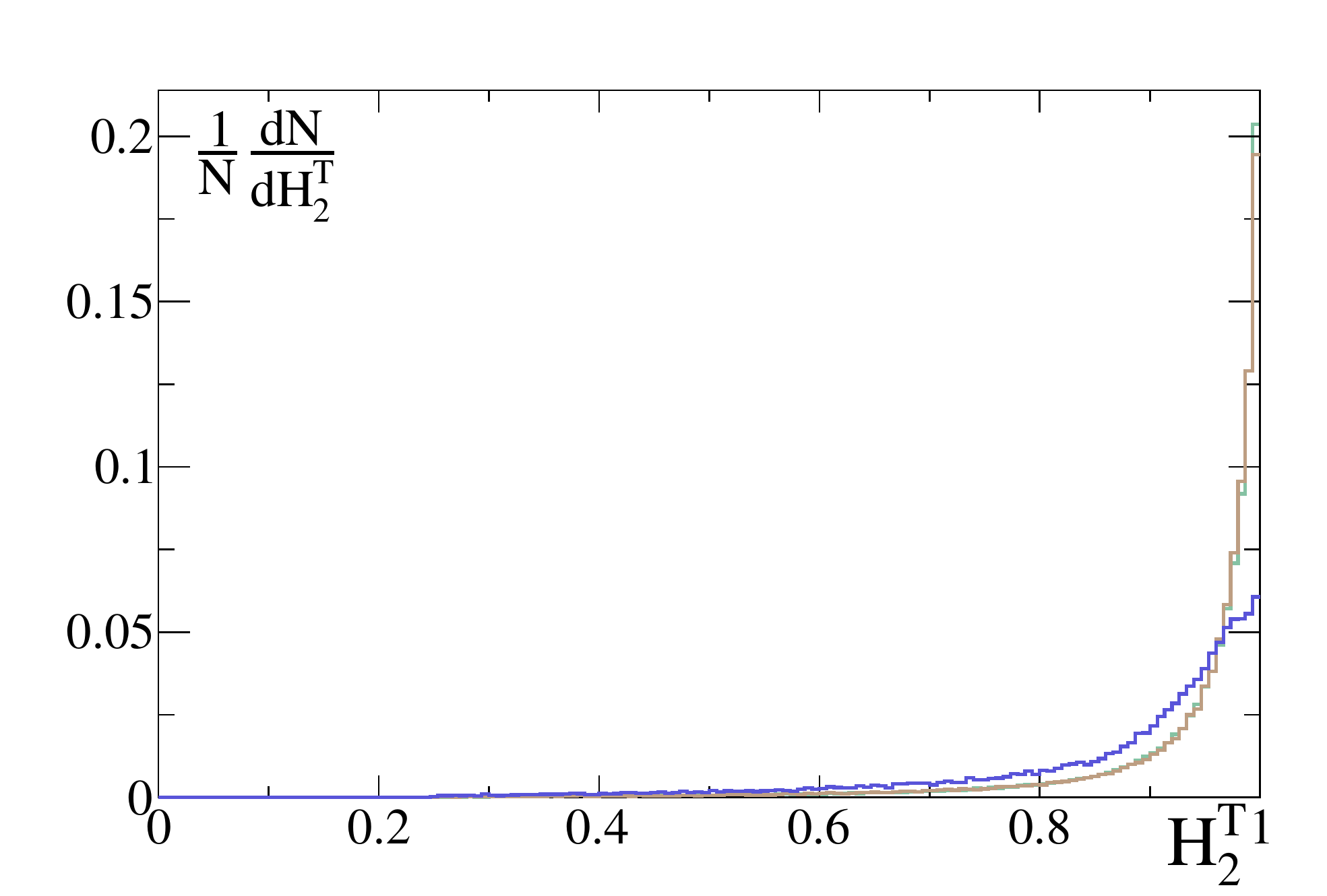} 
\includegraphics[width=0.24\textwidth]{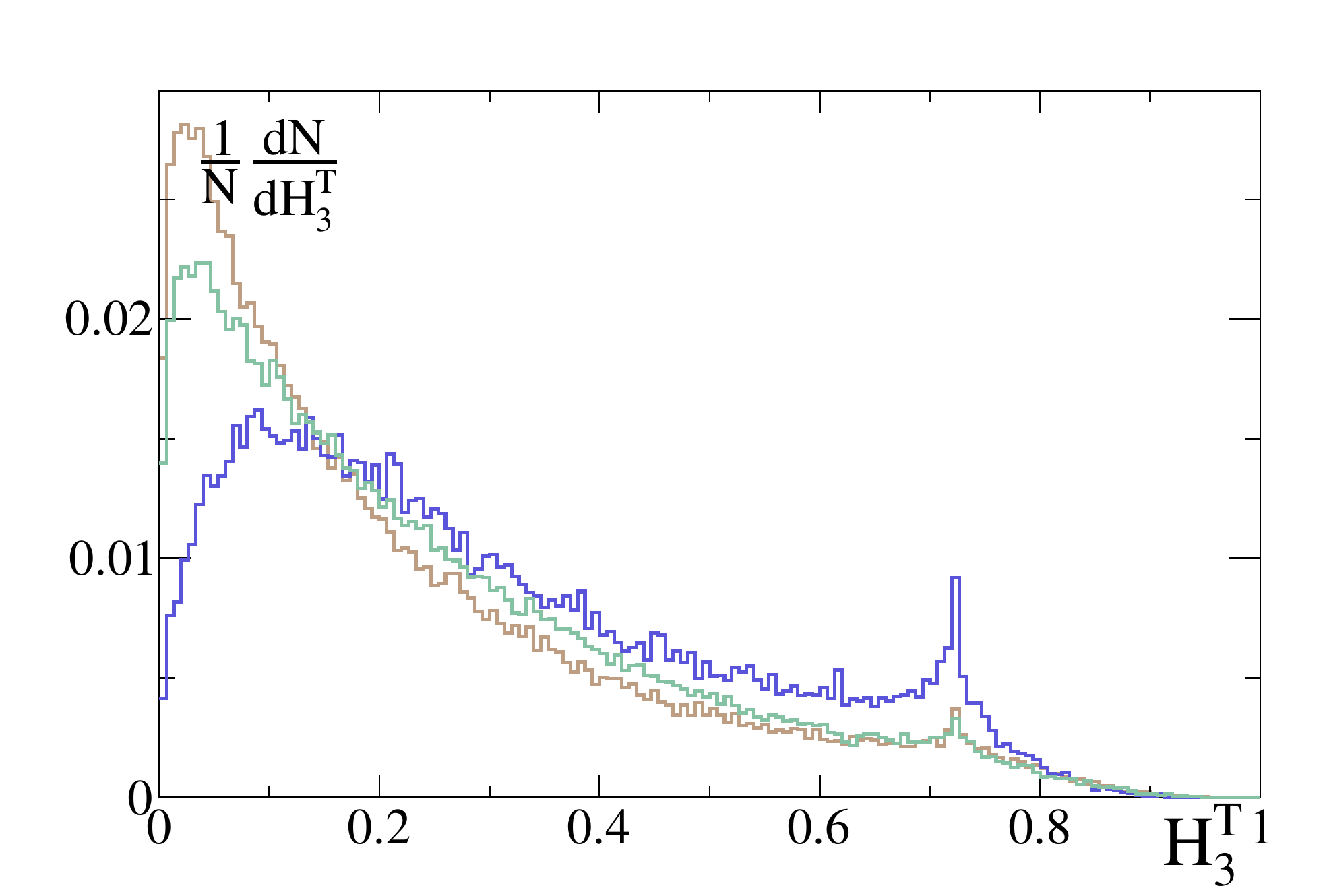} 
\includegraphics[width=0.24\textwidth]{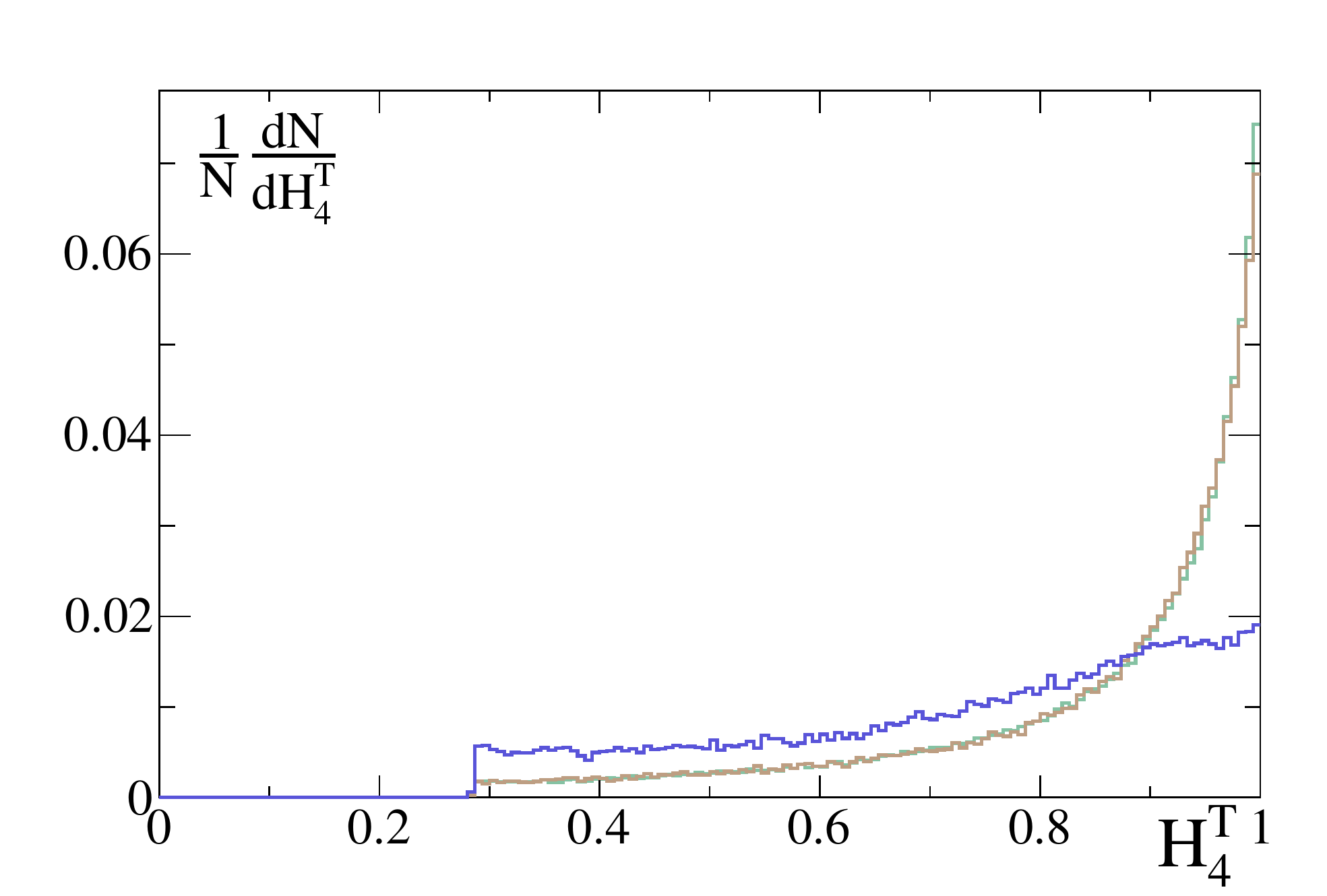} 
\includegraphics[width=0.24\textwidth]{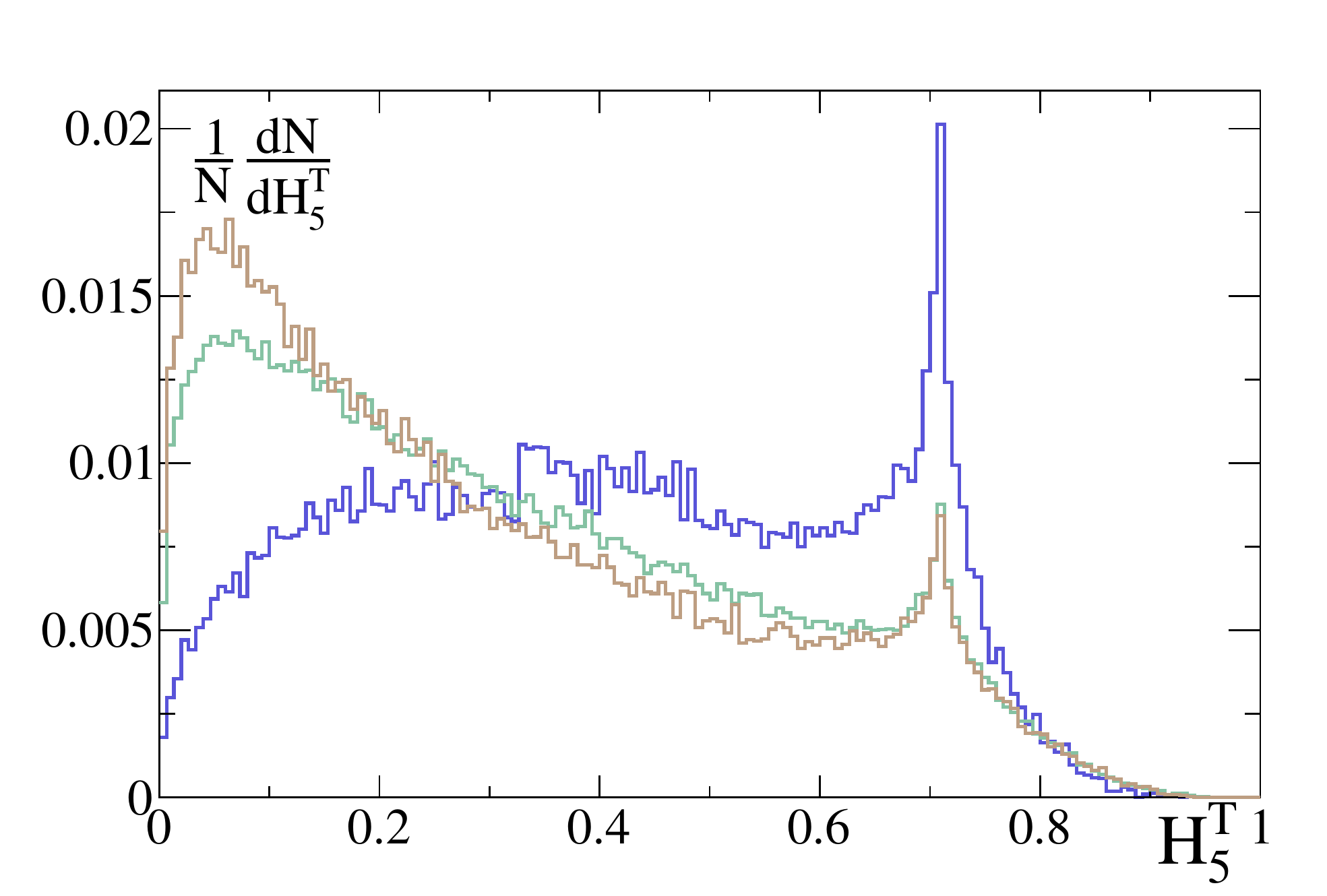} \\
\includegraphics[width=0.24\textwidth]{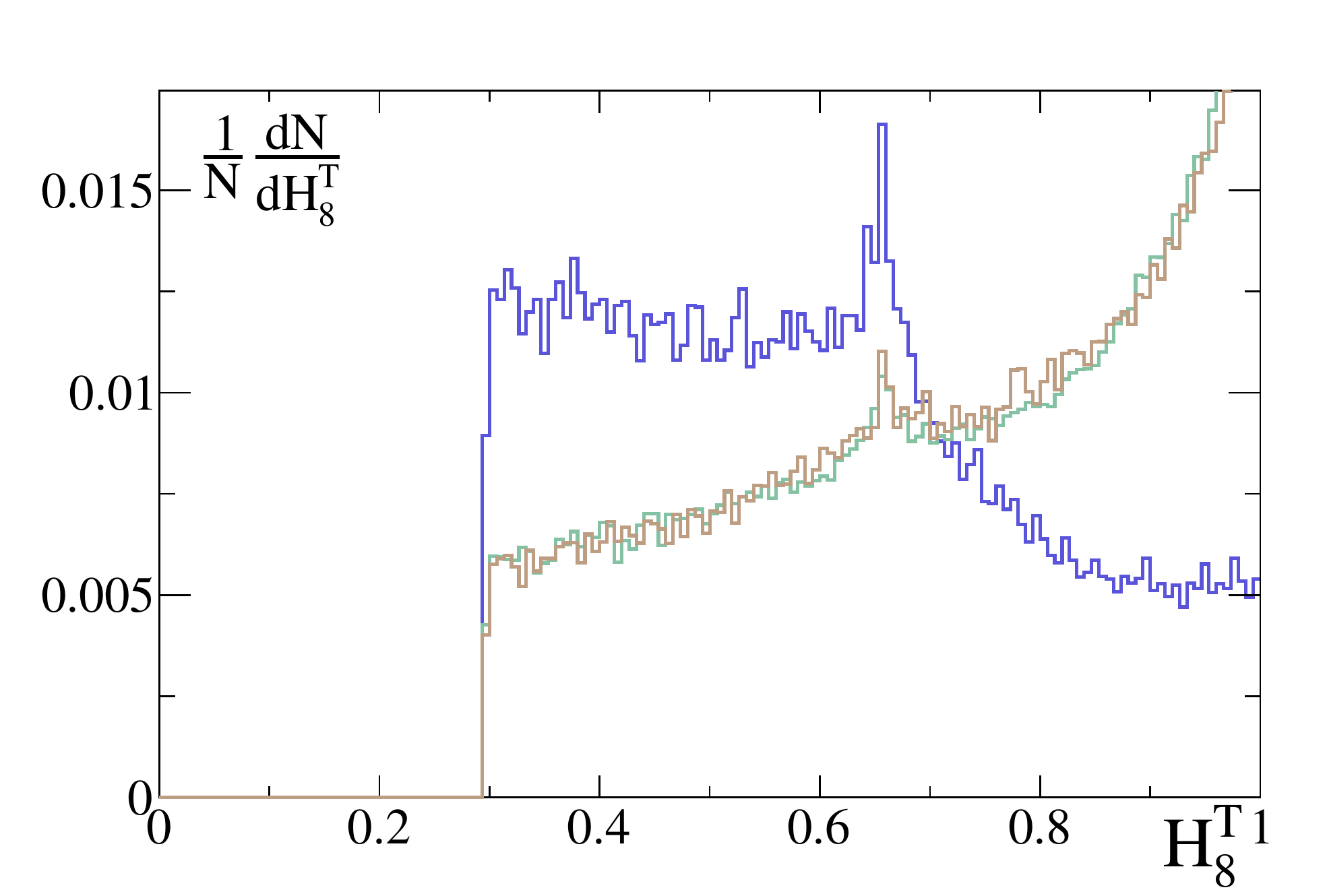} 
\includegraphics[width=0.24\textwidth]{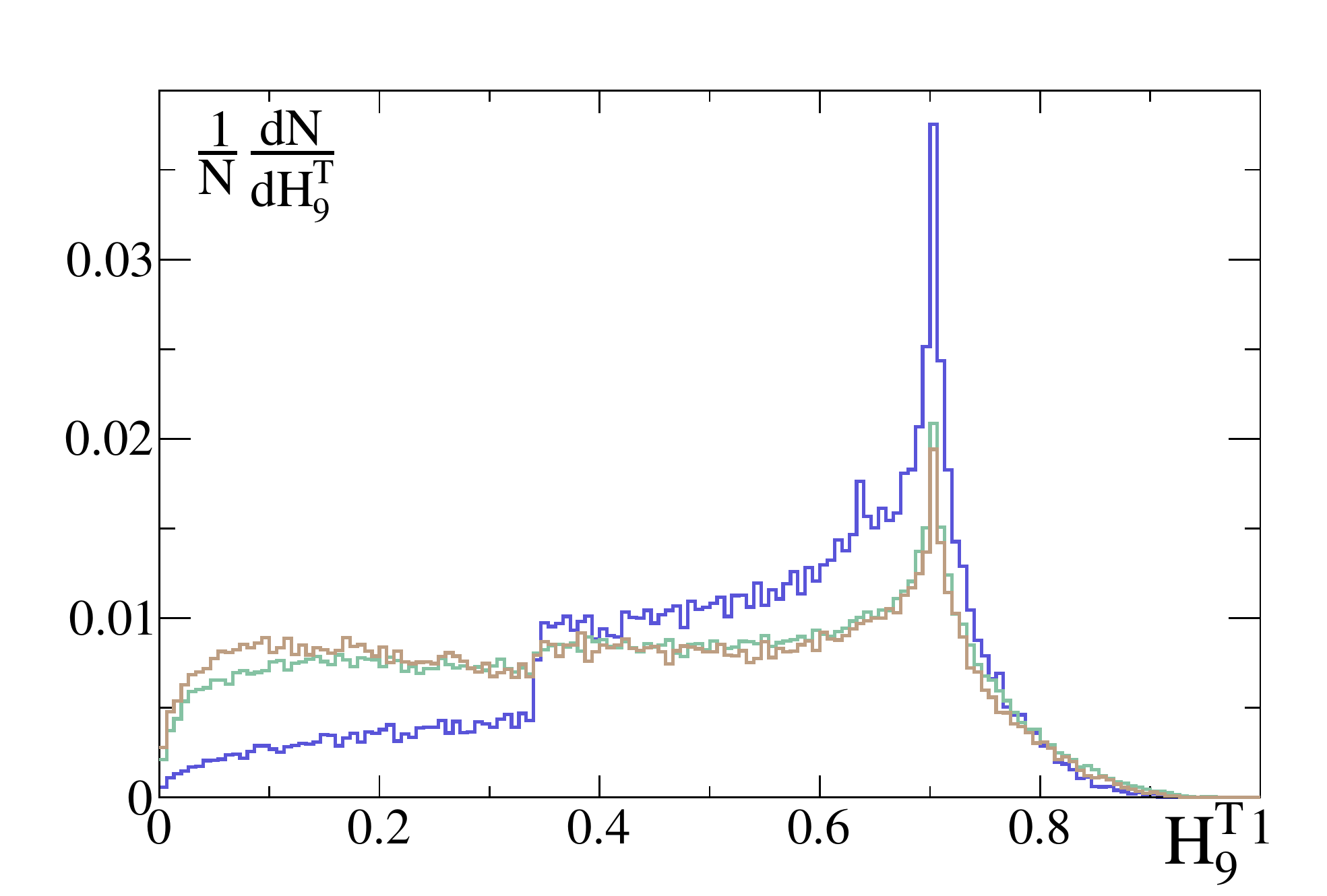} 
\includegraphics[width=0.24\textwidth]{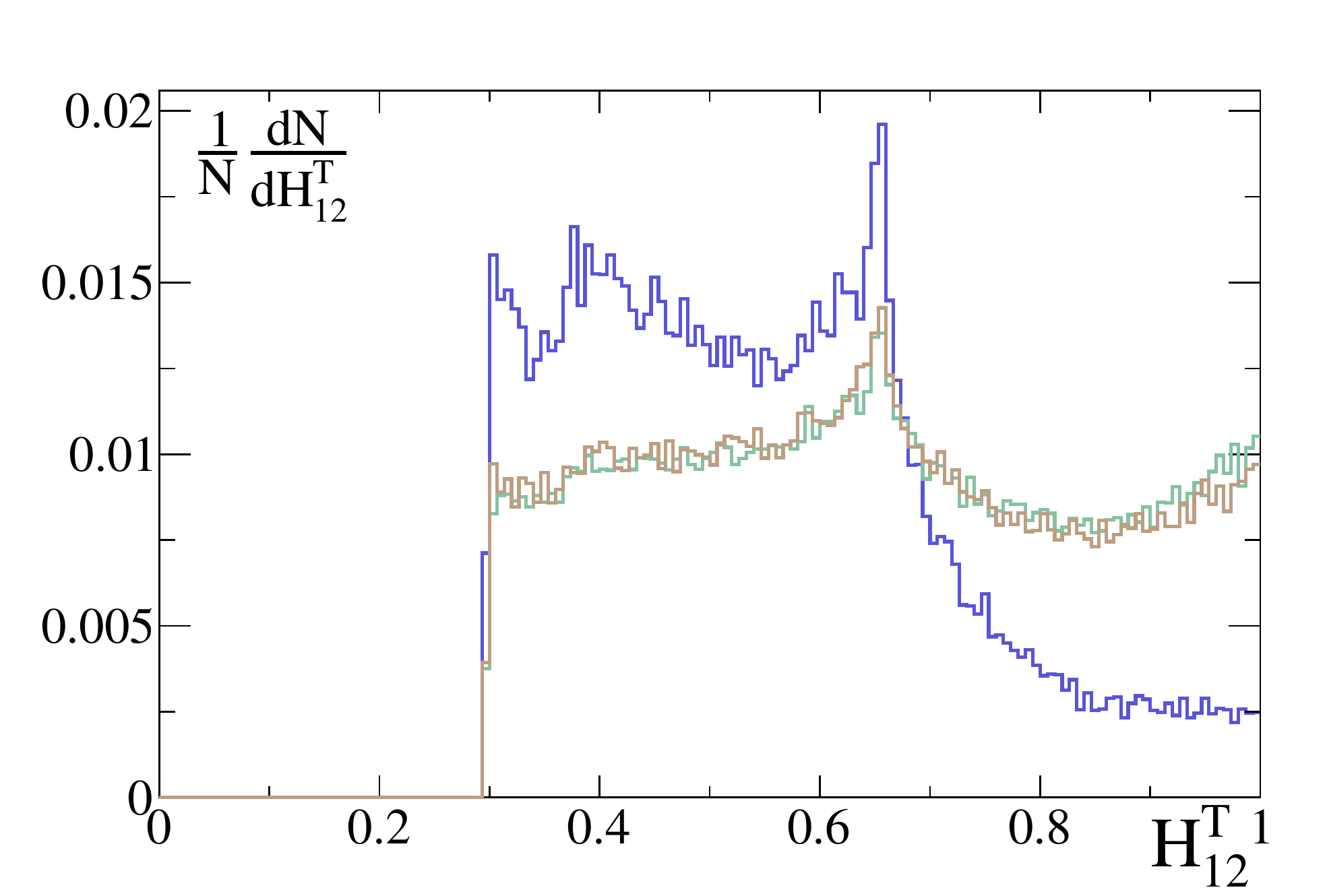} 
\includegraphics[width=0.24\textwidth]{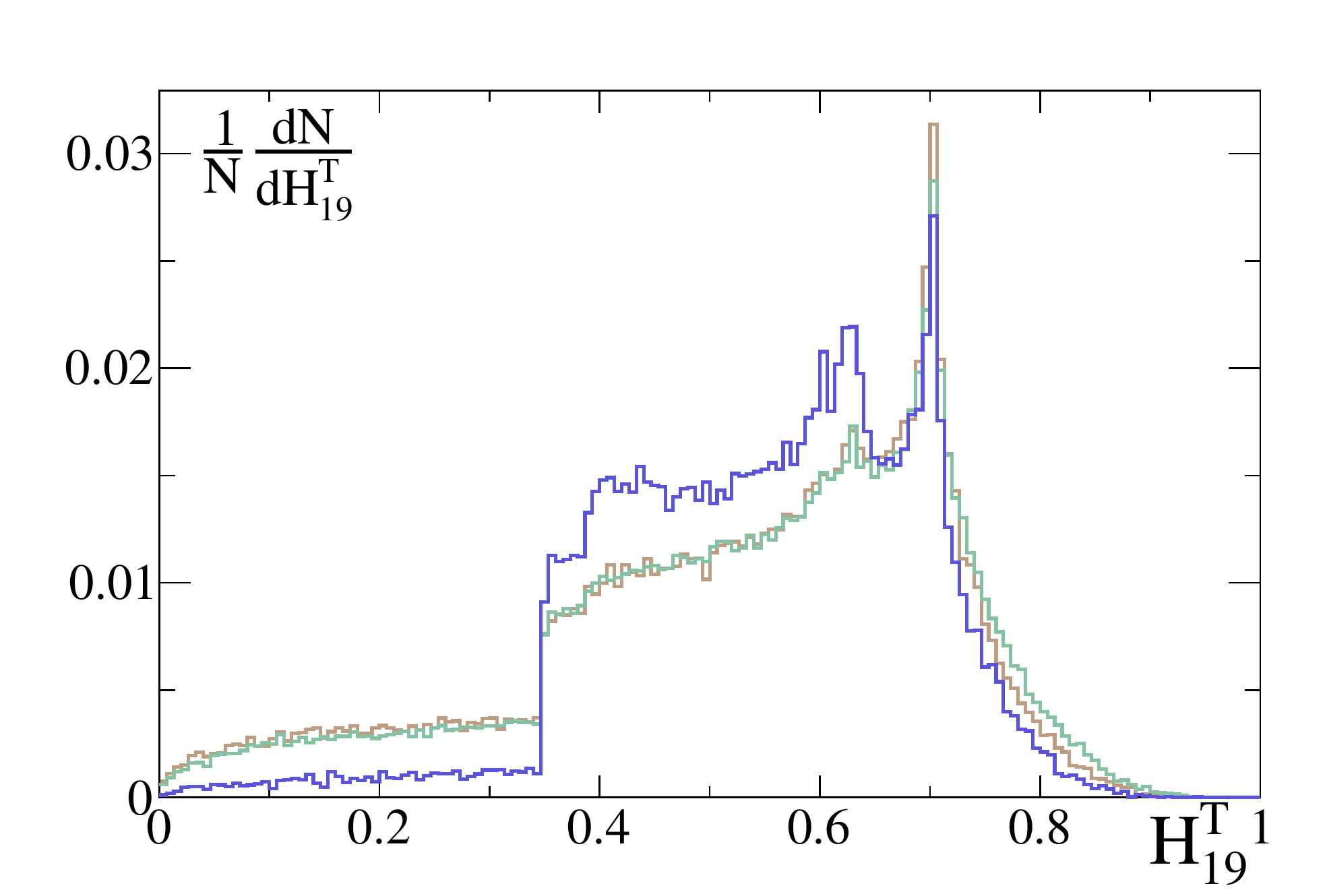} 
\caption{Normalized distributions of Fox--Wolfram moments computed
  from the two tagging jets only for $\ell = 2-5,8,9,12,19$ with a
  weight factors $W_{ij}^T$ for WBF H+1 jet signal (green), Z+2 jets 
  (brown) and $t\bar{t}$+1 jet (blue). All events pass the full set of QCD cuts
  Eqs.\eqref{eq:jj1}-\eqref{eq:jj3}.} 
\label{fig:all_moments2}
\end{figure}

If Fox--Wolfram moments computed from the two tagging jets alone
cannot replace geometric cuts altogether, the question becomes how
much they add after all the geometric cuts listed in
Tab.~\ref{tab:CFclassic}.  In Fig.~\ref{fig:all_moments2} we show a
selected set of transverse Fox--Wolfram moments $H^T$ for signal and
background events after applying all cuts including $\Delta y_{jj} >
4.4$ defined in Eq.\eqref{eq:jj3}.  At this stage the usual WBF cuts
on the tagging jets have been exhausted. Some of the even
$H^T_\ell$ distributions clearly distinguish top pair production on
the one hand from $H+$jets and $Z+$jets on the other. Unfortunately,
not even the large-$\ell$ moments will be able to clearly distinguish
between Higgs and $Z$ production. Any kind of improvement in $S/B$
will only arise because of a reduction of the top background. As an
example, following Fig.~\ref{fig:all_moments2} we can require
$H^T_{12} > 0.7$ to improve $S/B \sim 1/73$ to $1/57$, but without a
beneficial effect on $S/\sqrt{B}$.\bigskip

Summarizing the analysis of Fox--Wolfram moments computed from tagging
jets only, the moments will not replace kinematic cuts on the tagging
jet geometry altogether. However, they add useful information on the
Higgs signal and the $Z+$jets and $t\bar{t}$ backgrounds at different
stages of the analysis. Even after applying all the usual WBF tagging
jet cuts, the top pair background in particular can be further
suppressed just based on the tagging jet correlations phrased in terms
of $H_\ell$. Moreover, the different moments are less correlated than
one might have guessed from the toy model discussed in
Sec.~\ref{sec:moments}.

\section{All jets}
\label{sec:qcd}

\begin{figure}[!b]
\includegraphics[width=0.24\textwidth]{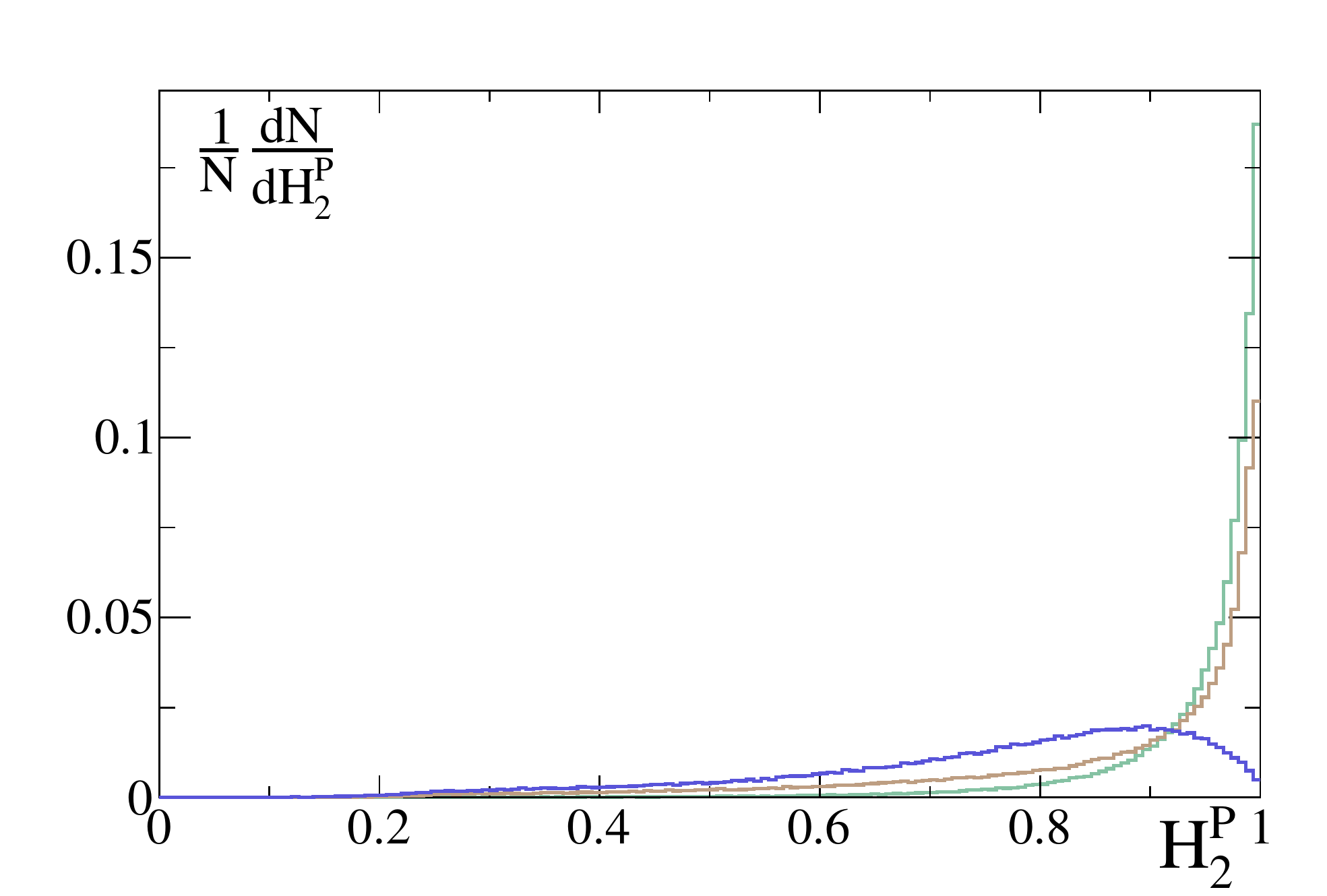}  
\includegraphics[width=0.24\textwidth]{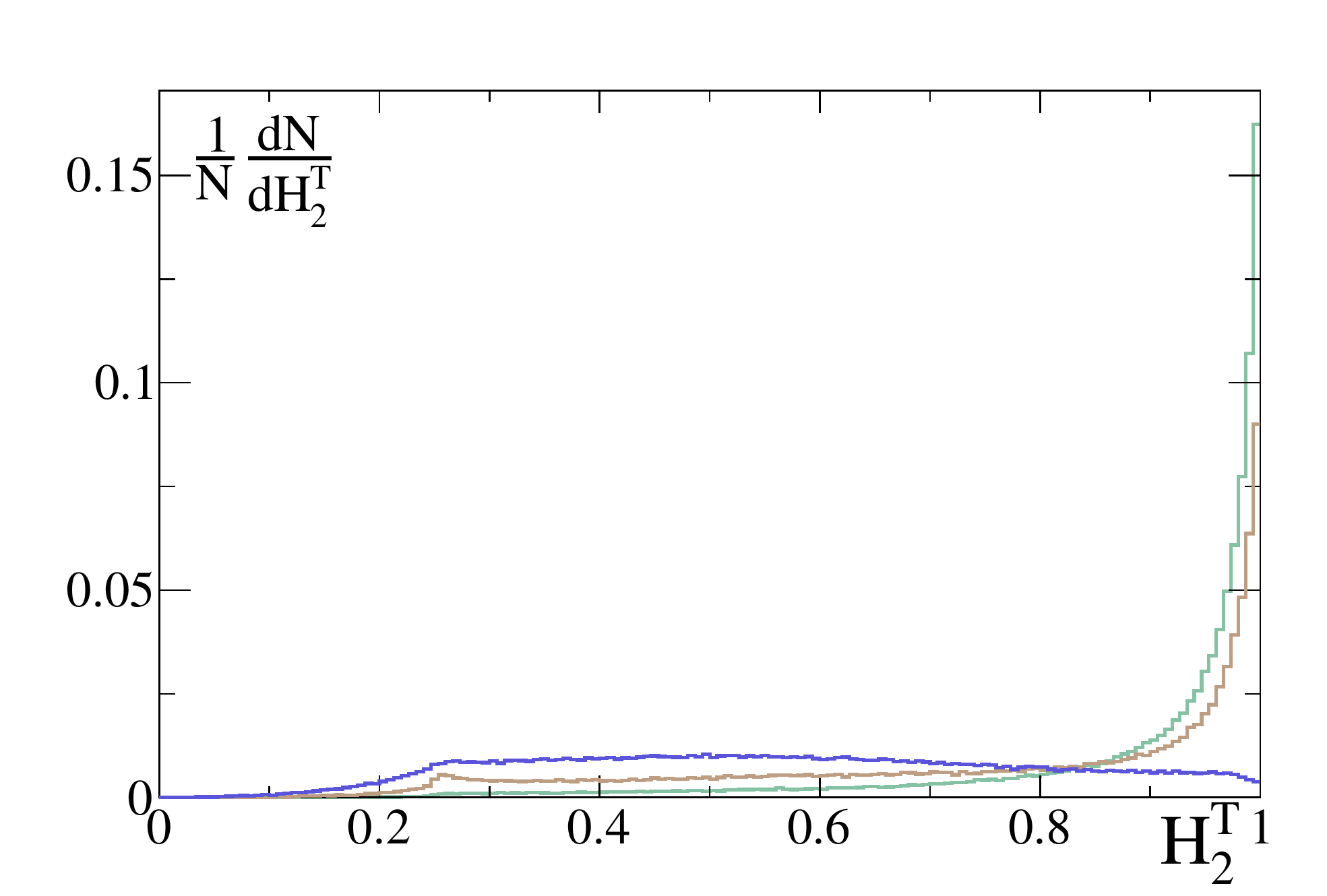} 
\includegraphics[width=0.24\textwidth]{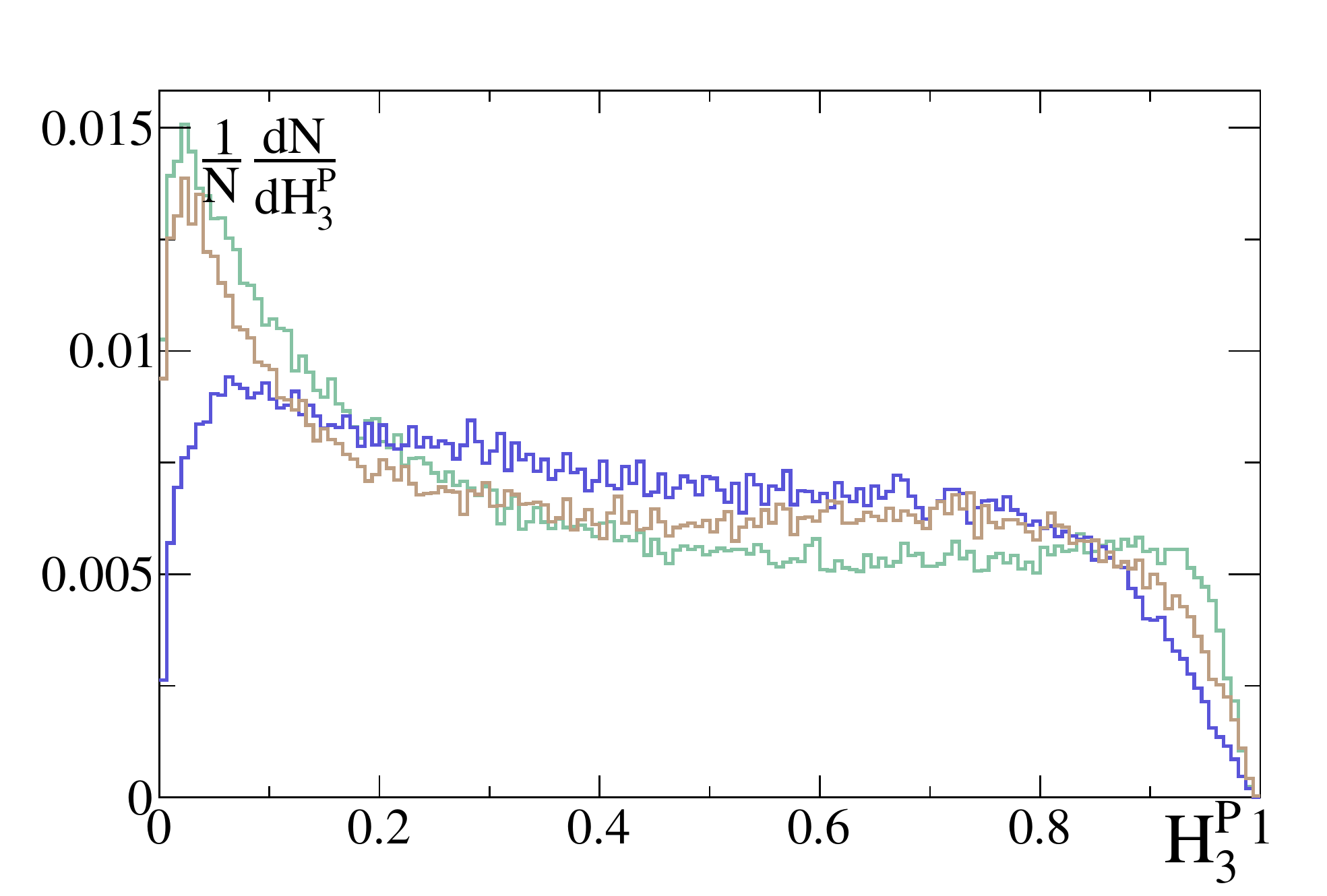} 
\includegraphics[width=0.24\textwidth]{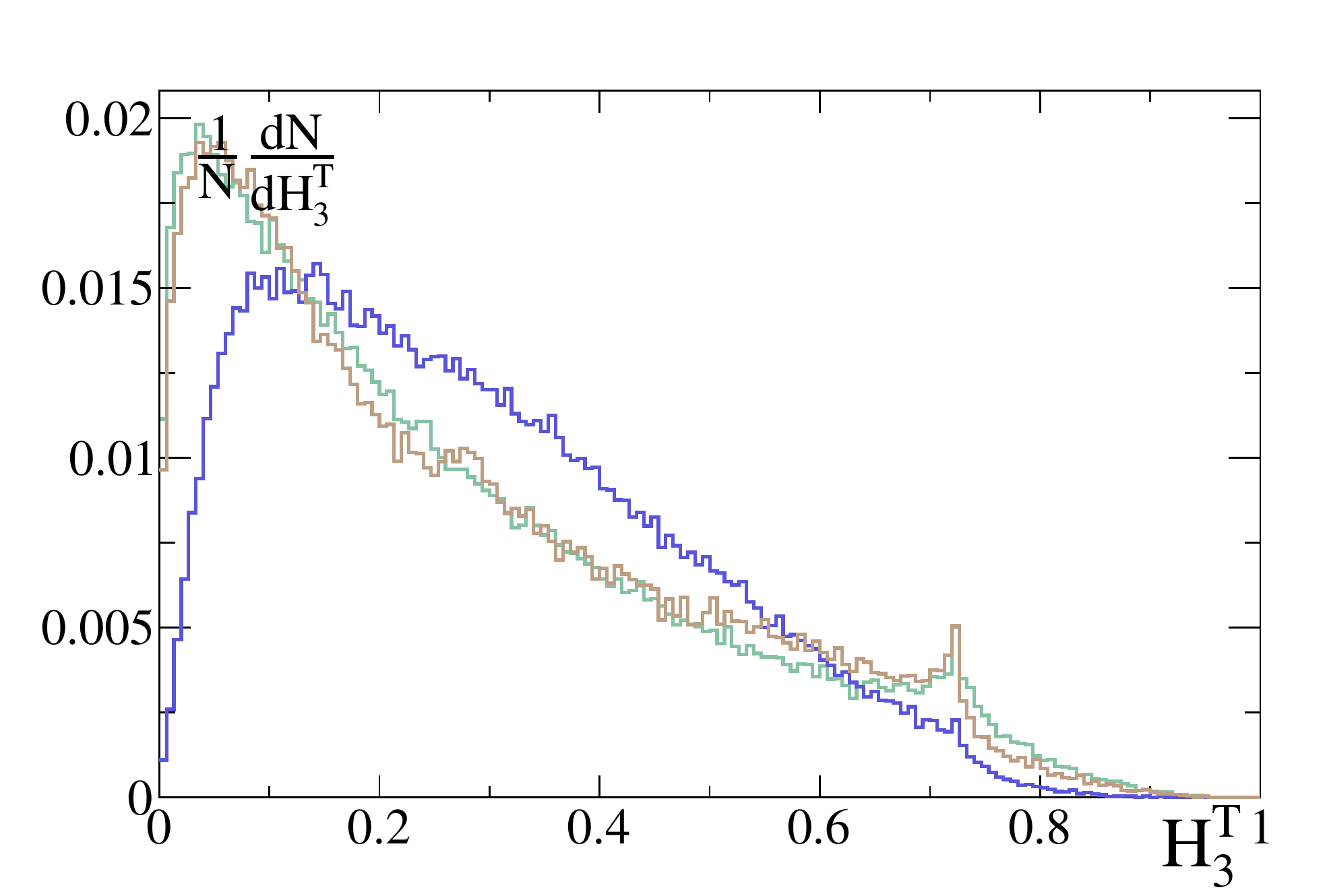} \\
\includegraphics[width=0.24\textwidth]{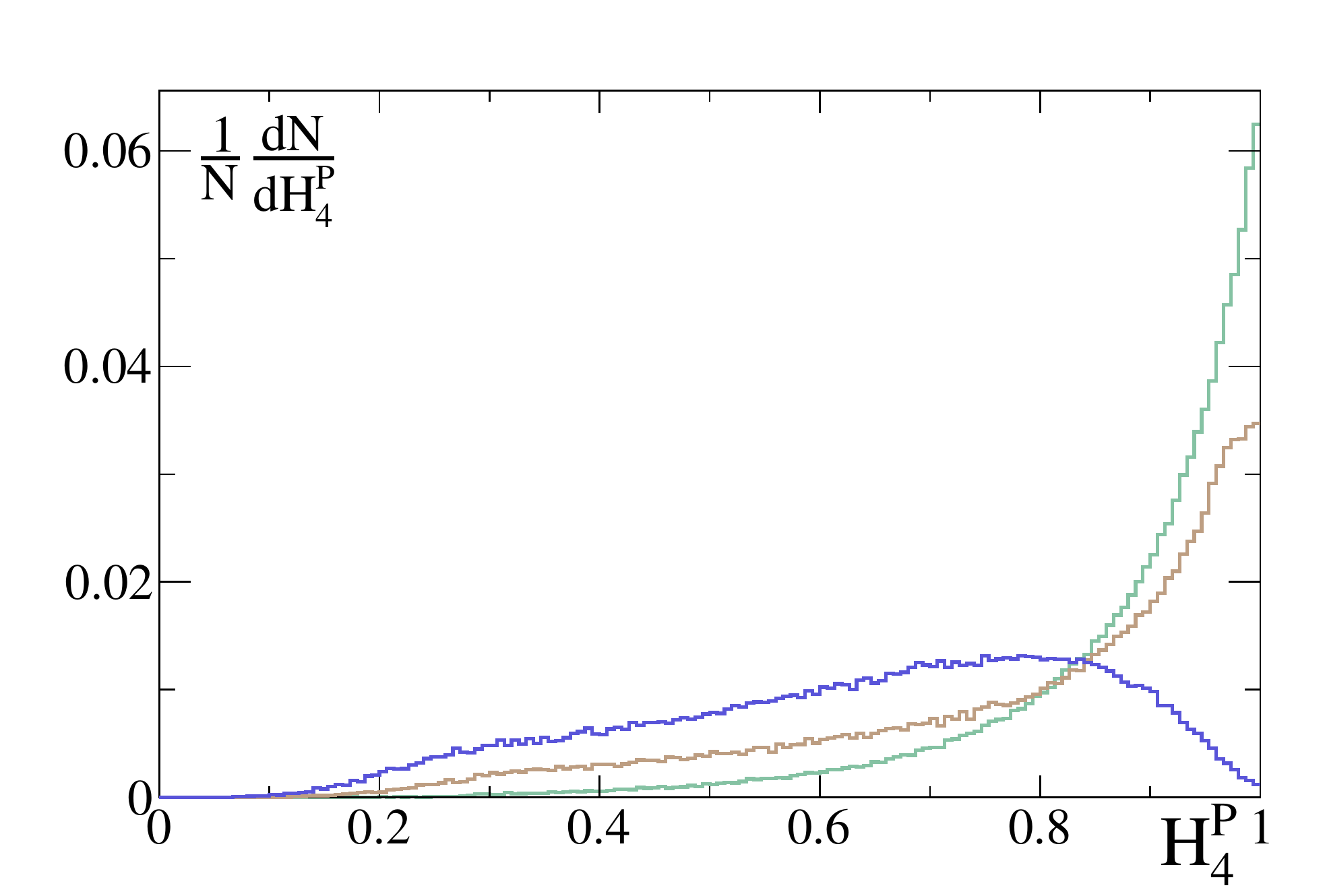} 
\includegraphics[width=0.24\textwidth]{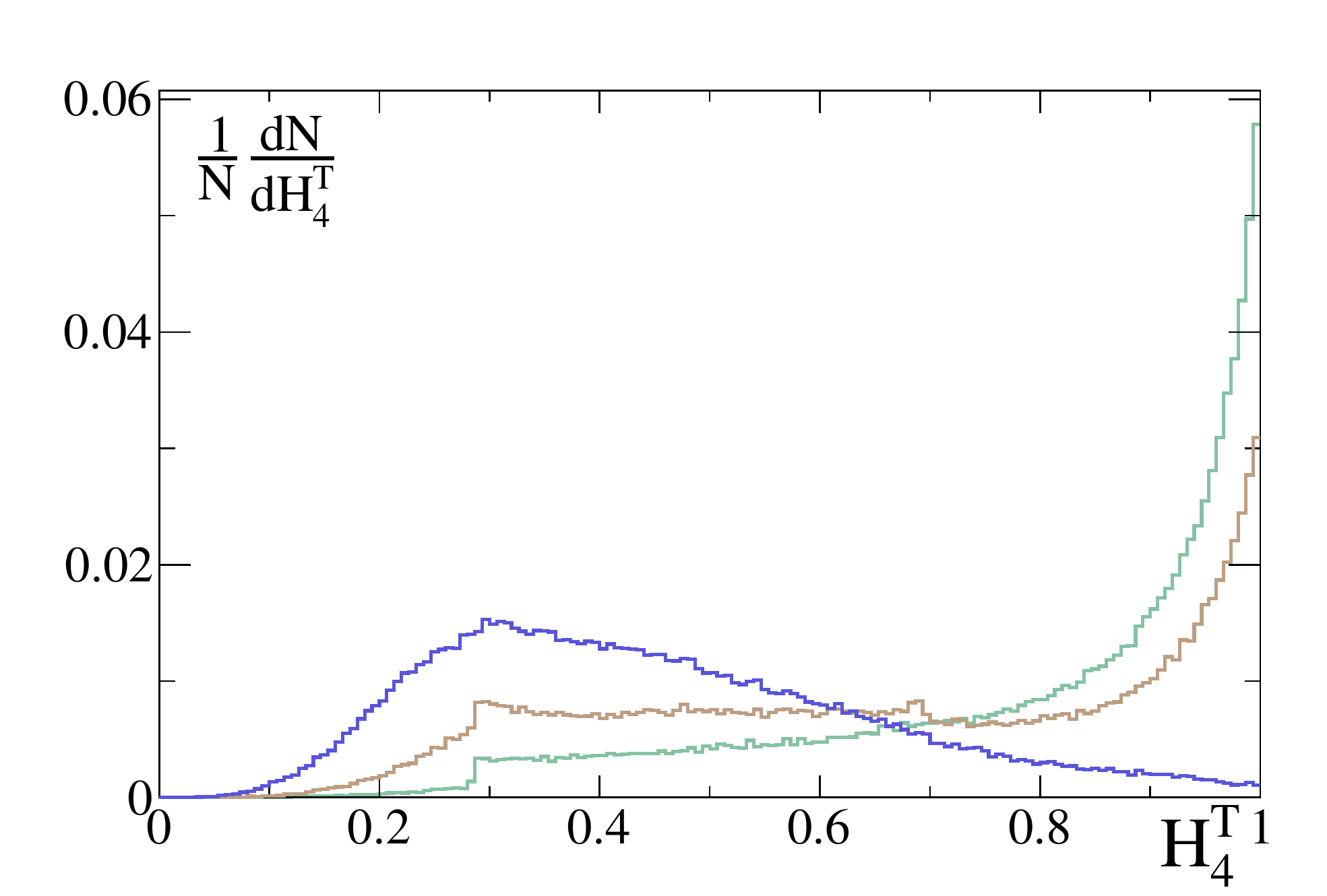} 
\includegraphics[width=0.24\textwidth]{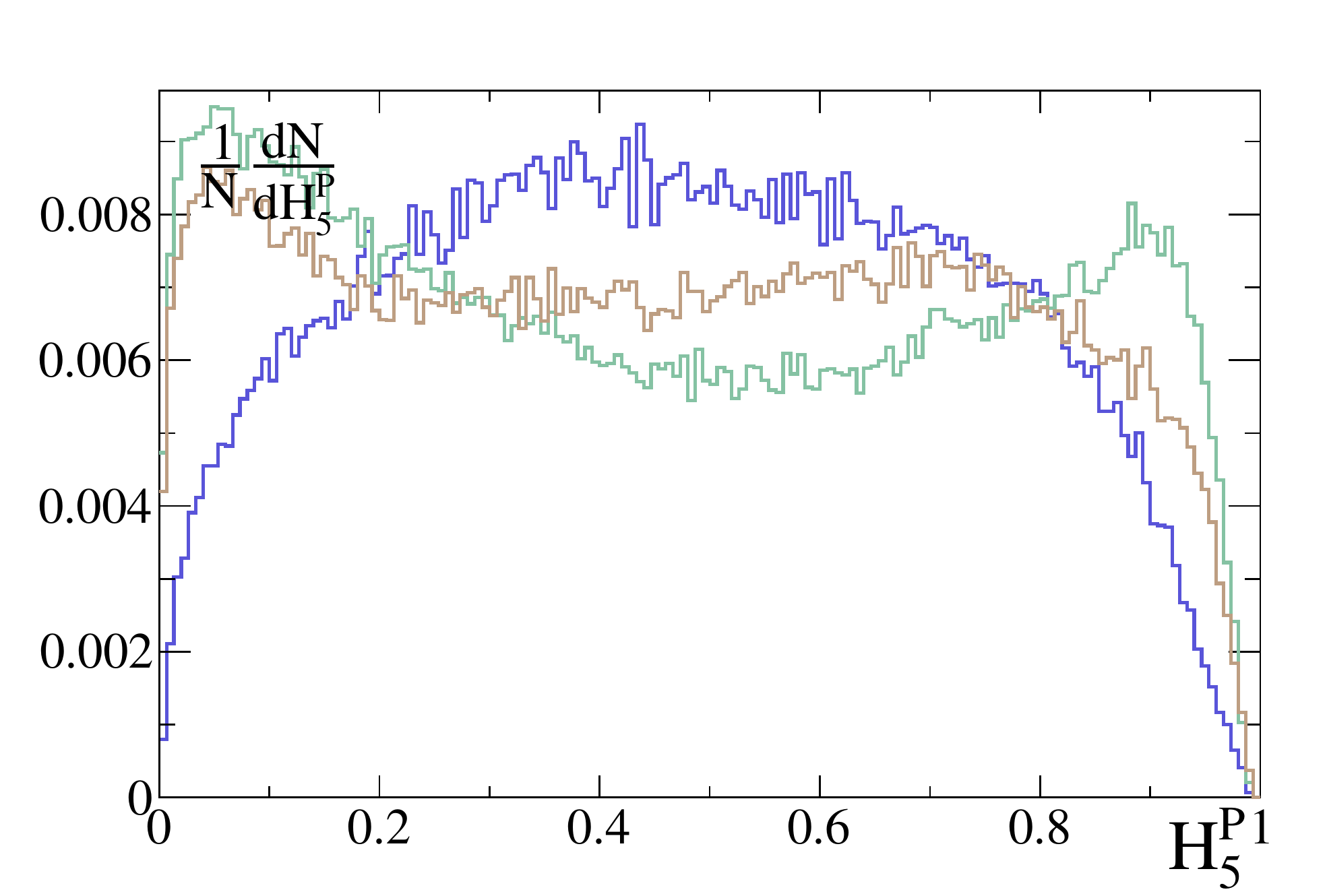} 
\includegraphics[width=0.24\textwidth]{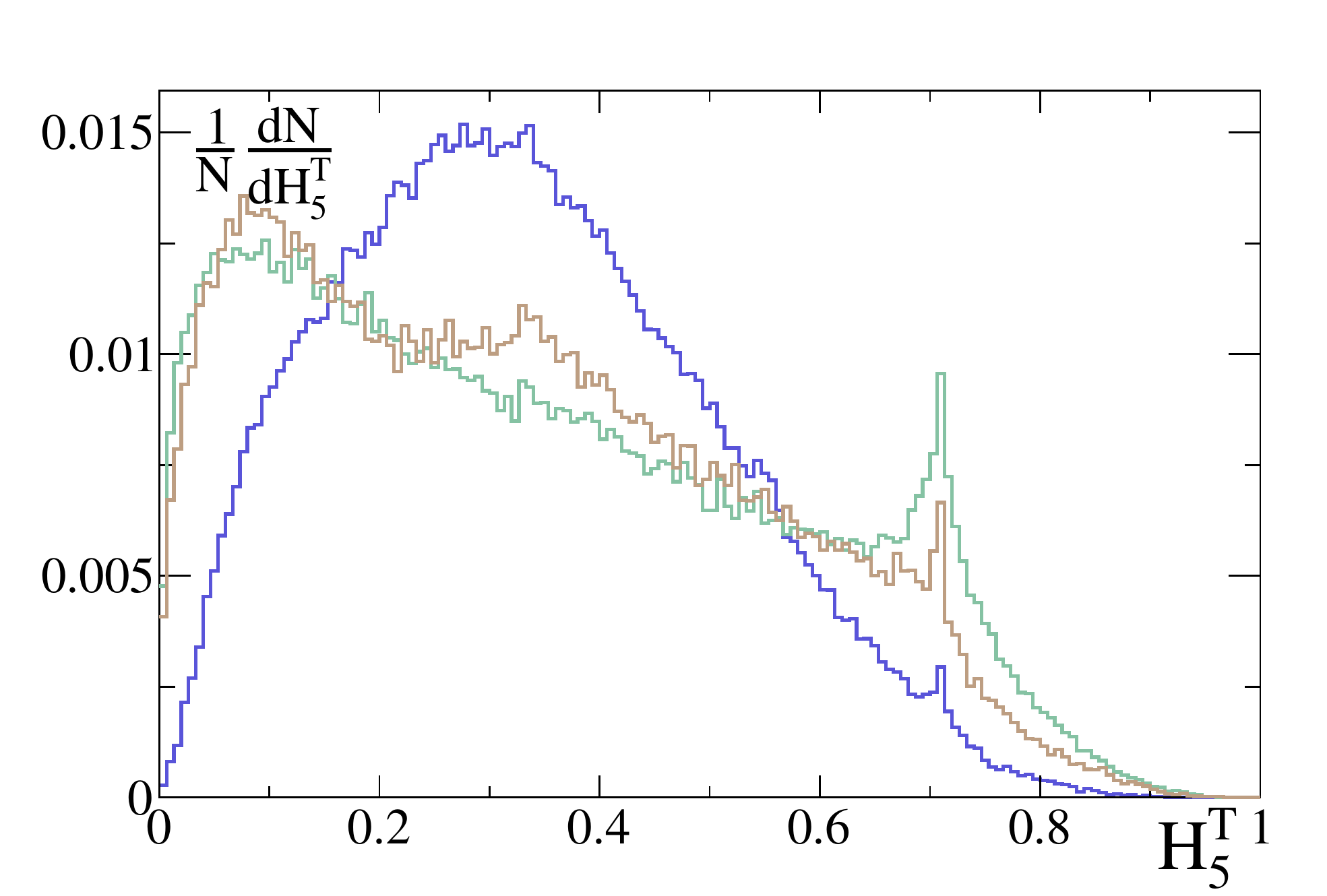} \\
\includegraphics[width=0.24\textwidth]{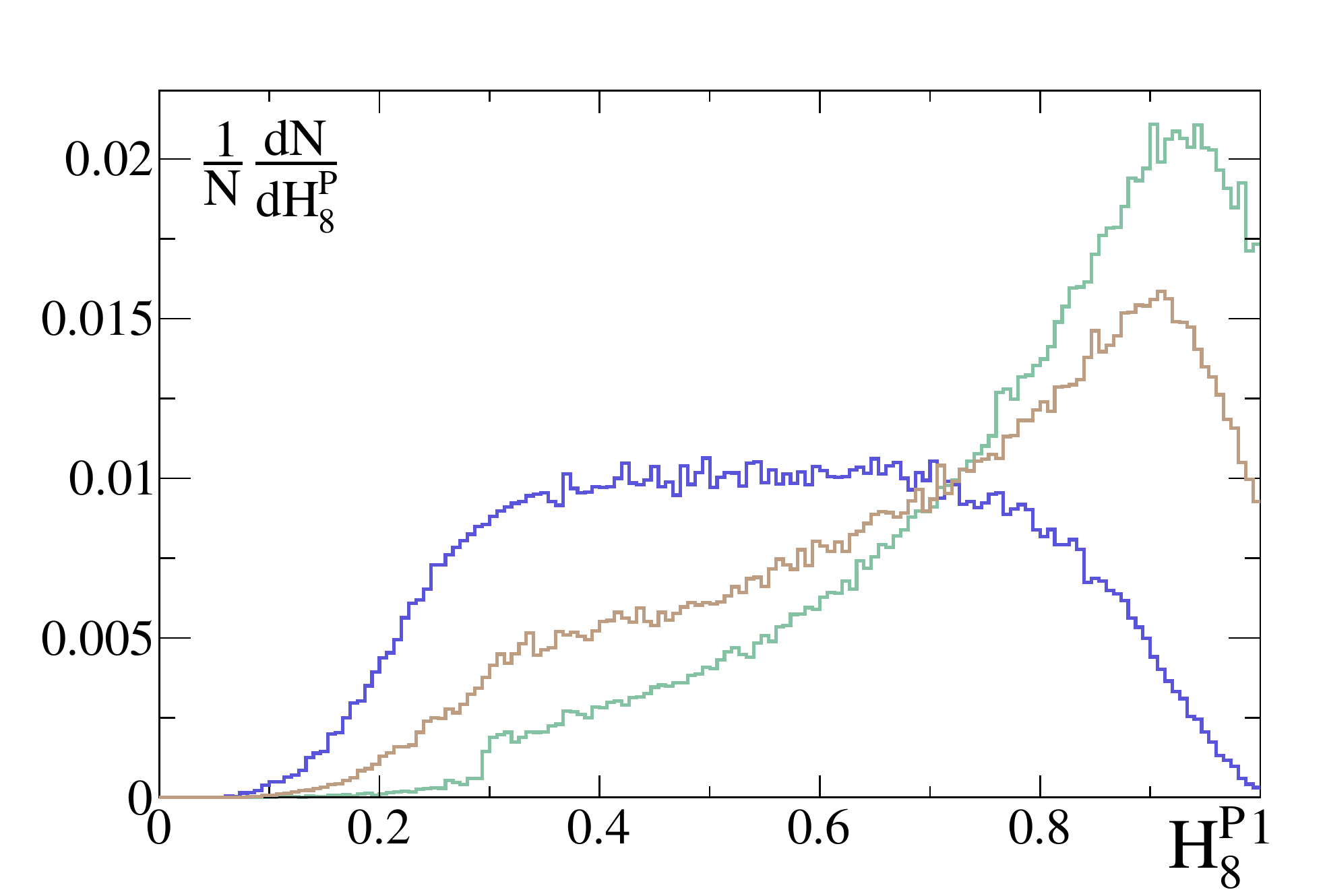} 
\includegraphics[width=0.24\textwidth]{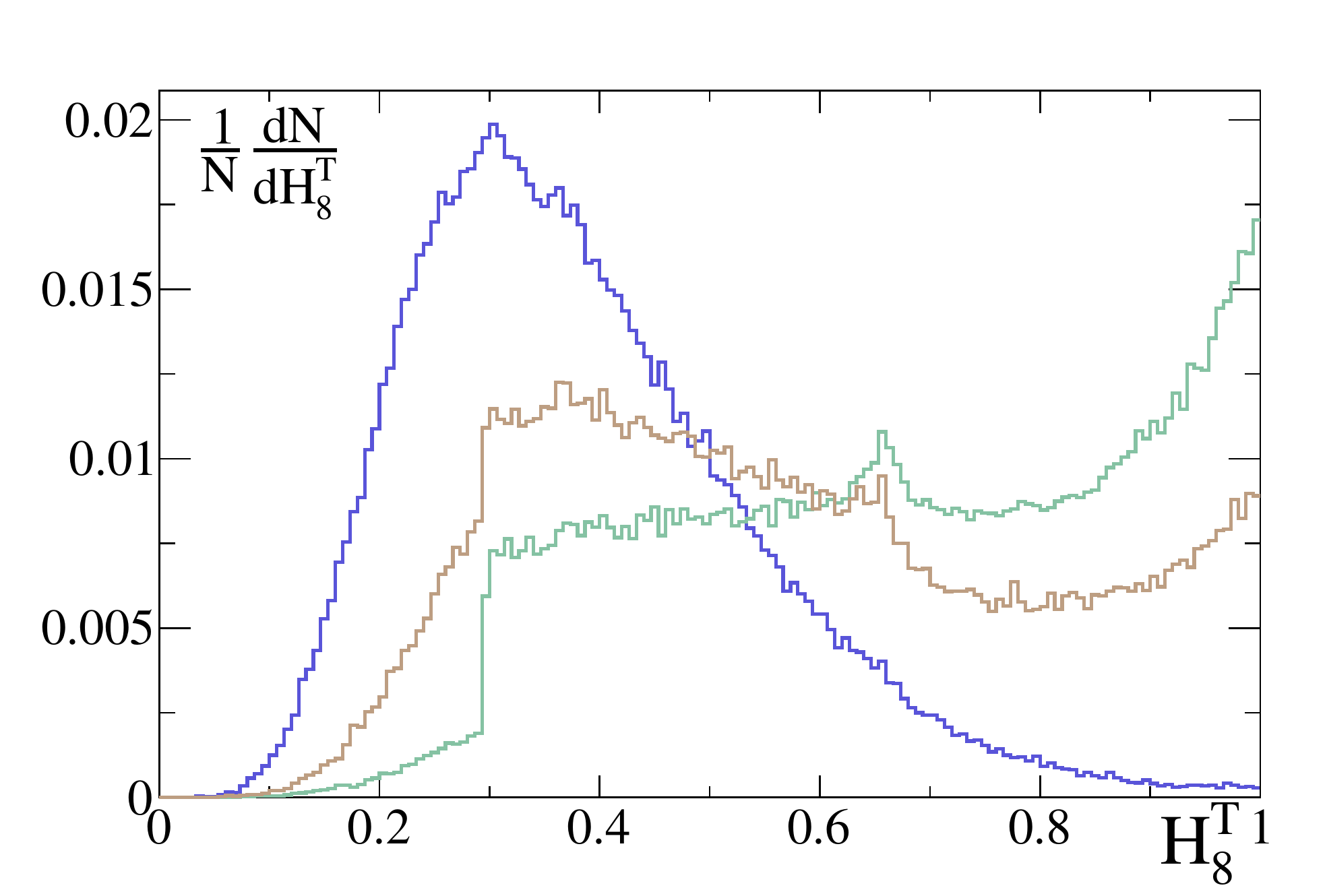} 
\includegraphics[width=0.24\textwidth]{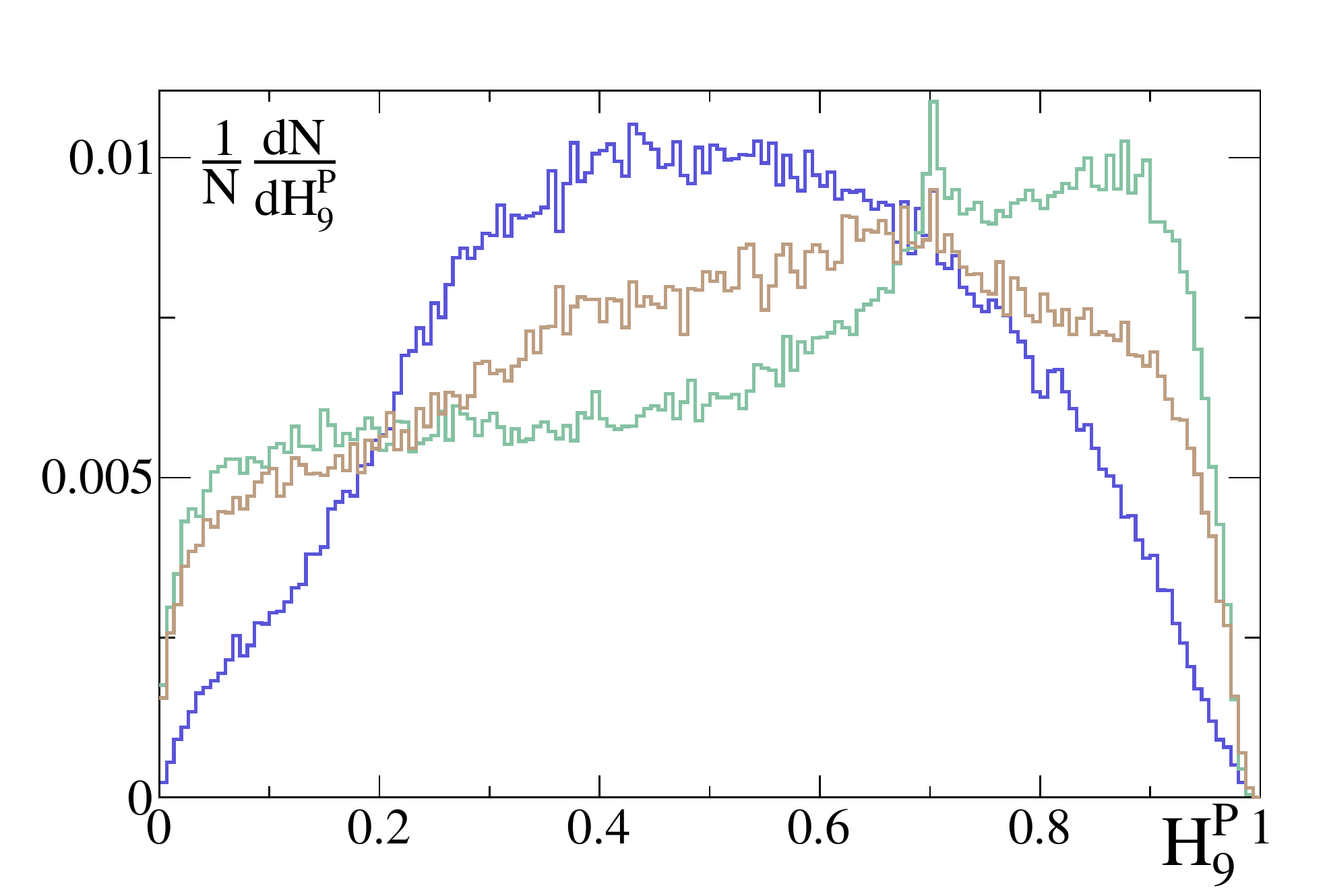} 
\includegraphics[width=0.24\textwidth]{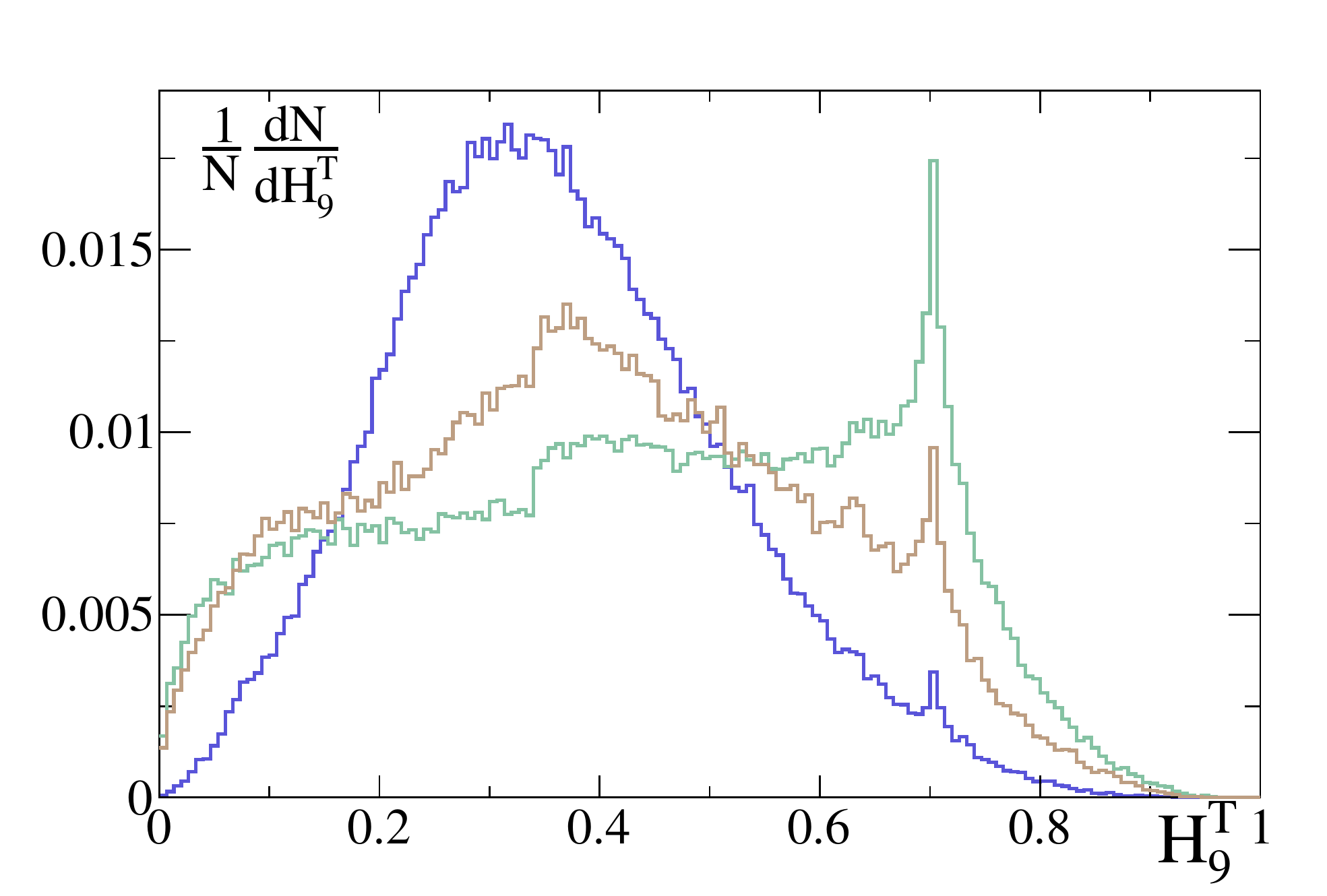} \\
\includegraphics[width=0.24\textwidth]{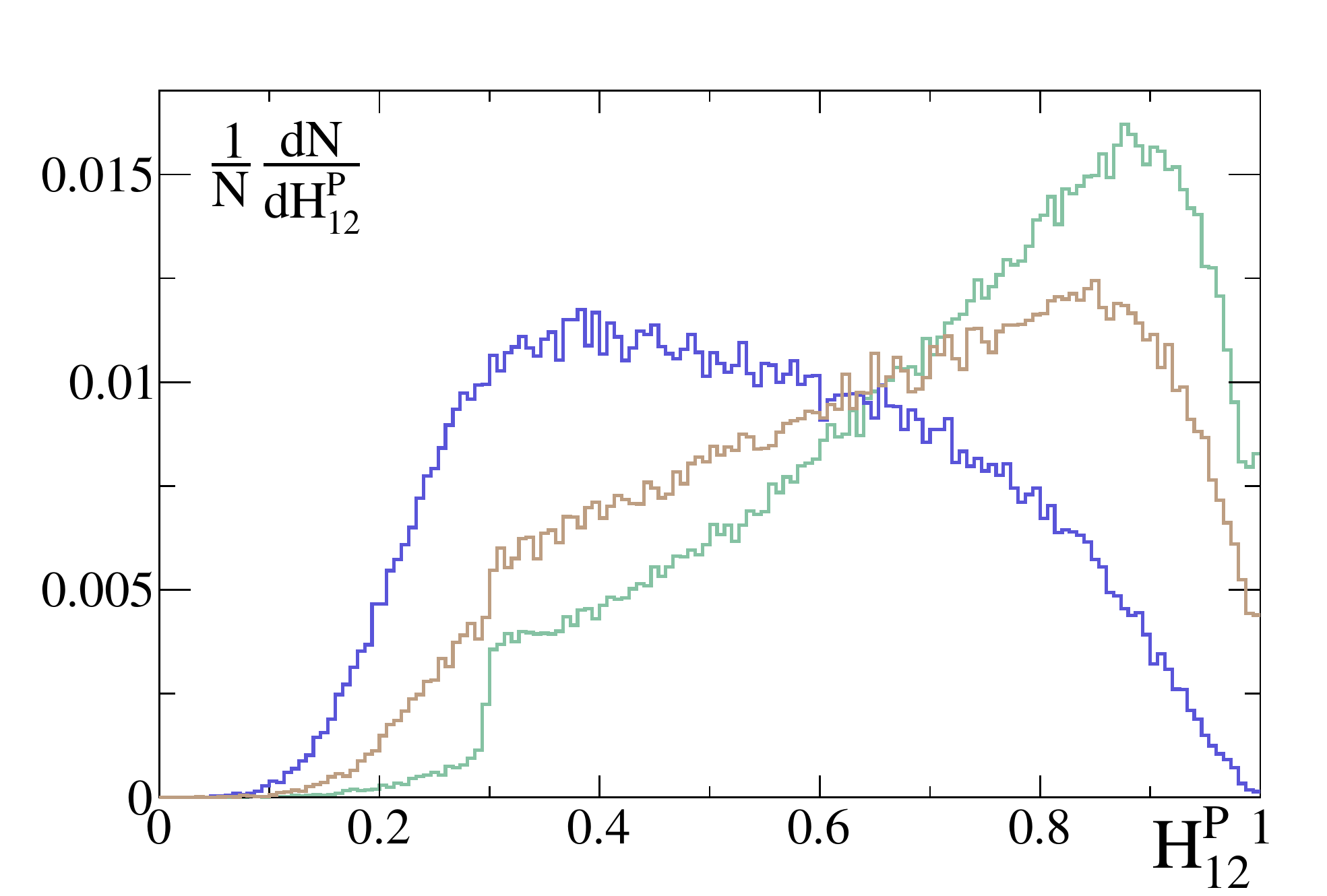} 
\includegraphics[width=0.24\textwidth]{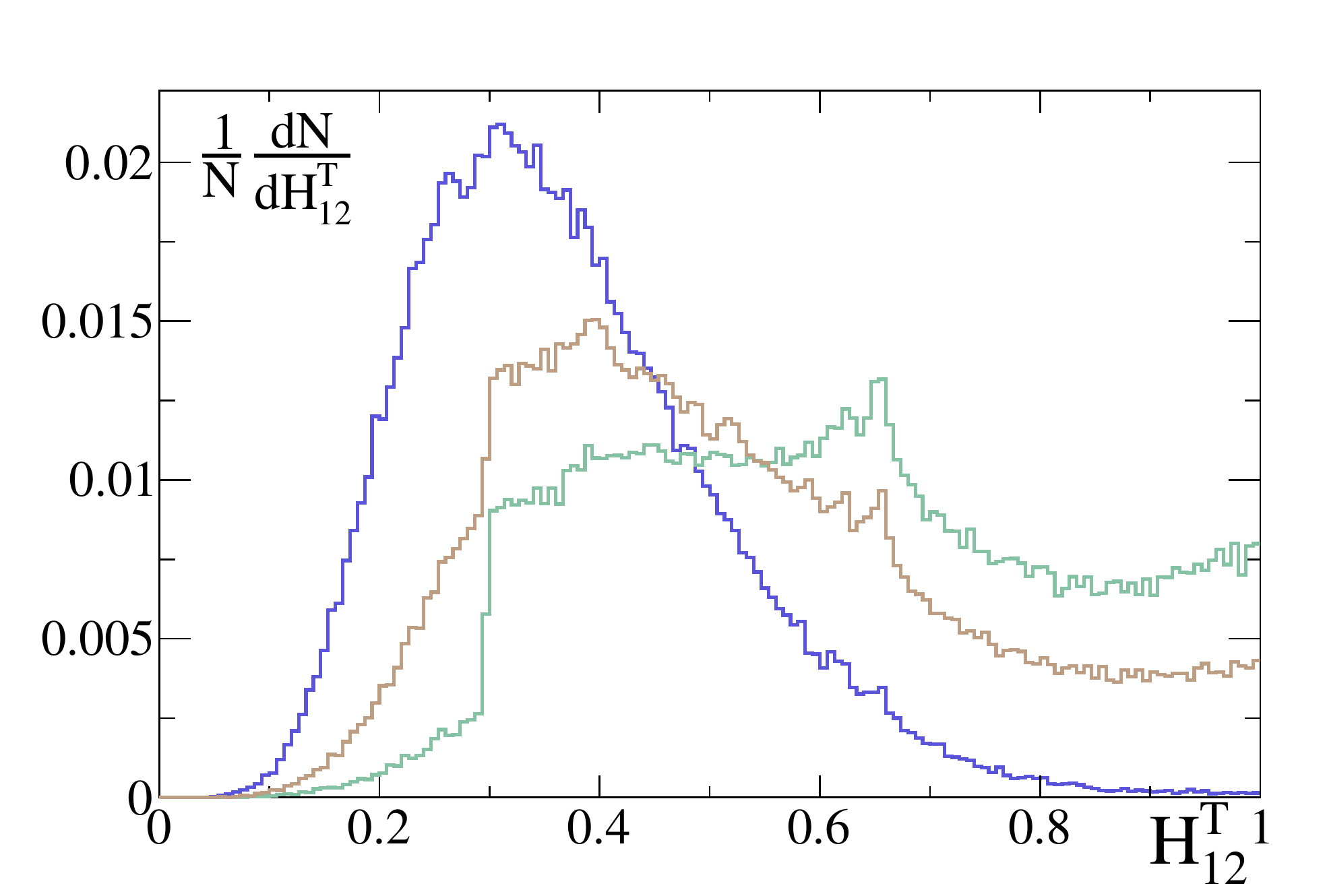} 
\includegraphics[width=0.24\textwidth]{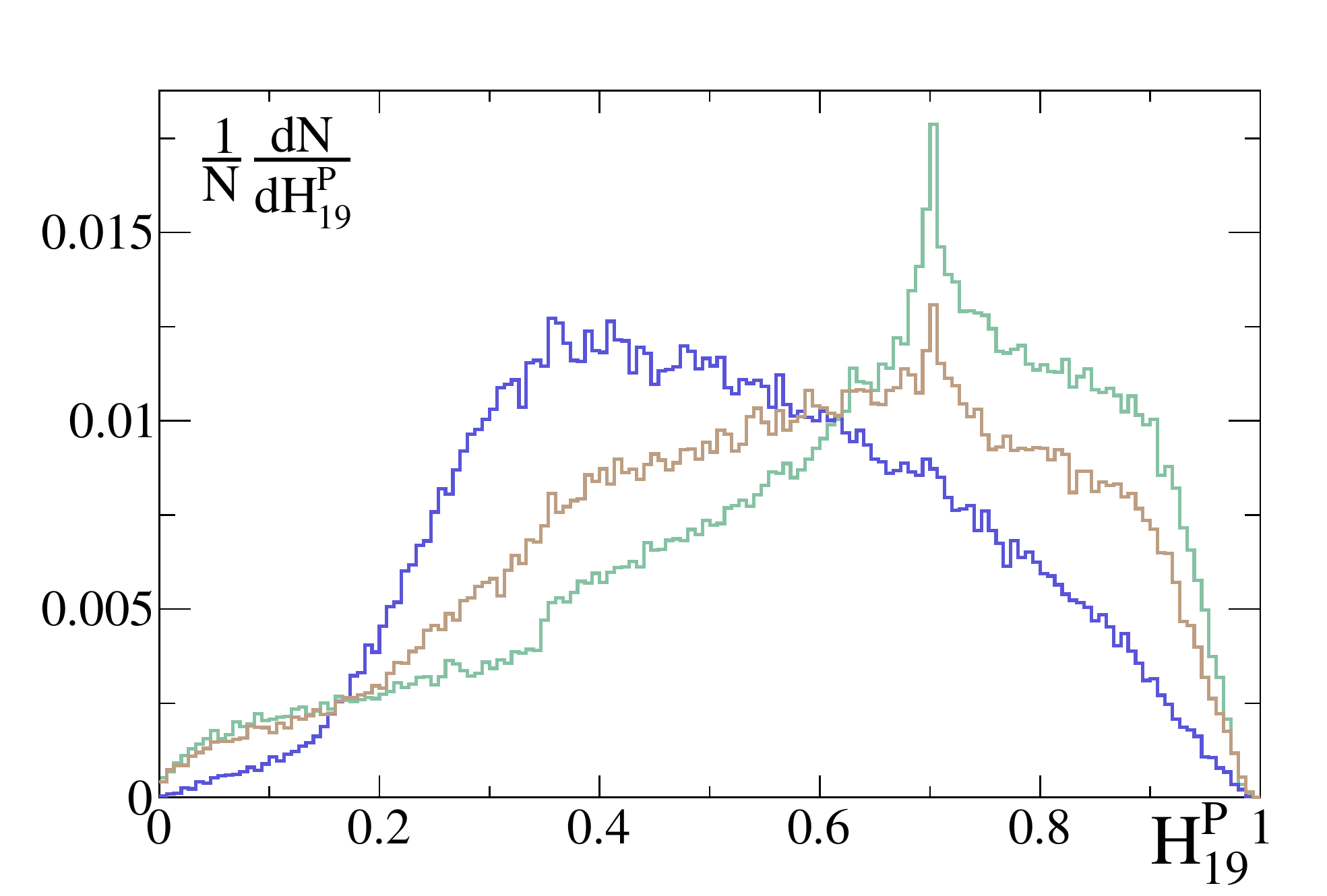} 
\includegraphics[width=0.24\textwidth]{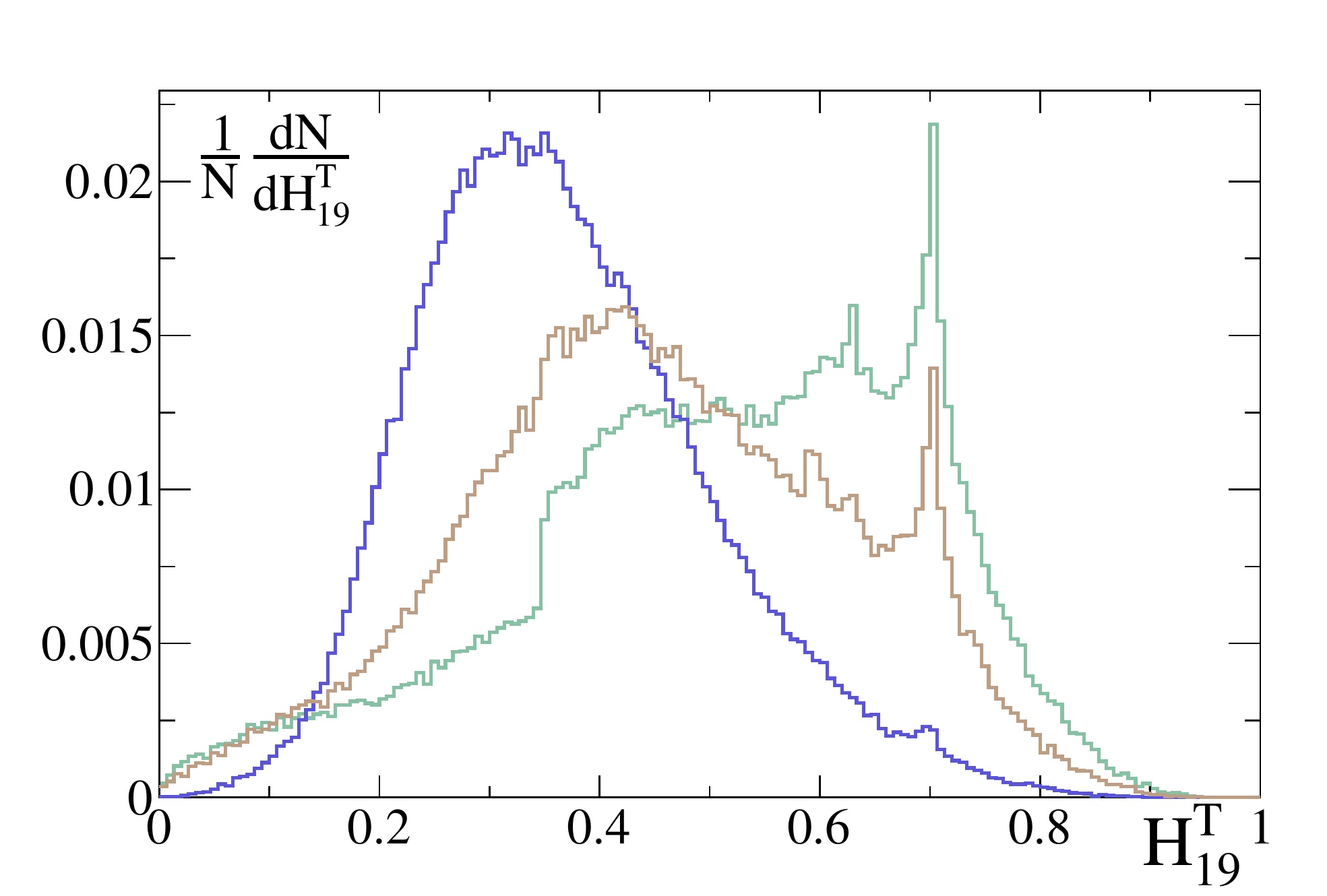} 
\caption{Normalized distributions on Fox--Wolfram moments computed
  from all available jets for $\ell = 2-5,8,9,12,19$ with weight
  factors $W_{ij}^p$ (left) and $W_{ij}^T$ (right) for WBF H+1 
  jet signal (green), Z+2 jets 
  (brown) and $t\bar{t}$+1 jet (blue).  All events pass
  the cuts Eq.\eqref{eq:jj1} and Eq.\eqref{eq:jj2}.}
\label{fig:all_moments_qcd}
\end{figure}

In addition to the hallmark tagging jets with a very large invariant mass and
sizeable transverse momentum the key to WBF Higgs analysis is the
reduced central jet activity. It can be understood in two different
ways.

First, weak boson fusion does not include any color correlations
between the two quark legs. In that sense, it consists of two distinct
deep inelastic scattering processes. QCD corrections involving a gluon
exchange between the two quark legs are exactly zero. The only source
of small virtual gluon contributions is the interference between the
two diagrams with exchanged tagging jets, but with hardly any common
phase space. Just like virtual gluon exchange central, real gluon
emission between the two tagging jets is strongly suppressed. QCD
radiation only occurs in the direction of the incoming and outgoing
quark legs~\cite{jetveto1}. 

Alternatively, we can study the jet radiation pattern by counting
additional jets. For any kind of background process the large value of
$m_{jj}$ generates an enhanced probability for central jet radiation,
leading to a Poisson distribution in the number of jets. For the WBF
signal the large values of $m_{jj}$ are natural, so the radiation
pattern remains staircase, with a significantly reduced radiation
probability for the first few emissions~\cite{jetveto2}.

Independent of the physics picture, the reduced radiation of
central jets is usually exploited by a central jet veto above a $p_T$
threshold around 20~GeV. The question is if we can make use of the
geometry of these additional jets instead of or before throwing them
away~\cite{durham}.\bigskip

Just like in Sec.~\ref{sec:tagging} we show a set of Fox--Wolfram
moments for the Higgs signal and the two backgrounds. The only
difference between Fig.~\ref{fig:all_moments_qcd} and
Fig.~\ref{fig:all_moments} is that instead of only the tagging jets we
now include all jets passing Eq.\eqref{eq:jj1} in the definition
of the moments, Eq.\eqref{eq:fwm_def3}.  All events pass the
$m_{jj}^\text{min}$ cut in Tab.~\ref{tab:CFclassic}.  Again, we show
the two different weights $H_\ell^p$ and $H_\ell^T$. The Fox--Wolfram
moments based on all jets are not very different from those based on
the tagging jets only, except that for even moments the additional
jets can in principle populate $H^T_\ell<0.3$ due to the central jet
activity as discussed in the toy example for three planar jets
at the end of Sec.~\ref{sec:moments}. Specifically
for the Higgs signal, the majority of events do not even have an
additional hard jet, so the two distributions are largely
identical. Even moments are still largely limited to $H^T_\ell > 0.3$
and show a clear peak towards $H^T_\ell \sim 1$. However, for $Z+$jets
some moments change, starting with $H^T_4$ and giving very visible
differences between the Higgs signal and the $Z$ background for
$H^T_8$ or $H^T_{10}$. This is not entirely unexpected, because a veto
on additional jets can also distinguish these two channels. Top pair
production with three relatively hard jets, two from the top decays
and one from QCD radiation, shows a distinct peak for example around
$H^T_8 \sim 0.3$.\bigskip

\begin{figure}[t]
\includegraphics[width=0.24\textwidth]{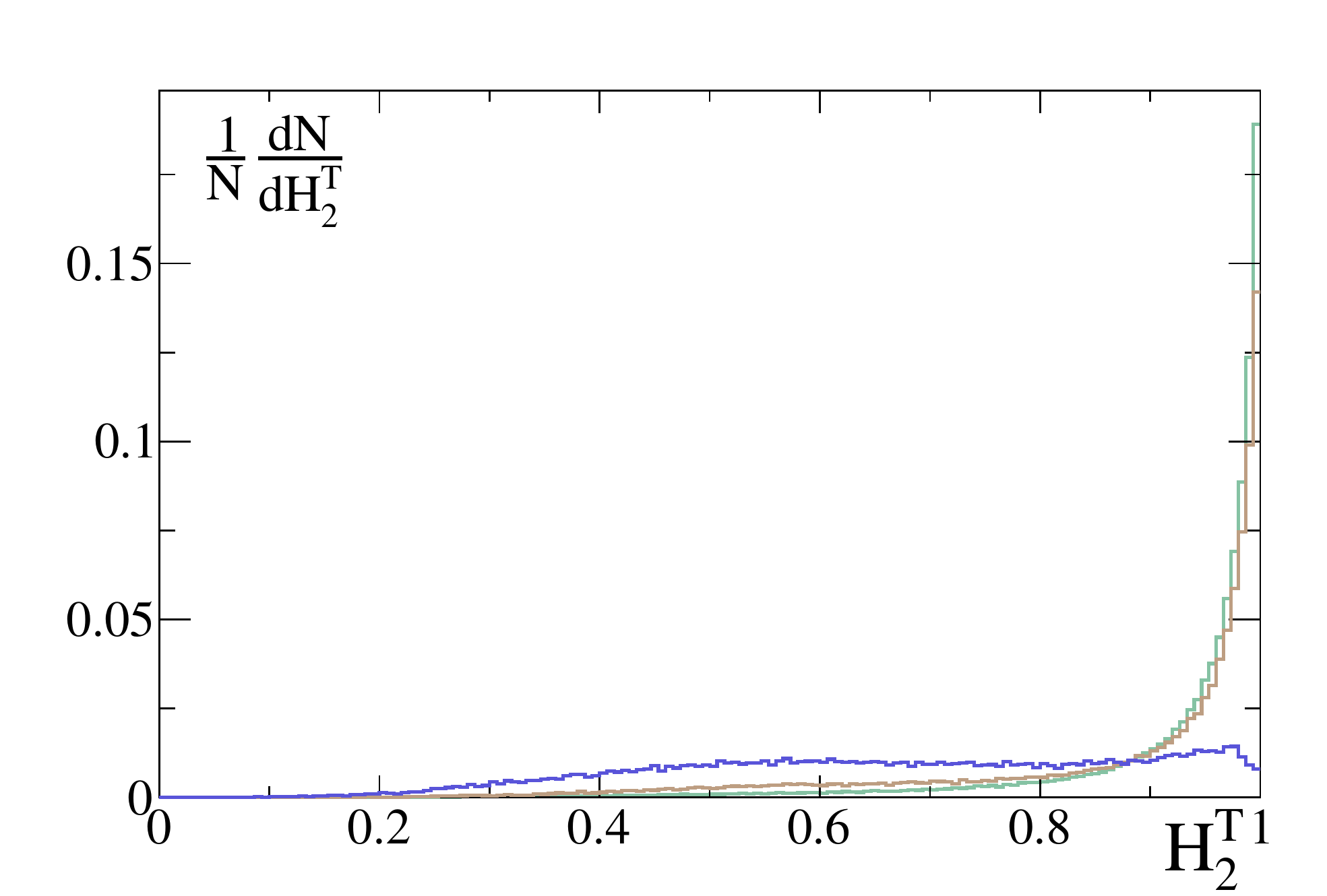} 
\includegraphics[width=0.24\textwidth]{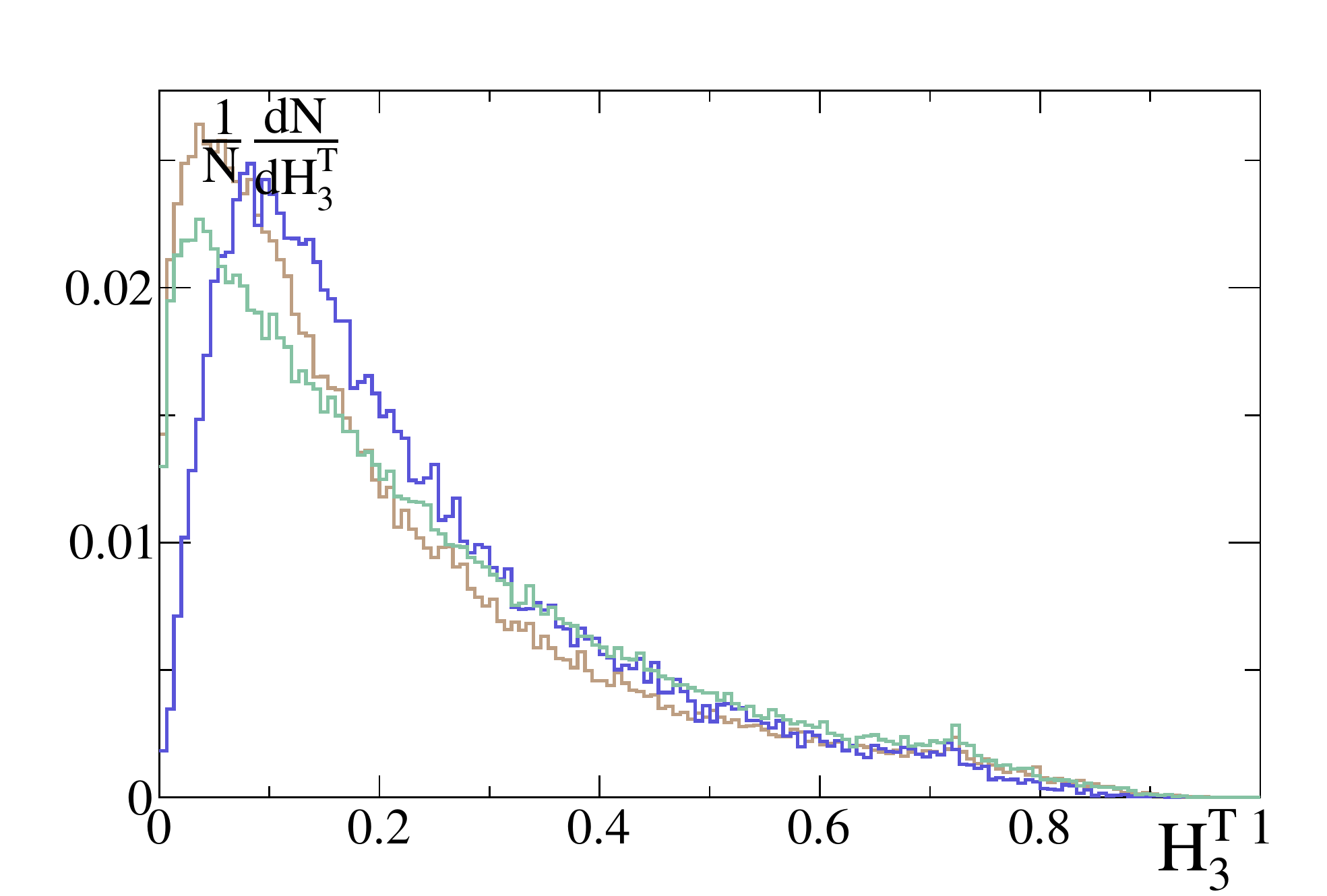} 
\includegraphics[width=0.24\textwidth]{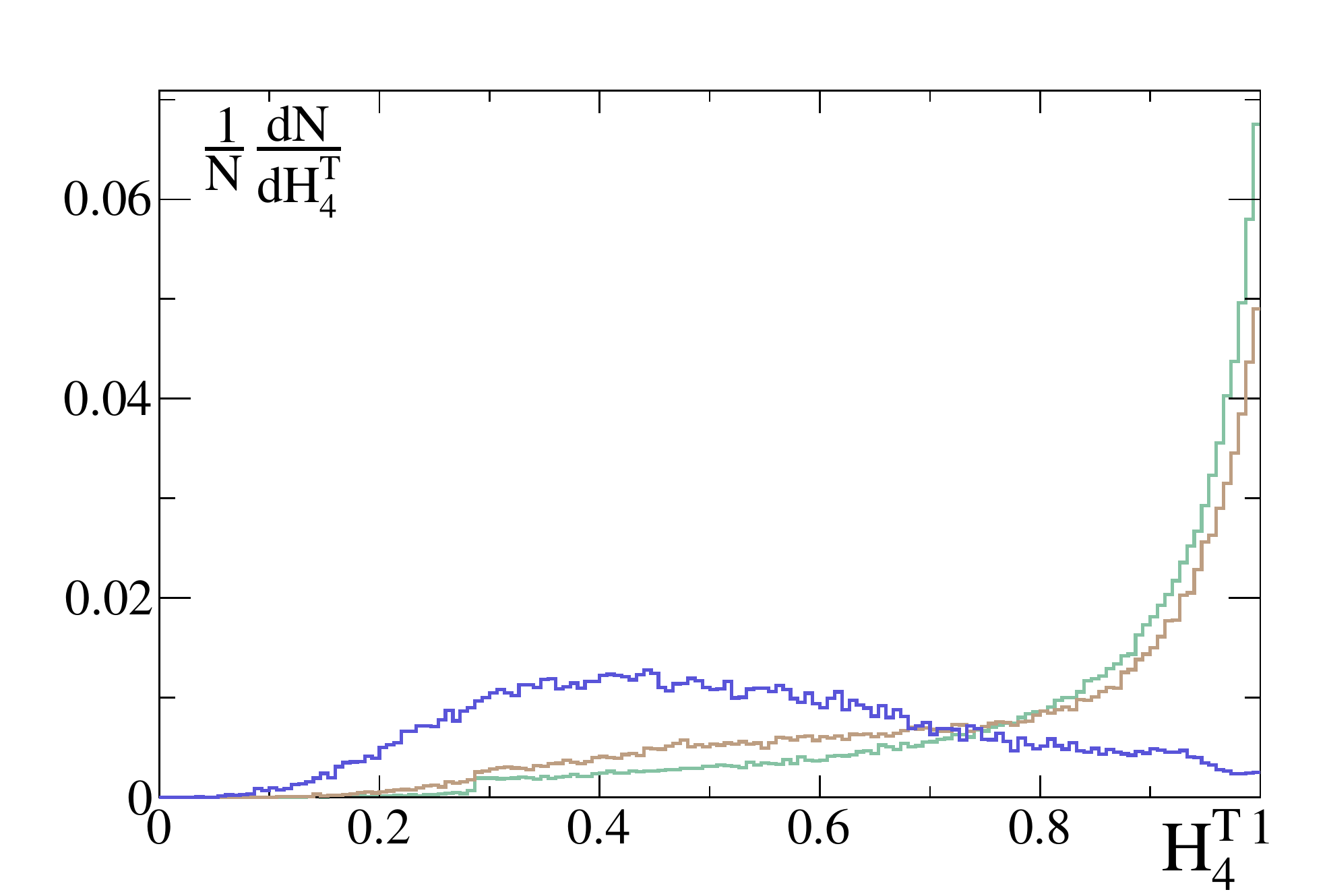} 
\includegraphics[width=0.24\textwidth]{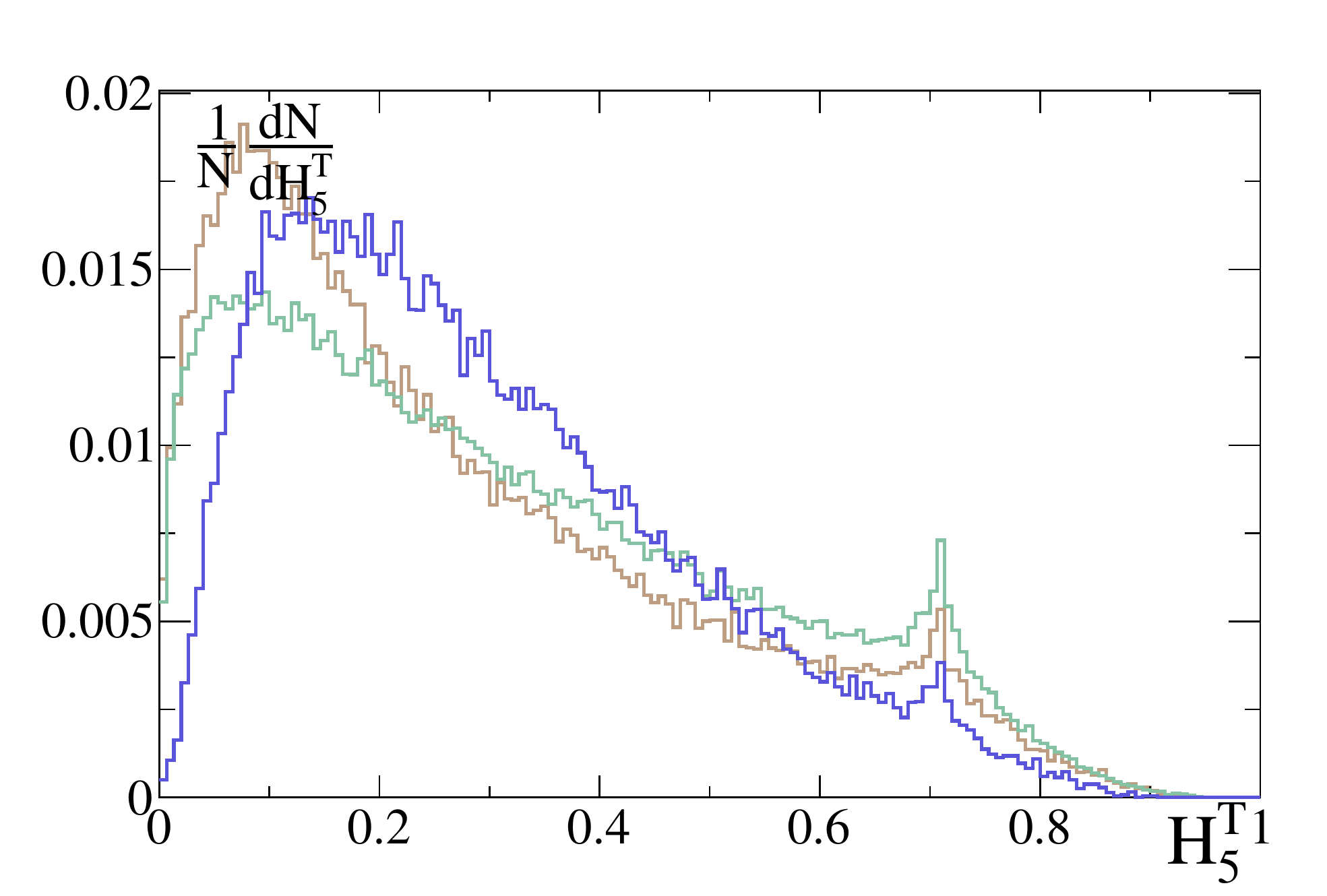} \\
\includegraphics[width=0.24\textwidth]{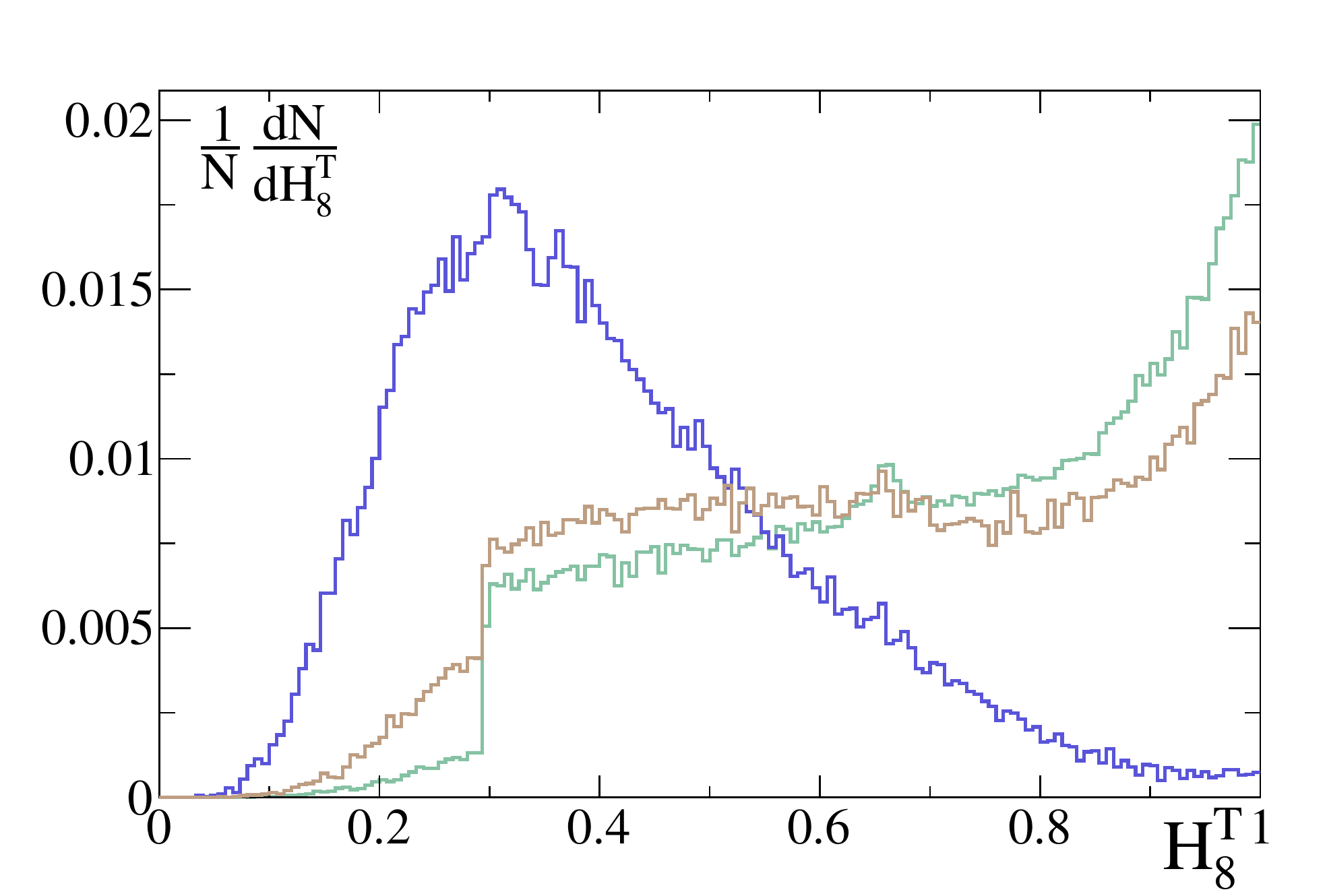} 
\includegraphics[width=0.24\textwidth]{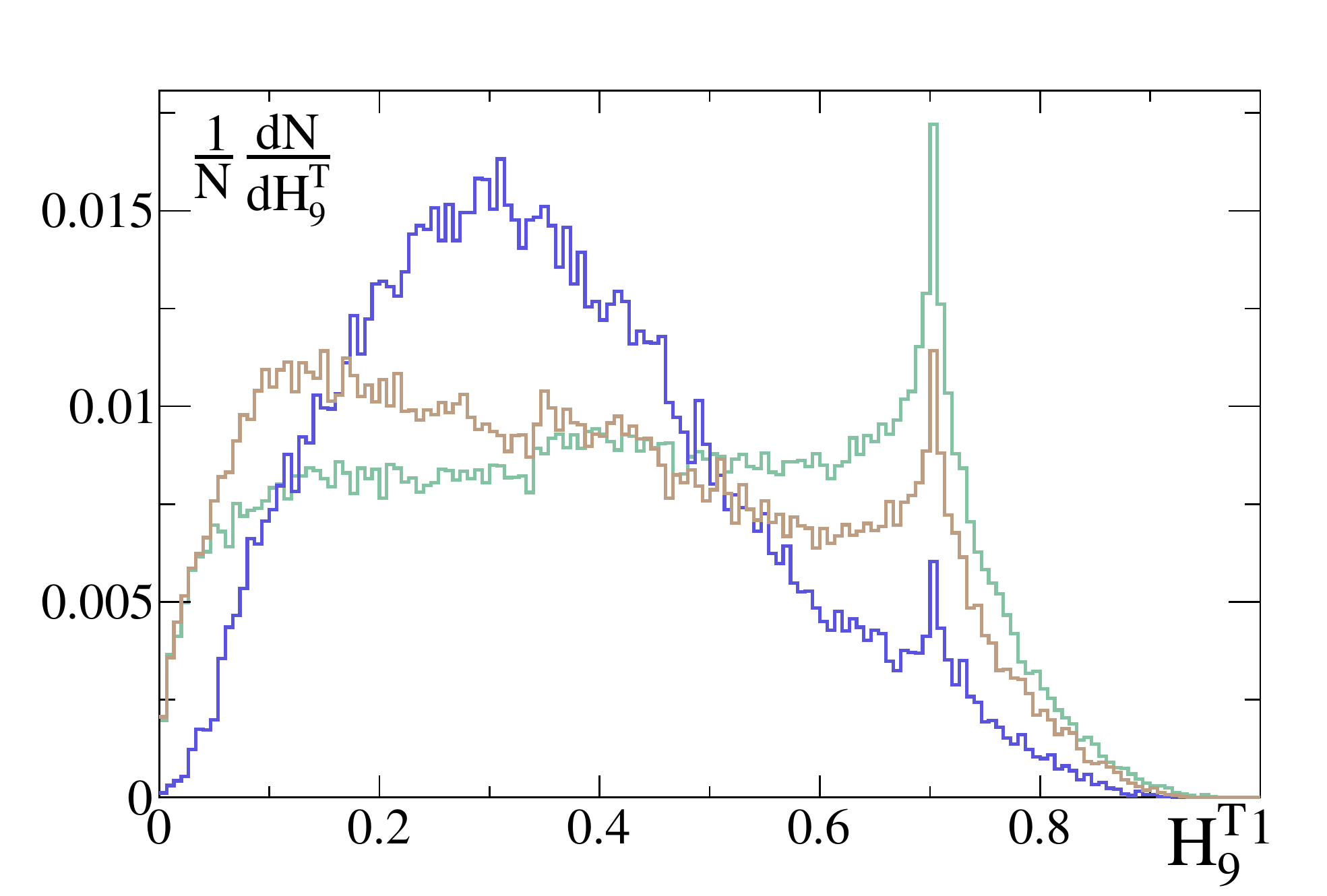} 
\includegraphics[width=0.24\textwidth]{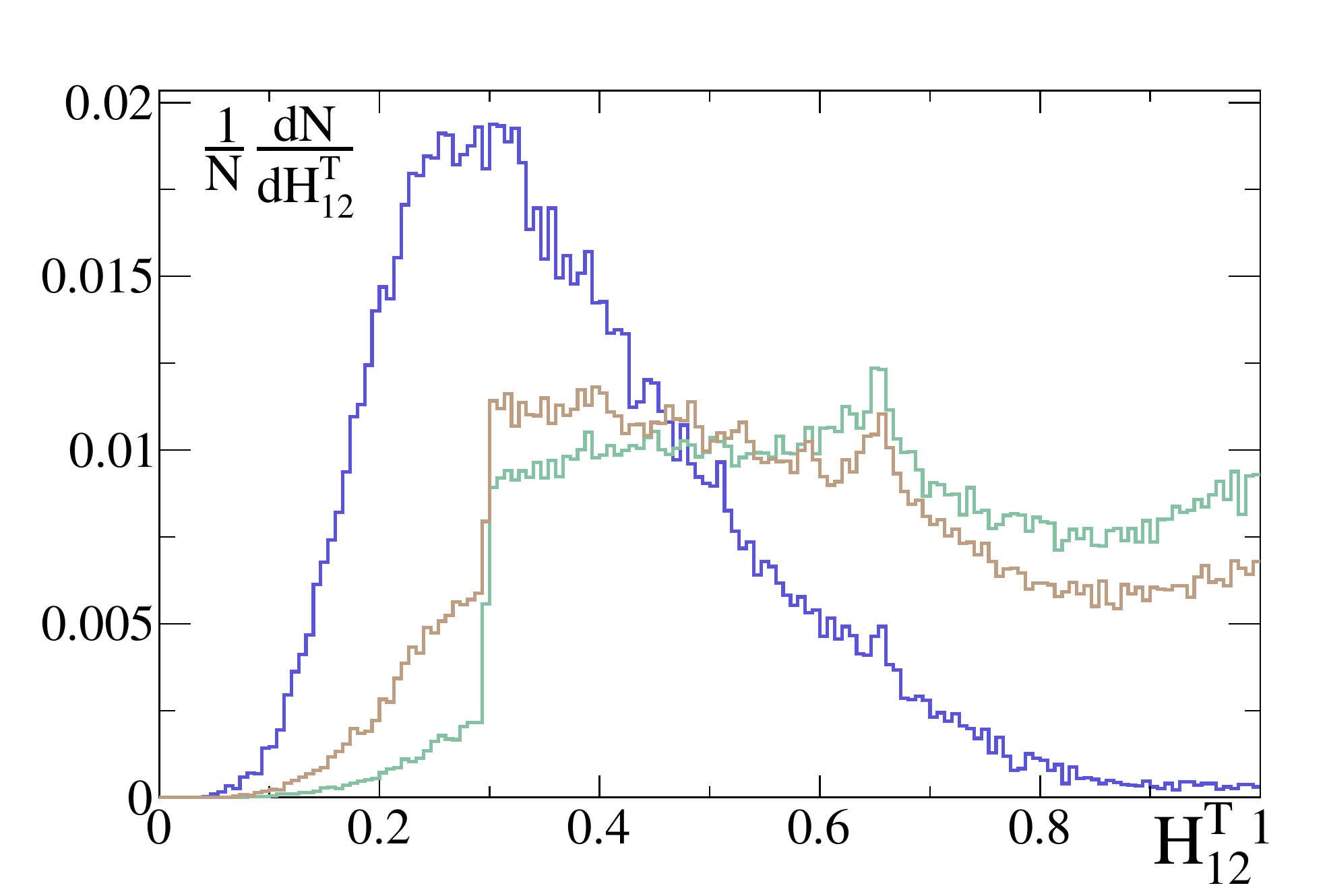} 
\includegraphics[width=0.24\textwidth]{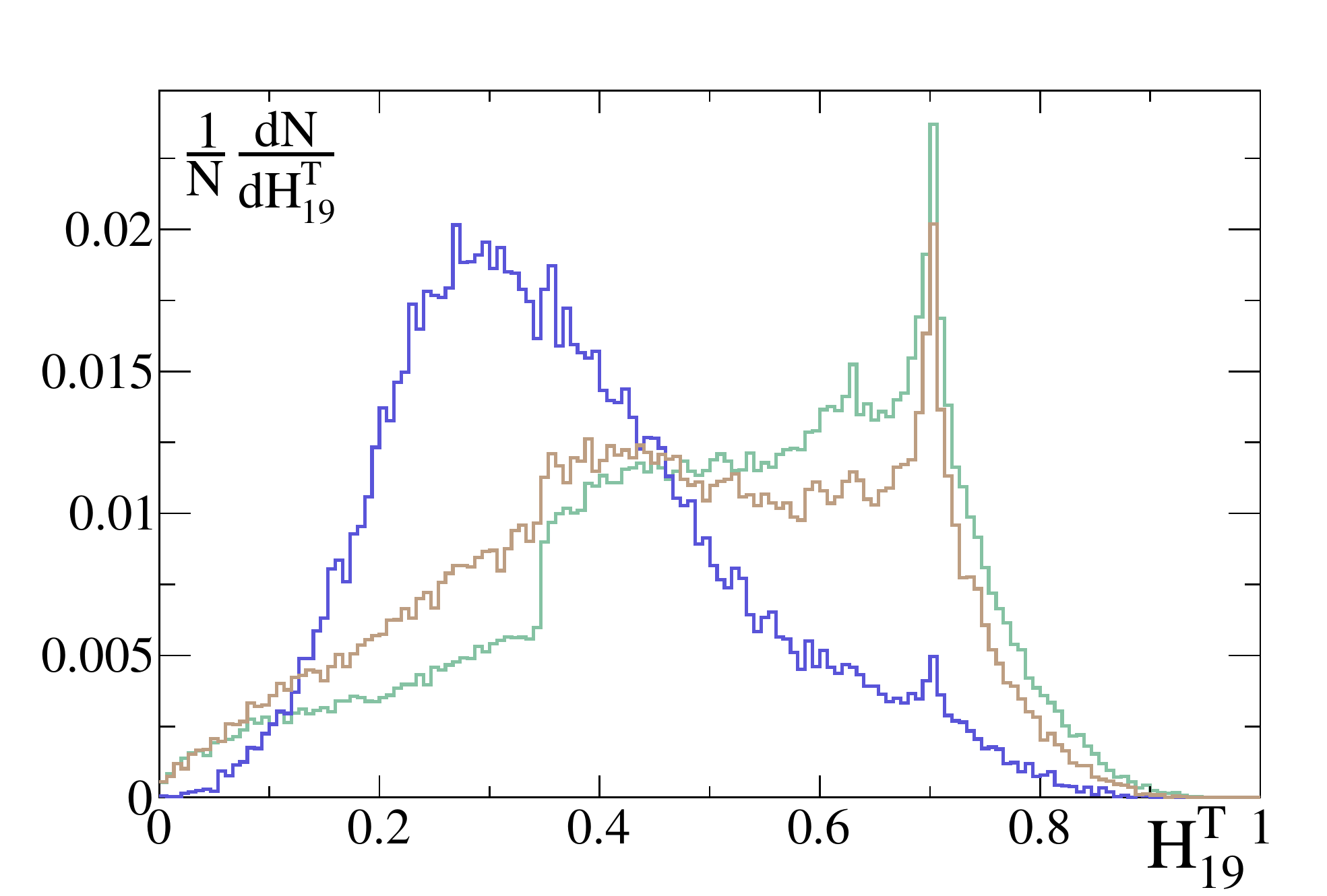} 
\caption{Normalized distributions of Fox--Wolfram moments computed
  from all jets for $\ell = 2-5,8,9,12,19$ with a weight factor
  $W_{ij}^T$ for WBF H+1 jet signal (green), Z+2 jets 
  (brown) and $t\bar{t}$+1 jet (blue). All events pass the full set of QCD cuts,
  Eqs.\eqref{eq:jj1}-\eqref{eq:jj3}.}
\label{fig:all_moments_qcd2}
\end{figure}

From Sec.~\ref{sec:tagging} we know that Fox--Wolfram moments are not
sufficiently effective to replace the tricky $\Delta y_{jj}$ cut in
the standard WBF analyses. In Fig.~\ref{fig:all_moments_qcd2} we show
the same moments with the transverse weight factor including all WBF
cuts Eqs.\eqref{eq:jj1}-\eqref{eq:jj3}. As expected, the sensitivity
to the differences in $H+$jets and $Z+$jets production is
reduced. Jets radiated off the hard WBF process and QCD jet radiation
become very similar at this stage. The most noticeable difference is
that the $Z+$jets background tends towards smaller even moments
without the sharp edge around $H^T_\ell = 0.3$.  This effect is
numerically limited because the usual weights $W_{ij}$ in the
Fox--Wolfram moments penalize jets with low (transverse) momentum, so
soft additional jets have relatively little impact.  
 This is different for the hard top decay jets, so we see
that high even moments $H^T_8$ or $H^T_{12}$ can be used to remove
events where the addition non-tagging jets have a visible impact. A
simple cut on one of these moments can improve $S/B$ from the value
1/73 quoted in Tab.~\ref{tab:CFclassic} to 1/50, keeping the majority
of signal events.\bigskip

\begin{figure}[t]
\includegraphics[width=0.24\textwidth]{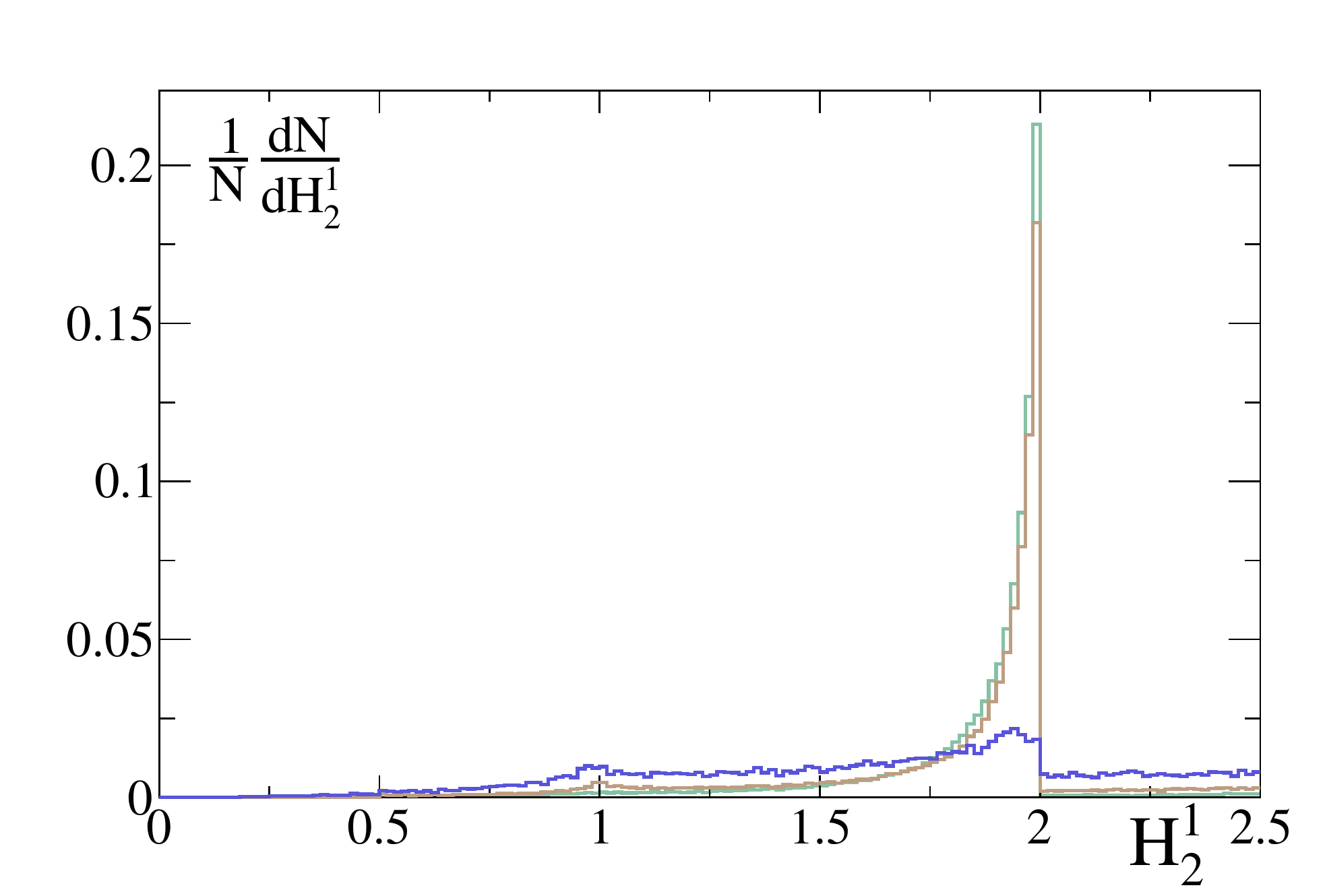} 
\includegraphics[width=0.24\textwidth]{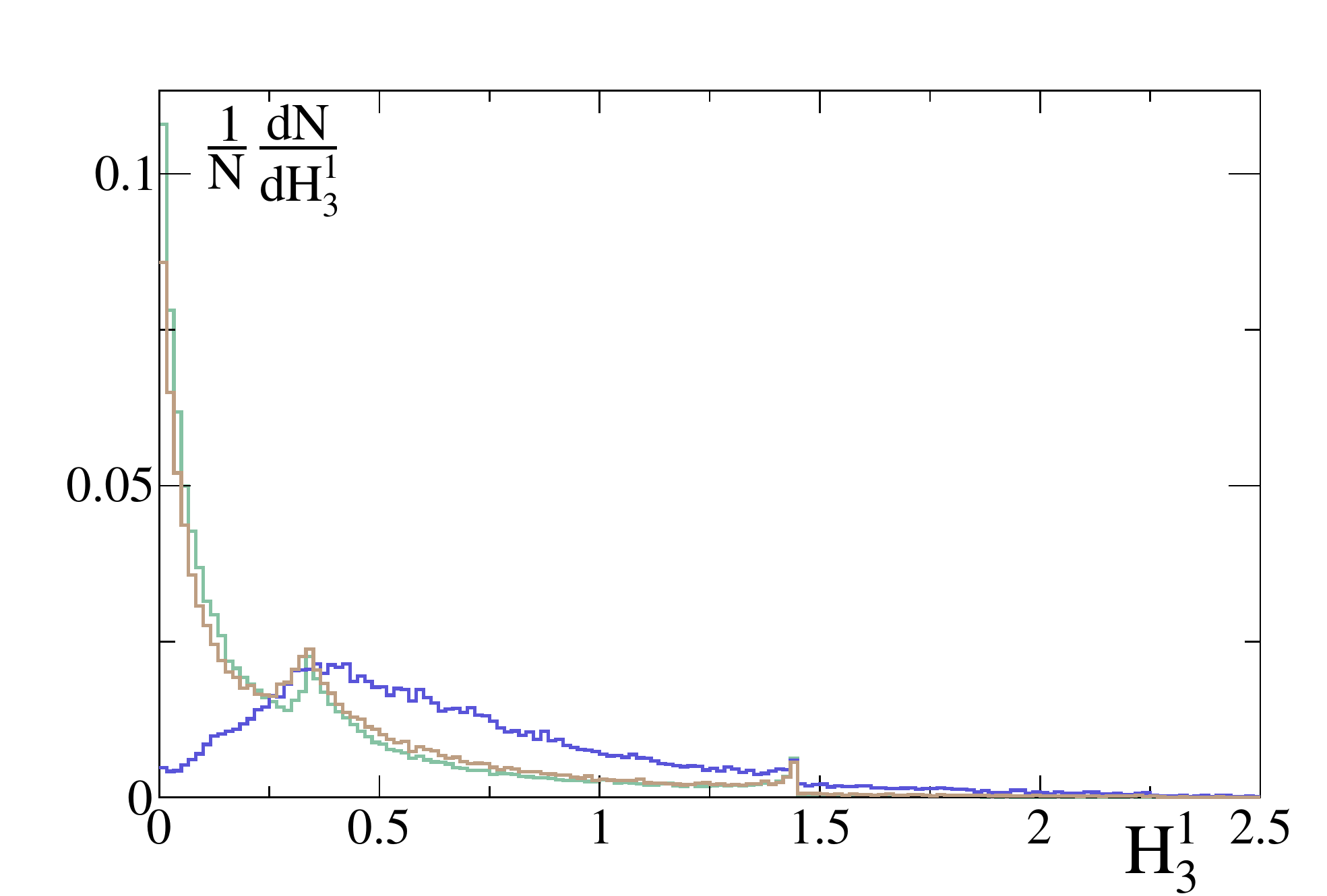} 
\includegraphics[width=0.24\textwidth]{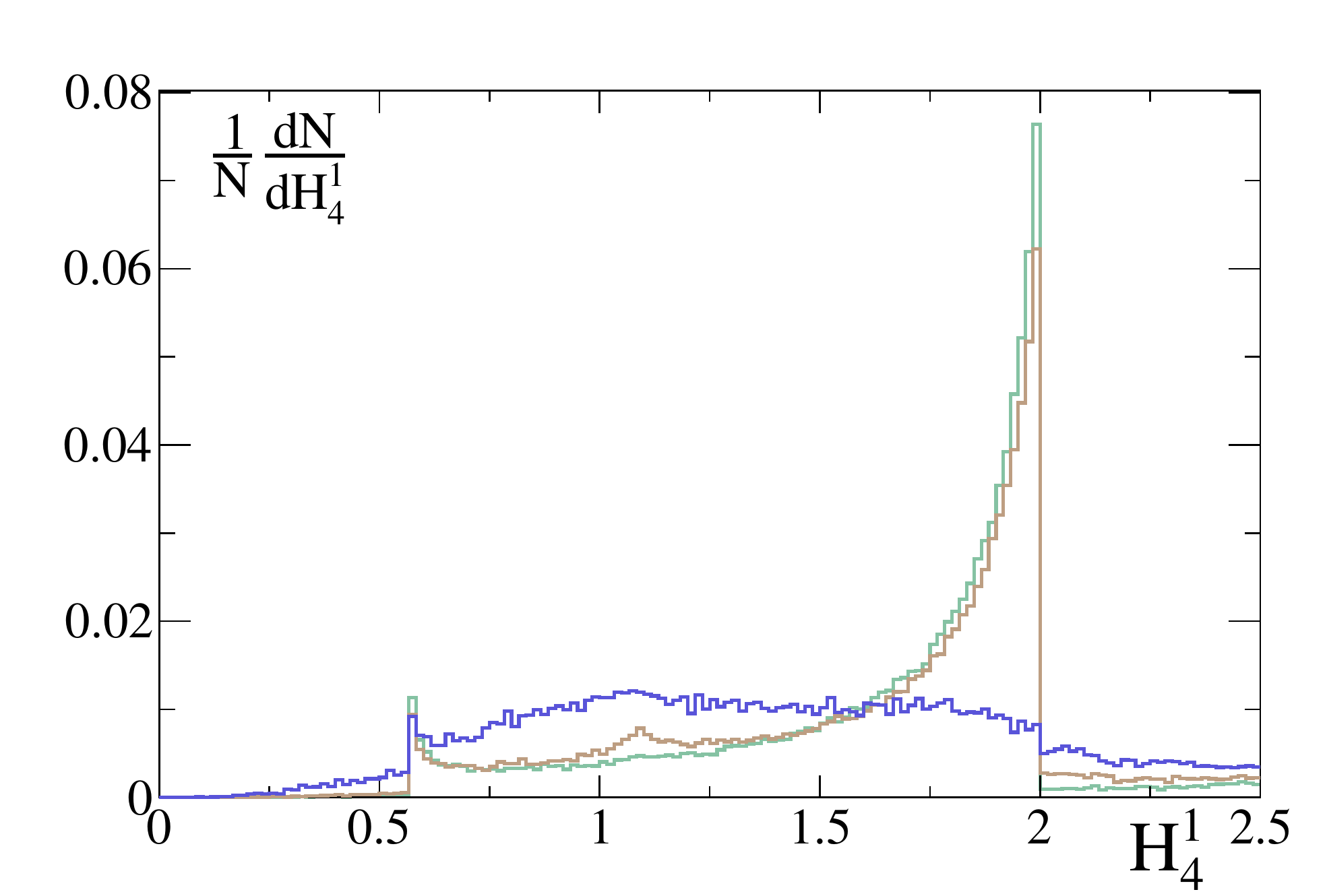} 
\includegraphics[width=0.24\textwidth]{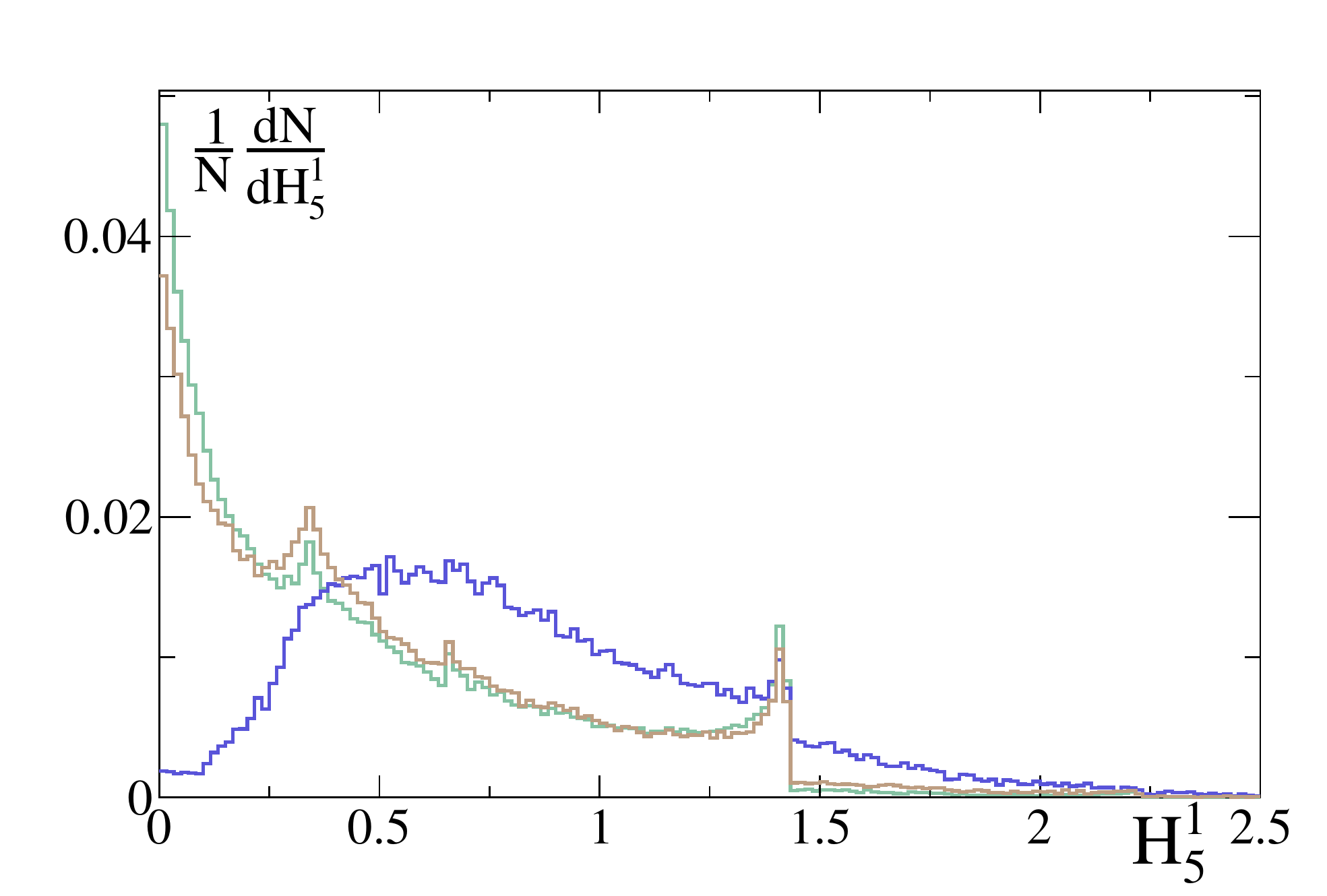} \\
\includegraphics[width=0.24\textwidth]{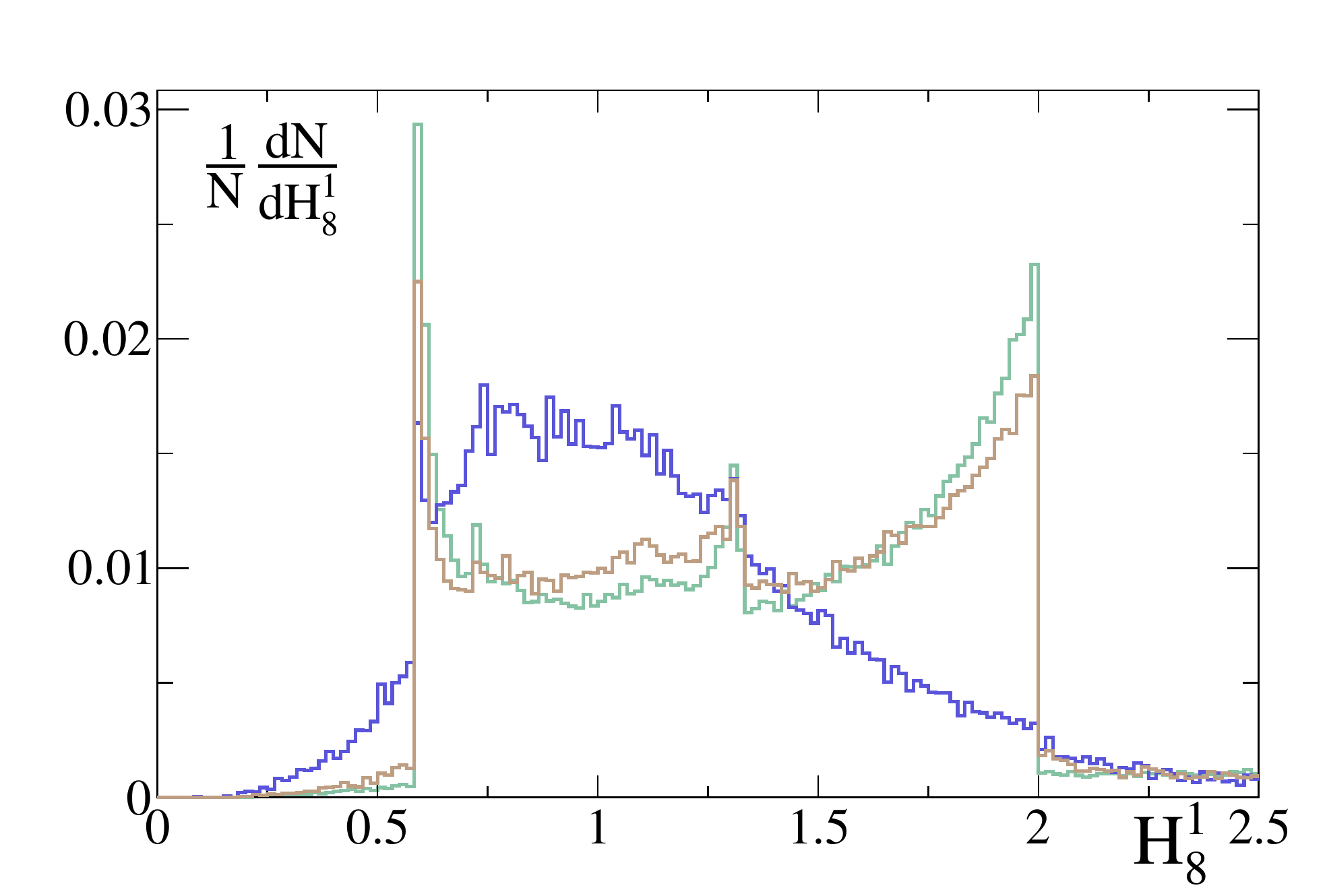} 
\includegraphics[width=0.24\textwidth]{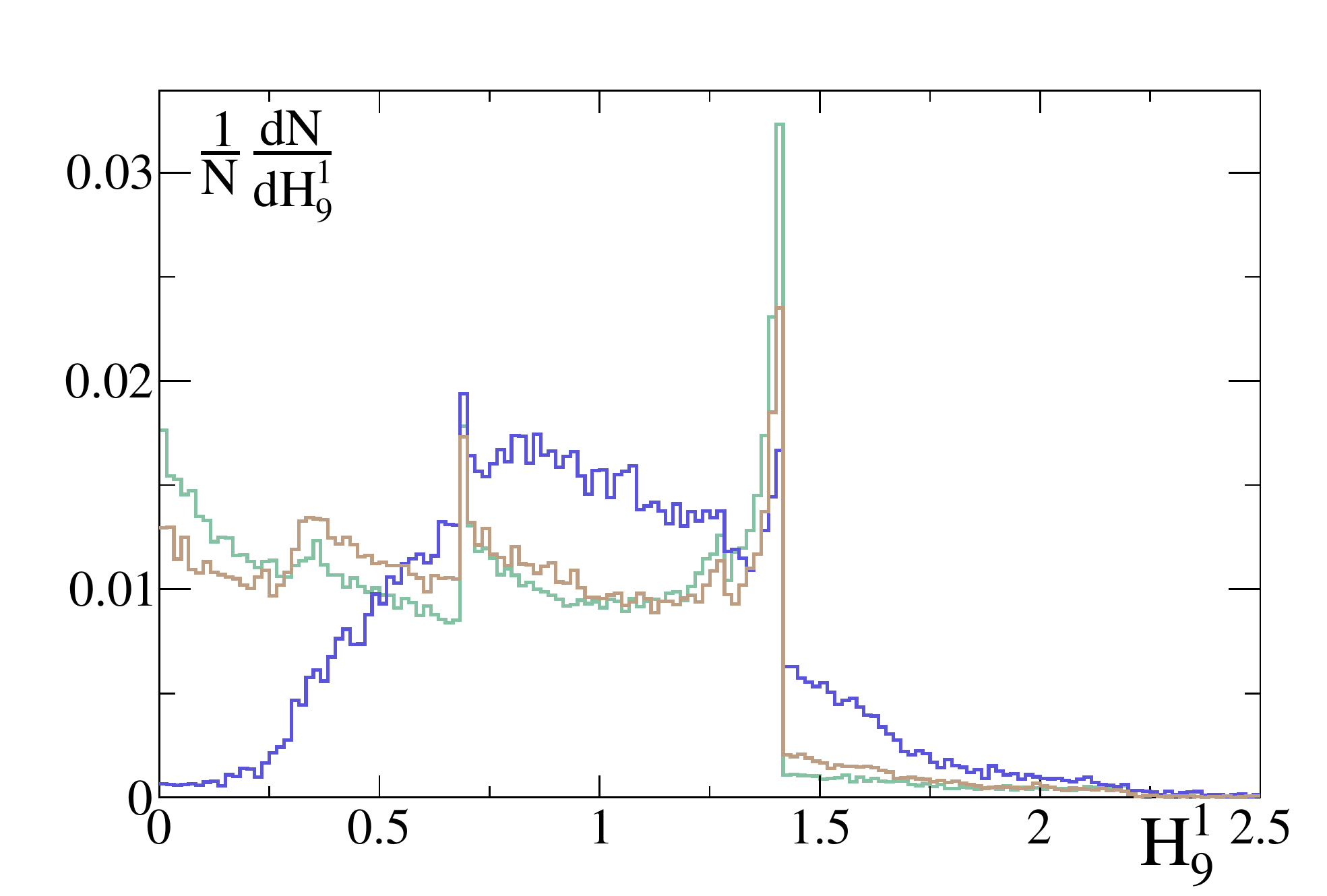} 
\includegraphics[width=0.24\textwidth]{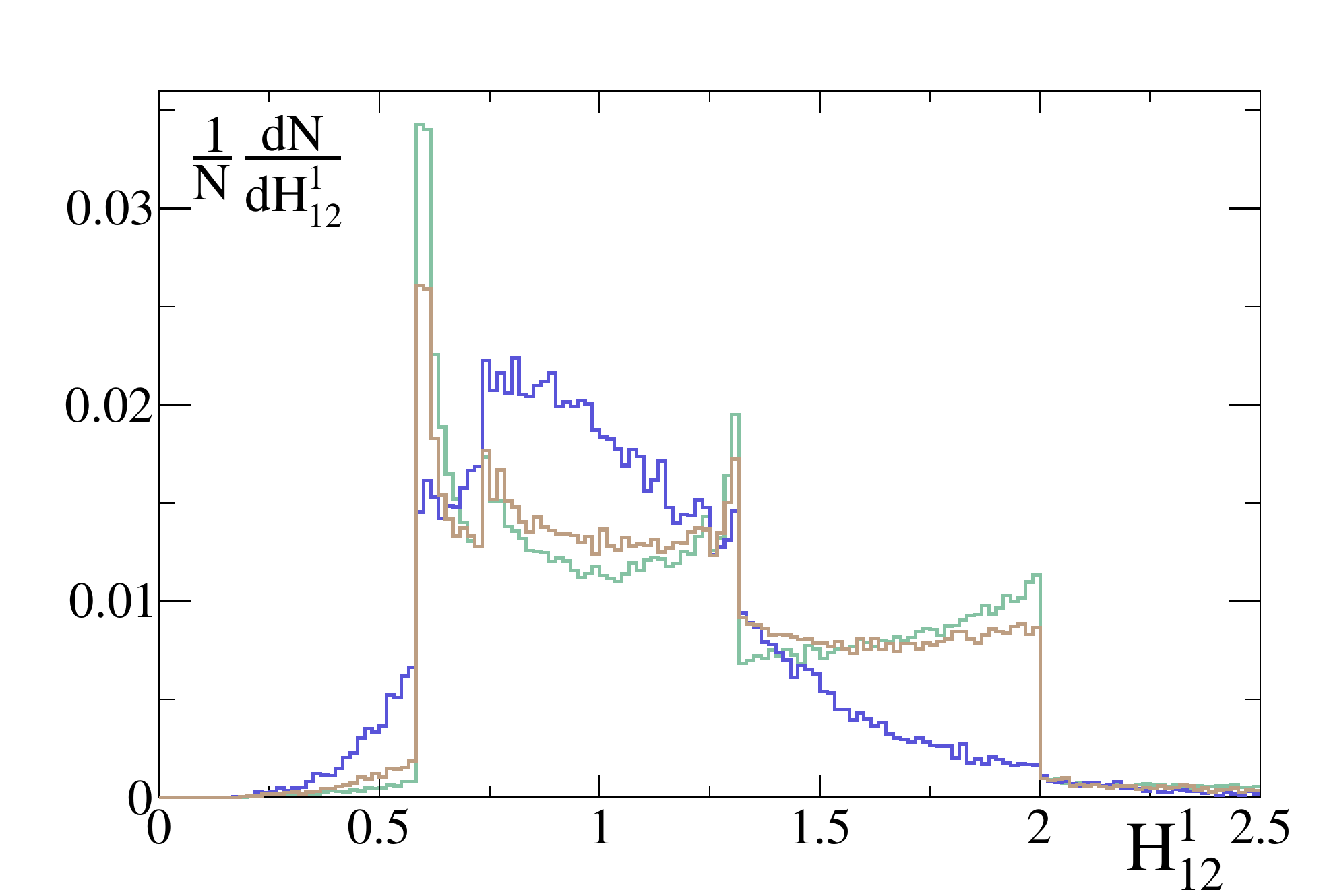} 
\includegraphics[width=0.24\textwidth]{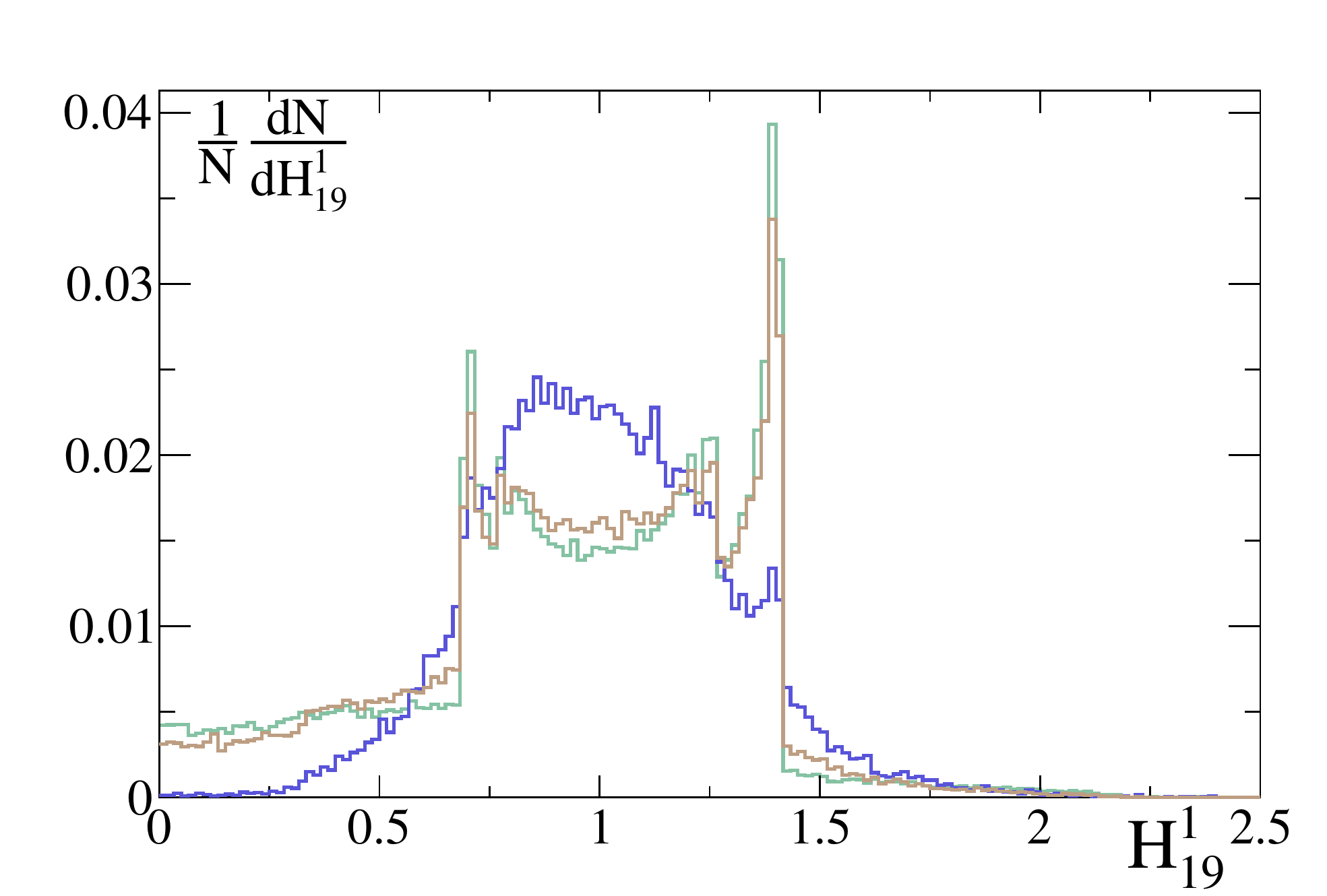} 
\caption{Normalized distributions of Fox--Wolfram moments computed
  from all jets for $\ell = 2-5,8,9,12,19$ with weight factor
  $W_{ij}^1$ for WBF H+1 jet signal (green), Z+2 jets 
  (brown) and $t\bar{t}$+1 jet (blue). All events pass the full set of QCD cuts,
  Eqs.\eqref{eq:jj1}-\eqref{eq:jj3}.}
\label{fig:all_moments_qcd3}
\end{figure}

To understand further how the low (transverse) momentum jets are
restricted by the presence of momentum dependent weights, we show
moments with unit weight $W^1_{ij}$ in
Fig.~\ref{fig:all_moments_qcd3}.  The most noticeable difference in
comparison to Fig.~\ref{fig:all_moments_qcd2} are the sharper peaks in
the $H+$jets and $Z+$jets distributions due to the uninhibited
presence of softer jet radiation.  This can be better understood in
the $H+$jets and $Z+$jets case, where there are typically only two
jets passing the full set of QCD cuts of
Eq.\eqref{eq:jj1}-\eqref{eq:jj3}, by referring to the two-jet toy model
of Sec.~\ref{sec:moments} in the case $r = 1$.  Here we see for
example that the sharp peaks at low $H^1_\ell$ correspond to small total
angles between the jets which are not dampened with a choice of unit
weight.

It is also worth emphasizing that because the moments are not defined
with respect to a preferred axis, they cannot select jets in a
specific region. Hence, the moments cannot be used to, for instance,
emphasize central jets in the absence of forward-backward jet
activity.  We verified this using $W^y_{ij}$ of Eq.\eqref{eq:weights}  as a weight that favors low
rapidity jets. While $H^T_\ell$ cannot replace a jet veto altogether,
it certainly includes useful information which will improve the WBF
analysis. As shown in Fig.~\ref{fig:moments_corr}, Fox--Wolfram
moments should be included as a somewhat correlated set of new
observables.

\section{Outlook}

In this first study we have shown that Fox--Wolfram
moments~\cite{fwm_orig} based on jets are very useful tools to improve
many LHC analyses benefiting from information about the QCD structure
of signal and background events. Our example process is
weak-boson-fusion Higgs production with tagging jets described by the
hard matrix element. Two leading backgrounds are $Z+$jets where the
tagging jets come form QCD radiation and top pair production, where at
least one of the tagging jets will be a top decay jet.\bigskip

For weak boson fusion the key questions are how much information we
can extract from the Fox--Wolfram moments before we apply a cut on the
rapidity separation of the two tagging jets or before we apply a jet
veto. We have shown that moments either based on the two tagging jets
alone or based on all jets in the event show distinctly different
features for the three signal and background processes.  We have
tested different weights entering the definition of the Fox--Wolfram
moments in Eq.~\eqref{eq:fwm_def3}. At least for the tagging jets a
weight based on transverse momenta is the most useful~\cite{field}. On
the other hand, for a study of the jet activity, alternative weights
might be helpful.  In addition, we have seen that the full set of odd
and even moments cannot be reduced to one or two representative
moments; they should be considered as a new class of correlated but
individually useful LHC observables.\bigskip

Clearly, we did not present a conclusive final study on Fox--Wolfram
moments at the LHC. Many aspects can and have to be improved, from the
physical objects entering the moments to the choice of weights or an
exhaustive study of their correlations. In an era where we are
becoming more and more confident in exploiting QCD features for many
LHC analyses Fox--Wolfram moments have the potential to play a key
role as universal analysis tools.

\acknowledgments

We would like to thank Peter Schichtel for
his technical help at many different stages of this project.



\begin{thebibliography}{99}

\bibitem{fwm_orig}
 G.~C.~Fox and S.~Wolfram,
  Phys.\ Rev.\ Lett.\  {\bf 41}, 1581 (1978).

\bibitem{field}
 R.~D.~Field, Y.~Kanev and M.~Tayebnejad,
  Phys.\ Rev.\ D {\bf 55}, 5685 (1997).

\bibitem{higgs}
 P.~W.~Higgs,
  Phys.\ Lett.\  {\bf 12}, 132 (1964);
 P.~W.~Higgs,
  Phys.\ Rev.\ Lett.\  {\bf 13}, 508 (1964);
 F.~Englert and R.~Brout,
  Phys.\ Rev.\ Lett.\  {\bf 13}, 321 (1964).

\bibitem{review}
 for LHC reviews \eg
 A.~Djouadi,
  Phys.\ Rept.\  {\bf 457}, 1 (2008);
 T.~Plehn,
  Lect.\ Notes Phys.\  {\bf 844}, 1 (2012)
  [arXiv:0910.4182 [hep-ph]].

\bibitem{jamie}
 D.~E.~Morrissey, T.~Plehn and T.~M.~P.~Tait,
  Phys.\ Rept.\  {\bf 515}, 1 (2012);
  H.~Dreiner, M.~Kr\"amer and J.~Tattersall,
  arXiv:1211.4981 [hep-ph].

\bibitem{atlas}
 G.~Aad {\it et al.}  [ATLAS Collaboration],
  Phys.\ Lett.\ B  {\bf{716}}, 1 (2012).

\bibitem{cms}
  S.~Chatrchyan {\it et al.}  [CMS Collaboration],
  Phys.\ Lett.\ B  {\bf{716}}, 30 (2012).

\bibitem{wbf_w}
 N.~Kauer, T.~Plehn, D.~Rainwater and D.~Zeppenfeld,
  Phys.\ Lett.\ B {\bf 503}, 113 (2001).

\bibitem{wbf_tau}
 D.~L.~Rainwater, D.~Zeppenfeld and K.~Hagiwara,
  Phys.\ Rev.\  D {\bf 59}, 014037 (1999);
 T.~Plehn, D.~L.~Rainwater and D.~Zeppenfeld,
  Phys.\ Rev.\  D {\bf 61}, 093005 (2000).

\bibitem{wbf_gamma}
 D.~L.~Rainwater and D.~Zeppenfeld,
  JHEP {\bf 9712}, 005 (1997);
 J.~R.~Andersen, C.~Englert and M.~Spannowsky,
  arXiv:1211.3011 [hep-ph].

\bibitem{tagging}
 R.~Kleiss and W.~J.~Stirling,
  Phys.\ Lett.\ B {\bf 200}, 193 (1988);
 U.~Baur and E.~W.~N.~Glover,
  Phys.\ Lett.\ B {\bf 252}, 683 (1990);
 V.~D.~Barger, K.~Cheung, T.~Han, J.~Ohnemus and D.~Zeppenfeld,
  Phys.\ Rev.\ D {\bf 44}, 1426 (1991).

\bibitem{jetveto1}
 D.~L.~Rainwater, R.~Szalapski and D.~Zeppenfeld,
  Phys.\ Rev.\ D {\bf 54}, 6680 (1996).

\bibitem{jetveto2}
 E.~Gerwick, T.~Plehn and S.~Schumann,
  Phys.\ Rev.\ Lett.\  {\bf 108}, 032003 (2012).

\bibitem{manchester}
 B.~E.~Cox, J.~R.~Forshaw and A.~D.~Pilkington,
  Phys.\ Lett.\ B {\bf 696}, 87 (2011).

\bibitem{event_shapes}
 for a nice overview of LHC applications see \eg
 A.~Banfi, G.~P.~Salam and G.~Zanderighi,
  JHEP {\bf 1006}, 038 (2010).

\bibitem{pythia}
 T.~Sjostrand, L.~Lonnblad, S.~Mrenna and P.~Z.~Skands,
  hep-ph/0308153.

\bibitem{wbf_ex}
 S.~Asai, G.~Azuelos, C.~Buttar, V.~Cavasinni, D.~Costanzo, K.~Cranmer, R.~Harper and K.~Jakobs {\it et al.},
  Eur.\ Phys.\ J.\ C {\bf 32S2}, 19 (2004).

\bibitem{sherpa}
 T.~Gleisberg, S.~H\"oche, F.~Krauss, M.~Sch\"onherr, S.~Schumann, F.~Siegert and J.~Winter,
  JHEP {\bf 0902}, 007 (2009).

\bibitem{ckkw} 
 S.~Catani, F.~Krauss, R.~Kuhn and B.~R.~Webber,
  JHEP {\bf 0111}, 063 (2001).

\bibitem{fastjet}
 M.~Cacciari, G.~P.~Salam and G.~Soyez,
  Eur.\ Phys.\ J.\ C {\bf 72}, 1896 (2012).

\bibitem{spin_wbf}
 T.~Plehn, D.~L.~Rainwater and D.~Zeppenfeld,
  Phys.\ Rev.\ Lett.\  {\bf 88}, 051801 (2002);
 C.~Ruwiedel, N.~Wermes and M.~Schumacher,
  Eur.\ Phys.\ J.\ C {\bf 51}, 385 (2007);
 C.~Englert, D.~Goncalves-Netto, K.~Mawatari and T.~Plehn,
  arXiv:1212.0843 [hep-ph].

\bibitem{higgs_couplings}
 see \eg  
 M.~Klute, R.~Lafaye, T.~Plehn, M.~Rauch and D.~Zerwas,
  Phys.\ Rev.\ Lett.\  {\bf 109}, 101801 (2012).

\bibitem{durham}
 for a similar idea see \eg
 C.~Englert, M.~Spannowsky and M.~Takeuchi,
  JHEP {\bf 1206}, 108 (2012).

\end{thebibliography}
\end{document}